\documentclass[a4paper,11pt]{article}
\usepackage[T2A]{fontenc}
\usepackage[utf8]{inputenc}
\usepackage[english]{babel}
\usepackage{amsmath}
\usepackage{amsfonts}
\usepackage{amssymb}
\usepackage{wasysym}
\usepackage{amsthm}
\usepackage{tikz}
\usepackage{tikz-cd}
\usetikzlibrary{decorations.markings}
\usetikzlibrary{positioning}
\usetikzlibrary{patterns,patterns.meta}
\usepackage{braids}
\usepackage{array}
\usepackage{ytableau}
\usepackage{longtable}
\usepackage{empheq}
\usepackage{multicol}
\usepackage{mathrsfs}
\usepackage{amssymb}
\usepackage{diagbox}
\usepackage{color}
\usepackage{physics}
\usepackage{ytableau}
\usepackage{cite}

\usepackage[most]{tcolorbox}

\usepackage{pdflscape}
\usepackage{array, longtable}
\newcolumntype{C}{>{$}c<{$}}

\usepackage{pb-diagram}
\usepackage{setspace}
\usepackage[hidelinks,colorlinks=true,unicode]{hyperref}
\hypersetup{
	linkcolor=teal,
	citecolor=teal,
	filecolor=magenta,
	urlcolor=teal,
}
\usepackage{url}
\usepackage[left=2.3cm,right=2.3cm,top=2.3cm,bottom=2.3cm,bindingoffset=0cm]{geometry}

\begin{document}
	
	\hfill MIPT/TH-6/24
	
	\hfill ITEP/TH-7/24
	
	\hfill IITP/TH-6/24
	
	\vskip 1.5in
	\begin{center}
		
		{\bf\Large Towards construction of superintegrable basis in matrix models}
		
		\vskip 0.2in
		\renewcommand{\thefootnote}{\fnsymbol{footnote}}
		{Batukhan Azheev$^{1}$\footnote[1]{e-mail: azheev.bd@phystech.edu} and Nikita Tselousov$^{1,2,3,4}$\footnote[2]{e-mail: tselousov.ns@phystech.edu}}
		\vskip 0.2in
		\renewcommand{\thefootnote}{\roman{footnote}}
		{\small{
				\textit{$^1$MIPT, 141701, Dolgoprudny, Russia}
				\vskip 0 cm
				\textit{$^2$NRC “Kurchatov Institute”, 123182, Moscow, Russia}
				\vskip 0 cm
				\textit{$^3$IITP RAS, 127051, Moscow, Russia}
				\vskip 0 cm
				\textit{$^4$ITEP, Moscow, Russia}
		}}
	\end{center}
	
	\vskip 0.2in
	\baselineskip 16pt
	
	\centerline{ABSTRACT}

	\bigskip
	
	{\footnotesize
		We develop methods for systematic construction of superintegrable polynomials in matrix/eigenvalue models. Our consideration is based on a tight connection of superintegrable property of Gaussian Hermitian model and $W_{1 + \infty}$ algebra in Fock representation. Motivated by this example, we propose a set of assumptions that may allow one to recover superintegrable polynomials. The main two assumptions are box adding/removing rule (Pierri rule) and existence of Hamiltonian for superintegrable polynomials. We detail our method in case of the Gaussian Hermitian model, and then apply it to the cubic Kontsevich model.
	}
	
	\bigskip
	
	\bigskip

	\ytableausetup{boxsize = 0.5em}
	
	\section{Introduction}
	\label{sec::intro}
	
	Matrix models \cite{Morozov:1994hh, Mironov:1993wi} are a rapidly developing area of modern mathematical physics. They represent zero-dimensional and therefore often exactly solvable examples of quantum field theories where one can study intriguing phenomena such as \textit{superintegrability} \cite{Mironov:2022fsr}:
	\begin{equation}
		\langle \, \chi \, \rangle \sim \chi( * )
	\end{equation} 
	Remarkably, quantum correlators of distinguished basis in fields $\chi$ turn out to be expressed via simple functions, or even more, via functions $\chi (*)$ themselves in a specific point. The classic example of the superintegrability phenomenon is the simple Gaussian Hermitian one-matrix model \cite{Mironov:2017och} (GH model), where integrals are taken over $N \times N$ Hermitian matrices $H$:
	\begin{equation}
		\label{correlators HGM}
		\langle \, f \, \rangle = \frac{1}{Z}\int dH  \, f(H) \, \exp\left( - \frac{1}{2} \tr H^2\right)
	\end{equation}
	The normalization factor $Z = \int dH \exp\left( - \frac{1}{2} \tr H^2\right)$ is chosen to satisfy $\langle \, 1 \, \rangle = 1$. Superintegrable basis $\chi$ in this case is given by well-known characters of linear groups -- Schur polynomials $S_{\lambda}$ that enumerated by Young diagrams $\lambda = [\lambda_1, \lambda_2, \ldots, \lambda_{l(\lambda)}]$, where $\lambda_i$ are positive integers obeying the non-increasing rule $\lambda_1 \geqslant \lambda_2 \geqslant \lambda_3 \geqslant \ldots \geqslant \lambda_{l(\lambda)} \geqslant 0$. All basic facts about Schur polynomials are listed in \cite{Macdonald}. Example of Young diagram is presented in the following picture:
	\begin{figure}[h!]
		\centering
		\begin{tikzpicture}
			\node(B) at (6,0) {$\begin{array}{c}
					\begin{tikzpicture}[scale=0.5]
						\foreach \i/\j in {0/0, 1/0, 2/0, 3/0, 4/0, 5/0, 6/0, 0/-1, 1/-1, 2/-1, 3/-1, 4/-1, 5/-1, 6/-1, 0/-2, 1/-2, 2/-2, 3/-2, 4/-2, 5/-2, 0/-3, 1/-3, 2/-3, 3/-3, 0/-4, 1/-4, 0/-5, 0/-6}
						{
							\draw[thick] (\i,\j) -- (\i+1,\j);
						}
						\foreach \i/\j in {0/0, 0/-1, 0/-2, 0/-3, 0/-4, 0/-5, 1/0, 1/-1, 1/-2, 1/-3, 1/-4, 1/-5, 2/0, 2/-1, 2/-2, 2/-3, 3/0, 3/-1, 3/-2, 4/0, 4/-1, 4/-2, 5/0, 5/-1, 6/0, 6/-1, 7/0}
						{
							\draw[thick] (\i,\j) -- (\i,\j-1);
						}
						\foreach \i/\j in {}
						{
							\draw[thick] (\i,\j) -- (\i-1,\j-1);
						}
					\end{tikzpicture}
				\end{array}$};
			\node[right] at (B.east) {$=\left[ 7, 6, 4, 2, 1, 1 \right]$};
			\node[left] at (B.west) {$\lambda = $};
		\end{tikzpicture}
		\label{fig:crystalYoung}
	\end{figure}
	
	Explicit superintegrable formula for this model reads:
	\begin{tcolorbox}
		\begin{equation}
			\langle \, S_{\lambda} \, \rangle = \frac{S_{\lambda}(N)}{S_{\lambda}(\delta_{k,1})} \cdot S_{\lambda} \left( \delta_{k,2}\right)
			\label{SI formula for GHM}
		\end{equation}
	\end{tcolorbox}
	where we understand Schur polynomials as functions of $p_k$ variables and our notation $S_{\lambda}(\delta_{k,1}) := S_{\lambda}(p_k \to \delta_{k,1})$,$S_{\lambda}(N) := S_{\lambda}(p_k \to N)$. Connection to symmetric functions of $N$ variables $x_i$ are given by Miwa transformation $p_k = \sum_{i=1}^{N} (x_{i})^k$. Schur polynomials for single-row Young diagrams $[n]$ can be computed from generating function:
	\begin{equation}
		\exp\left(\sum_{k=1} \frac{p_k z^k}{k}\right) = \sum_{n=1} z^{n} \cdot S_{[n]}(p_a)
	\end{equation}
	The other polynomials for higher Young diagram can be obtained from Jacobi-Trudi determinant formula:
	\begin{equation}
		S_{\lambda} = \det_{1 \leqslant i,j \leqslant l(\lambda)} S_{[\lambda_{i} + j - i]}
	\end{equation}
	Ratio of Schur polynomials in special points in formula \eqref{SI formula for GHM} is a fully factorizable quantity:
	\begin{equation}
		\frac{S_{\lambda}(N)}{S_{\lambda}(\delta_{k,1})} = \prod_{\Box \in \lambda} (N + j_{\Box} - i_{\Box})
	\end{equation}
	where product runs over all boxes in the Young diagram $\lambda$ and $(i_{\Box}, j_{\Box})$ are the vertical and horizontal coordinates of the box $\Box$. The quantity $\omega_{\Box} = N + j_{\Box} - i_{\Box}$ is called content of the box and it plays an important role in the superintegrable formula \eqref{SI formula for GHM}. 
	
	For this simple and well-studied model such as GH model one can offer the following explanation of superintegrability phenomenon. The first key observation is that full partition function of the model:
	\begin{equation}
		\label{partition func HGM}
		Z_{GH} \left( p_k \right) = \frac{1}{Z} \, \int dH \exp\left( - \frac{1}{2} \tr H^2 + \sum_{k = 1}^{\infty} \frac{p_k}{k} \tr H^k \right)
	\end{equation}
	admits $W$-representation \cite{Morozov:2009xk, Mironov:2021udg}:
	\begin{equation}
		Z_{GH}(p_k) = e^{\hat{W}} \cdot 1 
	\end{equation}
	Explicit form of $\hat{W}$ operator reads:
	\begin{equation}
		\hat{W} = \frac{1}{2} \sum_{a, b = 1}^{\infty}  (a+b-2) \, p_a p_b \frac{\partial }{\partial p_{a+b-2}} + a b \, p_{a+b+2} \frac{\partial}{\partial p_a} \frac{\partial}{\partial p_b}  + N \sum_{a = 1}^{\infty} a \, p_{a+2} \frac{\partial}{\partial p_a} + \frac{N^2 p_2}{2} + \frac{N p_1^2}{2}
	\end{equation}
	Note, that for simplicity of formulas we add extra factor $\frac{1}{k}$ in front of $p_k \tr H^{k}$ in the definition of partition function. The operator $\hat{W}$ has grading 2, that is explicitly seem from the following formula:
	\begin{equation}
		\Big[ \hat{D}, \hat{W} \Big] = 2 \hat{W} \hspace{20mm} \hat{D} = \sum_{k = 1}^{\infty} k \, p_k \frac{\partial}{\partial p_k}
	\end{equation}  
	Due to the positive grading 2 the image of operator $\hat{W}$ when acting on polynomials of grading $n$ is given by polynomials of grading $n+2$. Schur polynomials form \textit{distinguished basis} of graded polynomials where the action of $\hat{W}$ operator has very natural meaning:
	\begin{equation}
		\label{W action}
		\hat{W} \, S_{\lambda} = \frac{1}{2} \cdot \sum_{
			\begin{tikzpicture}[scale=0.15]
				\foreach \i/\j in {0/0, 0/-1, 1/0, 1/-1}
				{
					\draw[thick] (\i,\j) -- (\i+1,\j);
				}
				\foreach \i/\j in {0/0, 1/0, 2/0}
				{
					\draw[thick] (\i,\j) -- (\i,\j-1);
				}
				\foreach \i/\j in {}
				{
					\draw[thick] (\i,\j) -- (\i-1,\j-1);
				}
				\foreach \x/\y in {0/0, 1/0}
				{
					\draw[fill=pink] (\x,-\y) -- (\x+1,-\y) -- (\x+1,-\y-1) -- (\x,-\y-1) -- cycle;
				}
			\end{tikzpicture} \in \text{Add}_{[2]}(\lambda)} \omega_{\begin{tikzpicture}[scale=0.15]
				\foreach \i/\j in {0/0, 0/-1}
				{
					\draw[thick] (\i,\j) -- (\i+1,\j);
				}
				\foreach \i/\j in {0/0, 1/0}
				{
					\draw[thick] (\i,\j) -- (\i,\j-1);
				}
				\foreach \i/\j in {}
				{
					\draw[thick] (\i,\j) -- (\i-1,\j-1);
				}
				\foreach \x/\y in {0/0}
				{
					\draw[fill=pink] (\x,-\y) -- (\x+1,-\y) -- (\x+1,-\y-1) -- (\x,-\y-1) -- cycle;
				}
			\end{tikzpicture}_1} \omega_{\begin{tikzpicture}[scale=0.15]
				\foreach \i/\j in {0/0, 0/-1}
				{
					\draw[thick] (\i,\j) -- (\i+1,\j);
				}
				\foreach \i/\j in {0/0, 1/0}
				{
					\draw[thick] (\i,\j) -- (\i,\j-1);
				}
				\foreach \i/\j in {}
				{
					\draw[thick] (\i,\j) -- (\i-1,\j-1);
				}
				\foreach \x/\y in {0/0}
				{
					\draw[fill=pink] (\x,-\y) -- (\x+1,-\y) -- (\x+1,-\y-1) -- (\x,-\y-1) -- cycle;
				}
			\end{tikzpicture}_2} \, S_{\lambda + \begin{tikzpicture}[scale=0.15]
				\foreach \i/\j in {0/0, 0/-1, 1/0, 1/-1}
				{
					\draw[thick] (\i,\j) -- (\i+1,\j);
				}
				\foreach \i/\j in {0/0, 1/0, 2/0}
				{
					\draw[thick] (\i,\j) -- (\i,\j-1);
				}
				\foreach \i/\j in {}
				{
					\draw[thick] (\i,\j) -- (\i-1,\j-1);
				}
				\foreach \x/\y in {0/0, 1/0}
				{
					\draw[fill=pink] (\x,-\y) -- (\x+1,-\y) -- (\x+1,-\y-1) -- (\x,-\y-1) -- cycle;
				}
		\end{tikzpicture}} - \frac{1}{2} \cdot \sum_{
			\begin{tikzpicture}[scale=0.15]
				\foreach \i/\j in {0/0, 0/-1, 0/-2}
				{
					\draw[thick] (\i,\j) -- (\i+1,\j);
				}
				\foreach \i/\j in {0/0, 1/0, 0/-1, 1/-1}
				{
					\draw[thick] (\i,\j) -- (\i,\j-1);
				}
				\foreach \i/\j in {}
				{
					\draw[thick] (\i,\j) -- (\i-1,\j-1);
				}
				\foreach \x/\y in {0/0, 0/1}
				{
					\draw[fill=cyan] (\x,-\y) -- (\x+1,-\y) -- (\x+1,-\y-1) -- (\x,-\y-1) -- cycle;
				}
			\end{tikzpicture} \in \text{Add}_{[1,1]}(\lambda)} \omega_{\begin{tikzpicture}[scale=0.15]
				\foreach \i/\j in {0/0, 0/-1}
				{
					\draw[thick] (\i,\j) -- (\i+1,\j);
				}
				\foreach \i/\j in {0/0, 1/0}
				{
					\draw[thick] (\i,\j) -- (\i,\j-1);
				}
				\foreach \i/\j in {}
				{
					\draw[thick] (\i,\j) -- (\i-1,\j-1);
				}
				\foreach \x/\y in {0/0}
				{
					\draw[fill=cyan] (\x,-\y) -- (\x+1,-\y) -- (\x+1,-\y-1) -- (\x,-\y-1) -- cycle;
				}
			\end{tikzpicture}_1} \omega_{\begin{tikzpicture}[scale=0.15]
				\foreach \i/\j in {0/0, 0/-1}
				{
					\draw[thick] (\i,\j) -- (\i+1,\j);
				}
				\foreach \i/\j in {0/0, 1/0}
				{
					\draw[thick] (\i,\j) -- (\i,\j-1);
				}
				\foreach \i/\j in {}
				{
					\draw[thick] (\i,\j) -- (\i-1,\j-1);
				}
				\foreach \x/\y in {0/0}
				{
					\draw[fill=cyan] (\x,-\y) -- (\x+1,-\y) -- (\x+1,-\y-1) -- (\x,-\y-1) -- cycle;
				}
			\end{tikzpicture}_2} \, S_{\lambda +
			\begin{tikzpicture}[scale=0.15]
				\foreach \i/\j in {0/0, 0/-1, 0/-2}
				{
					\draw[thick] (\i,\j) -- (\i+1,\j);
				}
				\foreach \i/\j in {0/0, 1/0, 0/-1, 1/-1}
				{
					\draw[thick] (\i,\j) -- (\i,\j-1);
				}
				\foreach \i/\j in {}
				{
					\draw[thick] (\i,\j) -- (\i-1,\j-1);
				}
				\foreach \x/\y in {0/0, 0/1}
				{
					\draw[fill=cyan] (\x,-\y) -- (\x+1,-\y) -- (\x+1,-\y-1) -- (\x,-\y-1) -- cycle;
				}
		\end{tikzpicture}}
	\end{equation}
	
	Operator $\hat{W}$ acts on Schur polynomials by adding boxes to corresponding Young diagrams. In other words, on the r.h.s. appears only those polynomials which Young diagram is obtained by adding boxes to initial Young diagram $\lambda$ from the l.h.s. Definitions of sets $\text{Add}_{[2]}(\lambda)$ and $\text{Add}_{[2]}(\lambda)$ are clear from the following example:
	\begin{equation}
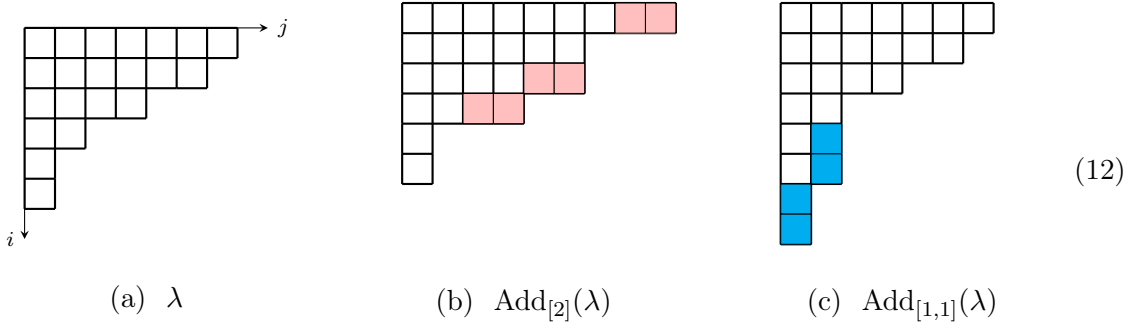
		\begin{array}{c}
			\begin{tikzpicture}[scale=0.4]
				\foreach \i/\j in {0/0, 1/0, 2/0, 3/0, 4/0, 5/0, 6/0, 0/-1, 1/-1, 2/-1, 3/-1, 4/-1, 5/-1, 6/-1, 0/-2, 1/-2, 2/-2, 3/-2, 4/-2, 5/-2, 0/-3, 1/-3, 2/-3, 3/-3, 0/-4, 1/-4, 0/-5, 0/-6}
				{
					\draw[thick] (\i,\j) -- (\i+1,\j);
				}
				\foreach \i/\j in {0/0, 0/-1, 0/-2, 0/-3, 0/-4, 0/-5, 1/0, 1/-1, 1/-2, 1/-3, 1/-4, 1/-5, 2/0, 2/-1, 2/-2, 2/-3, 3/0, 3/-1, 3/-2, 4/0, 4/-1, 4/-2, 5/0, 5/-1, 6/0, 6/-1, 7/0}
				{
					\draw[thick] (\i,\j) -- (\i,\j-1);
				}
				\foreach \i/\j in {}
				{
					\draw[thick] (\i,\j) -- (\i-1,\j-1);
				}
				\draw[-stealth] (0,0) -- (0,-7);
				\draw[-stealth] (0,0) -- (8,0);
				\node[left] at (0,-7) {$\scriptstyle i$};
				\node[right] at (8,0) {$\scriptstyle j$};
				\node at (4,-9) {(a) \ $\lambda$};
			\end{tikzpicture}
		\end{array}
		\hspace{10mm}
		\begin{array}{c}
			\begin{tikzpicture}[scale=0.4]
				\foreach \i/\j in {0/0, 1/0, 2/0, 3/0, 4/0, 5/0, 6/0, 0/-1, 1/-1, 2/-1, 3/-1, 4/-1, 5/-1, 6/-1, 0/-2, 1/-2, 2/-2, 3/-2, 4/-2, 5/-2, 0/-3, 1/-3, 2/-3, 3/-3, 0/-4, 1/-4, 0/-5, 0/-6, 7/0, 8/0, 7/-1, 8/-1, 4/-3, 5/-3, 2/-4, 3/-4}
				{
					\draw[thick] (\i,\j) -- (\i+1,\j);
				}
				\foreach \i/\j in {0/0, 0/-1, 0/-2, 0/-3, 0/-4, 0/-5, 1/0, 1/-1, 1/-2, 1/-3, 1/-4, 1/-5, 2/0, 2/-1, 2/-2, 2/-3, 3/0, 3/-1, 3/-2, 4/0, 4/-1, 4/-2, 5/0, 5/-1, 6/0, 6/-1, 7/0, 8/0, 9/0, 5/-2, 6/-2, 3/-3, 4/-3}
				{
					\draw[thick] (\i,\j) -- (\i,\j-1);
				}
				\foreach \i/\j in {}
				{
					\draw[thick] (\i,\j) -- (\i-1,\j-1);
				}
				\foreach \x/\y in {7/0, 8/0, 4/2, 5/2, 2/3, 3/3}
				{
					\draw[fill=pink] (\x,-\y) -- (\x+1,-\y) -- (\x+1,-\y-1) -- (\x,-\y-1) -- cycle;
				}
				\node at (4,-10) {(b) \ $\text{Add}_{[2]}(\lambda$)};
			\end{tikzpicture}
		\end{array}
		\hspace{10mm}
		\begin{array}{c}
			\begin{tikzpicture}[scale=0.4]
				\foreach \i/\j in {0/0, 1/0, 2/0, 3/0, 4/0, 5/0, 6/0, 0/-1, 1/-1, 2/-1, 3/-1, 4/-1, 5/-1, 6/-1, 0/-2, 1/-2, 2/-2, 3/-2, 4/-2, 5/-2, 0/-3, 1/-3, 2/-3, 3/-3, 0/-4, 1/-4, 0/-5, 0/-6, 1/-5, 1/-6, 0/-7, 0/-8}
				{
					\draw[thick] (\i,\j) -- (\i+1,\j);
				}
				\foreach \i/\j in {0/0, 0/-1, 0/-2, 0/-3, 0/-4, 0/-5, 1/0, 1/-1, 1/-2, 1/-3, 1/-4, 1/-5, 2/0, 2/-1, 2/-2, 2/-3, 3/0, 3/-1, 3/-2, 4/0, 4/-1, 4/-2, 5/0, 5/-1, 6/0, 6/-1, 7/0, 0/-6, 0/-7, 1/-6, 1/-7, 2/-4, 2/-5}
				{
					\draw[thick] (\i,\j) -- (\i,\j-1);
				}
				\foreach \i/\j in {}
				{
					\draw[thick] (\i,\j) -- (\i-1,\j-1);
				}
				\foreach \x/\y in {0/6, 0/7, 1/4, 1/5}
				{
					\draw[fill=cyan] (\x,-\y) -- (\x+1,-\y) -- (\x+1,-\y-1) -- (\x,-\y-1) -- cycle;
				}
				\node at (4,-10) {(c) \ $\text{Add}_{[1,1]}(\lambda$)};
			\end{tikzpicture}
		\end{array}
	\end{equation}
	
	Quantities $\omega_{\Box_{1,2}}$ in formula \eqref{W action} are contents $N+j_{\Box}-i_{\Box}$ of added boxes. For example:
	\begin{equation}
		\hat{W} \, S_{\begin{tikzpicture}[scale=0.15]
				\foreach \i/\j in {0/0, 0/-1, 1/0, 1/-1}
				{
					\draw[thick] (\i,\j) -- (\i+1,\j);
				}
				\foreach \i/\j in {0/0, 1/0, 2/0}
				{
					\draw[thick] (\i,\j) -- (\i,\j-1);
				}
				\foreach \i/\j in {}
				{
					\draw[thick] (\i,\j) -- (\i-1,\j-1);
				}
		\end{tikzpicture}} = \frac{1}{2}\,(N+2)(N+3) \, S_{\begin{tikzpicture}[scale=0.15]
				\foreach \i/\j in {0/0, 0/-1, 1/0, 1/-1, 2/0, 2/-1, 3/0, 3/-1}
				{
					\draw[thick] (\i,\j) -- (\i+1,\j);
				}
				\foreach \i/\j in {0/0, 1/0, 2/0, 3/0, 4/0}
				{
					\draw[thick] (\i,\j) -- (\i,\j-1);
				}
				\foreach \i/\j in {}
				{
					\draw[thick] (\i,\j) -- (\i-1,\j-1);
				}
				\foreach \x/\y in {2/0, 3/0}
				{
					\draw[fill=pink] (\x,-\y) -- (\x+1,-\y) -- (\x+1,-\y-1) -- (\x,-\y-1) -- cycle;
				}
		\end{tikzpicture}} + \frac{1}{2}\,(N-1)(N) \, S_{\begin{tikzpicture}[scale=0.15]
				\foreach \i/\j in {0/0, 0/-1, 0/-2, 1/0, 1/-1, 1/-2}
				{
					\draw[thick] (\i,\j) -- (\i+1,\j);
				}
				\foreach \i/\j in {0/0, 1/0, 0/-1, 1/-1, 2/0, 2/-1}
				{
					\draw[thick] (\i,\j) -- (\i,\j-1);
				}
				\foreach \i/\j in {}
				{
					\draw[thick] (\i,\j) -- (\i-1,\j-1);
				}
				\foreach \x/\y in {0/1, 1/1}
				{
					\draw[fill=pink] (\x,-\y) -- (\x+1,-\y) -- (\x+1,-\y-1) -- (\x,-\y-1) -- cycle;
				}
		\end{tikzpicture}} - \frac{1}{2}\,(N-1)(N-2) \, S_{\begin{tikzpicture}[scale=0.15]
				\foreach \i/\j in {0/0, 0/-1, 0/-2, 0/-3, 1/0, 1/-1}
				{
					\draw[thick] (\i,\j) -- (\i+1,\j);
				}
				\foreach \i/\j in {0/0, 1/0, 0/-1, 1/-1, 0/-2, 1/-2, 2/0}
				{
					\draw[thick] (\i,\j) -- (\i,\j-1);
				}
				\foreach \i/\j in {}
				{
					\draw[thick] (\i,\j) -- (\i-1,\j-1);
				}
				\foreach \x/\y in {0/1, 0/2}
				{
					\draw[fill=cyan] (\x,-\y) -- (\x+1,-\y) -- (\x+1,-\y-1) -- (\x,-\y-1) -- cycle;
				}
		\end{tikzpicture}}
	\end{equation}
	
	Orthogonal properties of Schur polynomials that are expressed in an elegant way via Cauchy identity \cite{Macdonald}:
	\begin{equation}
		\label{Cauchy formula}
		\sum_{\lambda} S_{\lambda} (p_a) \, S_{\lambda} (\bar{p}_a) = \exp\left( \sum_{k=1}^{\infty} \frac{p_k \, \bar{p}_k}{k} \right)
	\end{equation}
	The above relation allows one to express the partition function \eqref{partition func HGM} in the following form:
	\begin{align}
		\begin{aligned}
			Z_{GH} (p_k) = \Big\langle \exp\left( \sum_{k = 1}^{\infty} \frac{p_k}{k} \tr H^k\right) \Big\rangle = \sum_{\lambda} S_{\lambda}(p_k) \, \Big\langle S_{\lambda} \left( \bar{p}_k = \tr H^k \right) \Big\rangle = \\
			= \sum_{\lambda} S_{\lambda} (p_k) \, S_{\lambda} \left( \delta_{k,2} \right) \, \prod_{\Box \in \lambda} (N + j_{\Box} - i_{\Box})
		\end{aligned}
	\end{align}
	
	At this point we are equipped with the knowledge that exponent of $\hat{W}$ operator generates the whole partition function and the operator acts on Schur polynomials by adding two boxes - domino tiles. It can be seen, if the partition function is split into graded components:
	
	\begin{equation}
		Z_{GH}(p_k) = Z_{GH}^{(0)} + Z_{GH}^{(1)} + Z_{GH}^{(2)} + \ldots
	\end{equation}
	
	\begin{equation}
		Z_{GH}^{(2k)} = \frac{1}{k!}\left(\hat{W} \right)^{k} \cdot 1
	\end{equation}
	\begin{equation}
		Z_{GH}^{(2k+1)} = 0
	\end{equation}
	We compute small levels explicitly by action of the $\hat{W}$ operator:
	\begin{align}
		Z_{GH}^{(0)} = 1
	\end{align}
	\begin{align}
		Z_{GH}^{(2)} = \hat{W} \cdot 1 = \underbrace{\frac{1}{2}\,(N)(N+1)}_{\langle S_{\begin{tikzpicture}[scale=0.15]
					\foreach \i/\j in {0/0, 0/-1, 1/0, 1/-1}
					{
						\draw[thick] (\i,\j) -- (\i+1,\j);
					}
					\foreach \i/\j in {0/0, 1/0, 2/0}
					{
						\draw[thick] (\i,\j) -- (\i,\j-1);
					}
					\foreach \i/\j in {}
					{
						\draw[thick] (\i,\j) -- (\i-1,\j-1);
					}
			\end{tikzpicture}}\rangle} \, S_{\begin{tikzpicture}[scale=0.15]
				\foreach \i/\j in {0/0, 0/-1, 1/0, 1/-1}
				{
					\draw[thick] (\i,\j) -- (\i+1,\j);
				}
				\foreach \i/\j in {0/0, 1/0, 2/0}
				{
					\draw[thick] (\i,\j) -- (\i,\j-1);
				}
				\foreach \i/\j in {}
				{
					\draw[thick] (\i,\j) -- (\i-1,\j-1);
				}
		\end{tikzpicture}} + \underbrace{\left(- \frac{1}{2} \right) \,(N)(N-1)}_{\langle S_{\begin{tikzpicture}[scale=0.15]
					\foreach \i/\j in {0/0, 0/-1, 0/-2}
					{
						\draw[thick] (\i,\j) -- (\i+1,\j);
					}
					\foreach \i/\j in {0/0, 1/0, 0/-1, 1/-1}
					{
						\draw[thick] (\i,\j) -- (\i,\j-1);
					}
					\foreach \i/\j in {}
					{
						\draw[thick] (\i,\j) -- (\i-1,\j-1);
					}
			\end{tikzpicture}}\rangle} \, S_{\begin{tikzpicture}[scale=0.15]
				\foreach \i/\j in {0/0, 0/-1, 0/-2}
				{
					\draw[thick] (\i,\j) -- (\i+1,\j);
				}
				\foreach \i/\j in {0/0, 1/0, 0/-1, 1/-1}
				{
					\draw[thick] (\i,\j) -- (\i,\j-1);
				}
				\foreach \i/\j in {}
				{
					\draw[thick] (\i,\j) -- (\i-1,\j-1);
				}
		\end{tikzpicture}}
	\end{align}
	\begin{align}
		\begin{aligned}
			Z^{(4)}_{GH} = \frac{1}{2} \hat{W} \cdot Z^{(2)}_{GH} = \underbrace{\frac{1}{8} (N)(N+1)(N+2)(N+3)}_{\langle S_{\begin{tikzpicture}[scale=0.15]
						\foreach \i/\j in {0/0, 0/-1, 1/0, 1/-1, 2/0, 2/-1, 3/0, 3/-1}
						{
							\draw[thick] (\i,\j) -- (\i+1,\j);
						}
						\foreach \i/\j in {0/0, 1/0, 2/0, 3/0, 4/0}
						{
							\draw[thick] (\i,\j) -- (\i,\j-1);
						}
						\foreach \i/\j in {}
						{
							\draw[thick] (\i,\j) -- (\i-1,\j-1);
						}
				\end{tikzpicture}}\rangle} \, S_{\begin{tikzpicture}[scale=0.15]
					\foreach \i/\j in {0/0, 0/-1, 1/0, 1/-1, 2/0, 2/-1, 3/0, 3/-1}
					{
						\draw[thick] (\i,\j) -- (\i+1,\j);
					}
					\foreach \i/\j in {0/0, 1/0, 2/0, 3/0, 4/0}
					{
						\draw[thick] (\i,\j) -- (\i,\j-1);
					}
					\foreach \i/\j in {}
					{
						\draw[thick] (\i,\j) -- (\i-1,\j-1);
					}
			\end{tikzpicture}} + \underbrace{\left(-\frac{1}{8}\right)(N)(N-1)(N+1)(N+2)}_{\langle S_{\begin{tikzpicture}[scale=0.15]
						\foreach \i/\j in {0/0, 0/-1, 0/-2, 1/0, 1/-1, 2/0, 2/-1}
						{
							\draw[thick] (\i,\j) -- (\i+1,\j);
						}
						\foreach \i/\j in {0/0, 1/0, 0/-1, 1/-1, 2/0, 3/0}
						{
							\draw[thick] (\i,\j) -- (\i,\j-1);
						}
						\foreach \i/\j in {}
						{
							\draw[thick] (\i,\j) -- (\i-1,\j-1);
						}
				\end{tikzpicture}}\rangle} \, S_{\begin{tikzpicture}[scale=0.15]
					\foreach \i/\j in {0/0, 0/-1, 0/-2, 1/0, 1/-1, 2/0, 2/-1}
					{
						\draw[thick] (\i,\j) -- (\i+1,\j);
					}
					\foreach \i/\j in {0/0, 1/0, 0/-1, 1/-1, 2/0, 3/0}
					{
						\draw[thick] (\i,\j) -- (\i,\j-1);
					}
					\foreach \i/\j in {}
					{
						\draw[thick] (\i,\j) -- (\i-1,\j-1);
					}
			\end{tikzpicture}} + \\
			+ \underbrace{\left(\frac{1}{8}+\frac{1}{8}\right) (N)^2(N-1)(N+1)}_{\langle S_{\begin{tikzpicture}[scale=0.15]
						\foreach \i/\j in {0/0, 0/-1, 0/-2, 1/0, 1/-1, 1/-2}
						{
							\draw[thick] (\i,\j) -- (\i+1,\j);
						}
						\foreach \i/\j in {0/0, 1/0, 0/-1, 1/-1, 2/0, 2/-1}
						{
							\draw[thick] (\i,\j) -- (\i,\j-1);
						}
						\foreach \i/\j in {}
						{
							\draw[thick] (\i,\j) -- (\i-1,\j-1);
						}
				\end{tikzpicture}}\rangle} \, S_{\begin{tikzpicture}[scale=0.15]
					\foreach \i/\j in {0/0, 0/-1, 0/-2, 1/0, 1/-1, 1/-2}
					{
						\draw[thick] (\i,\j) -- (\i+1,\j);
					}
					\foreach \i/\j in {0/0, 1/0, 0/-1, 1/-1, 2/0, 2/-1}
					{
						\draw[thick] (\i,\j) -- (\i,\j-1);
					}
					\foreach \i/\j in {}
					{
						\draw[thick] (\i,\j) -- (\i-1,\j-1);
					}
			\end{tikzpicture}} + \underbrace{\left(-\frac{1}{8}\right) N(N+1)(N-1)(N-2)}_{\langle S_{\begin{tikzpicture}[scale=0.15]
						\foreach \i/\j in {0/0, 0/-1, 0/-2, 0/-3, 1/0, 1/-1}
						{
							\draw[thick] (\i,\j) -- (\i+1,\j);
						}
						\foreach \i/\j in {0/0, 1/0, 0/-1, 1/-1, 0/-2, 1/-2, 2/0}
						{
							\draw[thick] (\i,\j) -- (\i,\j-1);
						}
						\foreach \i/\j in {}
						{
							\draw[thick] (\i,\j) -- (\i-1,\j-1);
						}
				\end{tikzpicture}} \rangle} \, S_{\begin{tikzpicture}[scale=0.15]
					\foreach \i/\j in {0/0, 0/-1, 0/-2, 0/-3, 1/0, 1/-1}
					{
						\draw[thick] (\i,\j) -- (\i+1,\j);
					}
					\foreach \i/\j in {0/0, 1/0, 0/-1, 1/-1, 0/-2, 1/-2, 2/0}
					{
						\draw[thick] (\i,\j) -- (\i,\j-1);
					}
					\foreach \i/\j in {}
					{
						\draw[thick] (\i,\j) -- (\i-1,\j-1);
					}
			\end{tikzpicture}} + \\
			\underbrace{\frac{1}{8} (N)(N-1)(N-2)(N-3)}_{\langle S_{\begin{tikzpicture}[scale=0.15]
						\foreach \i/\j in {0/0, 0/-1, 0/-2, 0/-3, 0/-4}
						{
							\draw[thick] (\i,\j) -- (\i+1,\j);
						}
						\foreach \i/\j in {0/0, 1/0, 0/-1, 1/-1, 0/-2, 1/-2, 0/-3, 1/-3}
						{
							\draw[thick] (\i,\j) -- (\i,\j-1);
						}
						\foreach \i/\j in {}
						{
							\draw[thick] (\i,\j) -- (\i-1,\j-1);
						}
				\end{tikzpicture}} \rangle} \, S_{\begin{tikzpicture}[scale=0.15]
					\foreach \i/\j in {0/0, 0/-1, 0/-2, 0/-3, 0/-4}
					{
						\draw[thick] (\i,\j) -- (\i+1,\j);
					}
					\foreach \i/\j in {0/0, 1/0, 0/-1, 1/-1, 0/-2, 1/-2, 0/-3, 1/-3}
					{
						\draw[thick] (\i,\j) -- (\i,\j-1);
					}
					\foreach \i/\j in {}
					{
						\draw[thick] (\i,\j) -- (\i-1,\j-1);
					}
			\end{tikzpicture}}
		\end{aligned}
	\end{align}
	
	Now we better understand the superintegrable formula \eqref{SI formula for GHM} due to the following consequences of the above formulas:
	\begin{itemize}
		\item The overall product over boxes of content function $\omega_{\Box} = N + j_{\Box} - i_{\Box}$ in formula \eqref{SI formula for GHM} is generated step by step due to factors in r.h.s. of $\hat{W}$ operator \footnote{Note, that $\hat{W}$-operator may act on Schur polynomial of arbitrary Young diagram, however for GH model relevant only diagrams of even size.} action \eqref{W action}.
		\item Correlator of Schur polynomial $\langle S_{\lambda} \rangle$ has non-vanishing value if and only if the diagram $\lambda$ can be constructed only with domino-tiles $\begin{tikzpicture}[scale=0.15]
			\foreach \i/\j in {0/0, 0/-1, 1/0, 1/-1}
			{
				\draw[thick] (\i,\j) -- (\i+1,\j);
			}
			\foreach \i/\j in {0/0, 1/0, 2/0}
			{
				\draw[thick] (\i,\j) -- (\i,\j-1);
			}
			\foreach \i/\j in {}
			{
				\draw[thick] (\i,\j) -- (\i-1,\j-1);
			}
		\end{tikzpicture}$ and $\begin{tikzpicture}[scale=0.15]
			\foreach \i/\j in {0/0, 0/-1, 0/-2}
			{
				\draw[thick] (\i,\j) -- (\i+1,\j);
			}
			\foreach \i/\j in {0/0, 1/0, 0/-1, 1/-1}
			{
				\draw[thick] (\i,\j) -- (\i,\j-1);
			}
			\foreach \i/\j in {}
			{
				\draw[thick] (\i,\j) -- (\i-1,\j-1);
			}
		\end{tikzpicture}$ and starting from empty diagram $\varnothing$. More accurately, the number of ways to construct diagram $\lambda$ from domino-tiles (note, that the vertical tile is weighted with $(-1)$) is equal to $2^{\frac{|\lambda|}{2}} \cdot (\frac{|\lambda|}{2})! \cdot S_{\lambda}(\delta_{k,2})$.
	\end{itemize}
	
	After all, a logical question follows: \textit{how can one guess that Schur polynomials should arise as a superintegrable basis of correlators?} In particular case of Gaussian Hermitian model the answer is known. We formulate it in two steps:
	\begin{itemize}
		\item $\hat{W}$ operator as differential operator in $p_k$ variables belongs to an infinite dimensional algebra $W_{1 + \infty}$ \cite{Pope:1989sr, Awata:1994tf} in Fock representation that acts on space of polynomials $\mathbb{C}[ p_1, p_2, \ldots]$.
		\item Commutative family of operators $\hat{\psi}_a$  of $W_{1 + \infty}$  in Fock representation have Schur polynomials as a common set of eigenfunctions:
		\begin{equation}
			\hat{\psi}_a \, S_{\lambda} = \mathcal{E}_{\lambda,a} \, S_{\lambda}
		\end{equation}
		\textit{Schur polynomials are distinguished by the underlying algebra} to which $\hat{W}$ belongs.
		Due to Schur polynomials one can say that $W_{1 + \infty}$ algebra in Fock representation acts on the Hilbert space, where basis vectors enumerated by Young diagrams. 
	\end{itemize}
	
	We do not give here details of $W_{1 + \infty}$ representation theory  and list only some basic formulas concerning Fock representation. We work with generators $\hat{e}_{k}, \hat{\psi}_k, \hat{f}_k$, where $k = 0,1,2,\ldots$, that are rising operators, Cartan-like operators and lowering operators respectively. Among them there is a special small set of operators $\hat{e}_0, \hat{\psi}_3, \hat{f}_0$ that have the following form in Fock representation:
	\begin{equation}
		\hat{e}_0 = p_1 \hspace{25mm} \hat{f}_{0} = \frac{\partial}{\partial p_1}
	\end{equation}
	\begin{equation}
		\label{Ham for Schurs}
		\hat{\psi}_3 = \frac{1}{2} \sum_{a, b = 1}^{\infty}  (a+b) \, p_a p_b \frac{\partial }{\partial p_{a+b}} + a b \, p_{a+b} \frac{\partial}{\partial p_a} \frac{\partial}{\partial p_b}  + N \sum_{a = 1}^{\infty} a \, p_{a} \frac{\partial}{\partial p_a} 
	\end{equation}
	The higher operators can be expressed via only these three operators by the following rules:
	\begin{equation}
		\label{higher e and f via psi}
		\hat{e}_k = \left( \text{ad}_{\hat{\psi}_3} \right)^k \hat{e}_0 \hspace{25mm} \hat{f}_k = (-)^k \left( \text{ad}_{\hat{\psi}_3} \right)^k \hat{f}_0
	\end{equation}
	\begin{equation}
		\hat{\psi}_{a+b} = \Big[ \hat{e}_a, \hat{f}_b \Big]
	\end{equation}
	Action of small set operators $\hat{e}_0, \hat{\psi}_3, \hat{f}_0$ on Schur polynomials obeys the following formulas:
	\begin{align}
		\begin{aligned}
			\hat{e}_0 \, S_{\lambda} = \sum_{\begin{tikzpicture}[scale=0.15]
					\foreach \i/\j in {0/0, 0/-1}
					{
						\draw[thick] (\i,\j) -- (\i+1,\j);
					}
					\foreach \i/\j in {0/0, 1/0}
					{
						\draw[thick] (\i,\j) -- (\i,\j-1);
					}
					\foreach \i/\j in {}
					{
						\draw[thick] (\i,\j) -- (\i-1,\j-1);
					}
					\foreach \x/\y in {0/0}
					{
						\draw[fill=pink] (\x,-\y) -- (\x+1,-\y) -- (\x+1,-\y-1) -- (\x,-\y-1) -- cycle;
					}
				\end{tikzpicture} \in \text{Add}(\lambda)} S_{\lambda + \begin{tikzpicture}[scale=0.15]
					\foreach \i/\j in {0/0, 0/-1}
					{
						\draw[thick] (\i,\j) -- (\i+1,\j);
					}
					\foreach \i/\j in {0/0, 1/0}
					{
						\draw[thick] (\i,\j) -- (\i,\j-1);
					}
					\foreach \i/\j in {}
					{
						\draw[thick] (\i,\j) -- (\i-1,\j-1);
					}
					\foreach \x/\y in {0/0}
					{
						\draw[fill=pink] (\x,-\y) -- (\x+1,-\y) -- (\x+1,-\y-1) -- (\x,-\y-1) -- cycle;
					} 
			\end{tikzpicture}}  \\
			\hat{f}_0 \, S_{\lambda} = \sum_{\begin{tikzpicture}[scale=0.15]
					\foreach \i/\j in {0/0, 0/-1}
					{
						\draw[thick] (\i,\j) -- (\i+1,\j);
					}
					\foreach \i/\j in {0/0, 1/0}
					{
						\draw[thick] (\i,\j) -- (\i,\j-1);
					}
					\foreach \i/\j in {}
					{
						\draw[thick] (\i,\j) -- (\i-1,\j-1);
					}
					\foreach \x/\y in {0/0}
					{
						\draw[fill=teal] (\x,-\y) -- (\x+1,-\y) -- (\x+1,-\y-1) -- (\x,-\y-1) -- cycle;
					}
				\end{tikzpicture} \in \text{Rem}(\lambda)} S_{\lambda - \begin{tikzpicture}[scale=0.15]
					\foreach \i/\j in {0/0, 0/-1}
					{
						\draw[thick] (\i,\j) -- (\i+1,\j);
					}
					\foreach \i/\j in {0/0, 1/0}
					{
						\draw[thick] (\i,\j) -- (\i,\j-1);
					}
					\foreach \i/\j in {}
					{
						\draw[thick] (\i,\j) -- (\i-1,\j-1);
					}
					\foreach \x/\y in {0/0}
					{
						\draw[fill=teal] (\x,-\y) -- (\x+1,-\y) -- (\x+1,-\y-1) -- (\x,-\y-1) -- cycle;
					}
			\end{tikzpicture}}
		\end{aligned}
		\label{Pierri rules}
	\end{align}
	\begin{equation}
		\hat{\psi}_3 \, S_{\lambda} = \left( \sum_{\Box \in \lambda} \omega_{\Box} \right) \cdot S_{\lambda}
	\end{equation}
	where as usual $\omega_{\Box} = N + j_{\Box} - i_{\Box}$. Operator \eqref{Ham for Schurs} is a cut-and-join operator \cite{Mironov:2009cj}, however, we call it Hamiltonian of Schur polynomials, due to connection to quantum integrable systems\footnote{$\beta$-deformed version of this operator coincides with quantum Hamiltonian of Calogero-Moser many body integrable system with trigonometric interaction \cite{CSMbook} }. Formulas \eqref{Pierri rules} for Schur polynomials are called Pierri rules \cite{Macdonald}. The sums in the above runs over all possible ways of adding/removing boxes in Young diagram. We provide example of $\text{Add}(\lambda)$ and $\text{Rem}(\lambda)$:
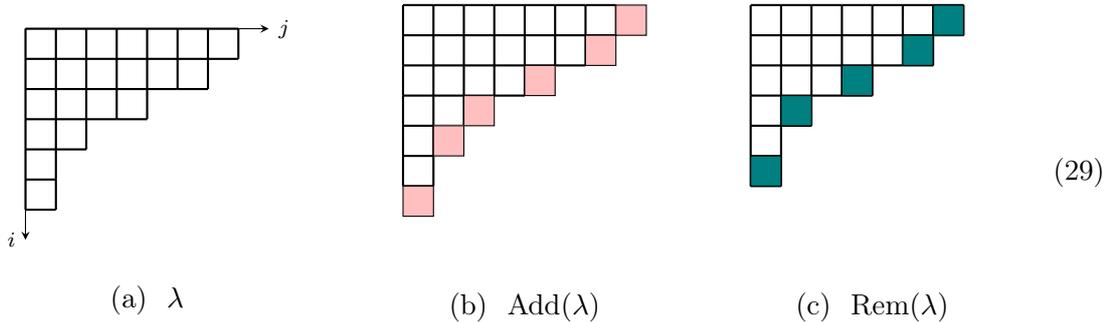
	\begin{equation}
		\label{Add Rem definition Schur}
		\begin{array}{c}
			\begin{tikzpicture}[scale=0.4]
				\foreach \i/\j in {0/0, 1/0, 2/0, 3/0, 4/0, 5/0, 6/0, 0/-1, 1/-1, 2/-1, 3/-1, 4/-1, 5/-1, 6/-1, 0/-2, 1/-2, 2/-2, 3/-2, 4/-2, 5/-2, 0/-3, 1/-3, 2/-3, 3/-3, 0/-4, 1/-4, 0/-5, 0/-6}
				{
					\draw[thick] (\i,\j) -- (\i+1,\j);
				}
				\foreach \i/\j in {0/0, 0/-1, 0/-2, 0/-3, 0/-4, 0/-5, 1/0, 1/-1, 1/-2, 1/-3, 1/-4, 1/-5, 2/0, 2/-1, 2/-2, 2/-3, 3/0, 3/-1, 3/-2, 4/0, 4/-1, 4/-2, 5/0, 5/-1, 6/0, 6/-1, 7/0}
				{
					\draw[thick] (\i,\j) -- (\i,\j-1);
				}
				\foreach \i/\j in {}
				{
					\draw[thick] (\i,\j) -- (\i-1,\j-1);
				}
				\draw[-stealth] (0,0) -- (0,-7);
				\draw[-stealth] (0,0) -- (8,0);
				\node[left] at (0,-7) {$\scriptstyle i$};
				\node[right] at (8,0) {$\scriptstyle j$};
				\node at (4,-9) {(a) \ $\lambda$};
			\end{tikzpicture}
		\end{array}
		\hspace{10mm}
		\begin{array}{c}
			\begin{tikzpicture}[scale=0.4]
				\foreach \i/\j in {0/0, 1/0, 2/0, 3/0, 4/0, 5/0, 6/0, 0/-1, 1/-1, 2/-1, 3/-1, 4/-1, 5/-1, 6/-1, 0/-2, 1/-2, 2/-2, 3/-2, 4/-2, 5/-2, 0/-3, 1/-3, 2/-3, 3/-3, 0/-4, 1/-4, 0/-5, 0/-6}
				{
					\draw[thick] (\i,\j) -- (\i+1,\j);
				}
				\foreach \i/\j in {0/0, 0/-1, 0/-2, 0/-3, 0/-4, 0/-5, 1/0, 1/-1, 1/-2, 1/-3, 1/-4, 1/-5, 2/0, 2/-1, 2/-2, 2/-3, 3/0, 3/-1, 3/-2, 4/0, 4/-1, 4/-2, 5/0, 5/-1, 6/0, 6/-1, 7/0}
				{
					\draw[thick] (\i,\j) -- (\i,\j-1);
				}
				\foreach \i/\j in {}
				{
					\draw[thick] (\i,\j) -- (\i-1,\j-1);
				}
				\foreach \x/\y in {7/0, 6/1, 4/2, 2/3, 1/4, 0/6}
				{
					\draw[fill=pink] (\x,-\y) -- (\x+1,-\y) -- (\x+1,-\y-1) -- (\x,-\y-1) -- cycle;
				}
				\node at (4,-10) {(b) \ $\text{Add}(\lambda$)};
			\end{tikzpicture}
		\end{array}
		\hspace{10mm}
		\begin{array}{c}
			\begin{tikzpicture}[scale=0.4]
				\foreach \i/\j in {0/0, 1/0, 2/0, 3/0, 4/0, 5/0, 6/0, 0/-1, 1/-1, 2/-1, 3/-1, 4/-1, 5/-1, 6/-1, 0/-2, 1/-2, 2/-2, 3/-2, 4/-2, 5/-2, 0/-3, 1/-3, 2/-3, 3/-3, 0/-4, 1/-4, 0/-5, 0/-6}
				{
					\draw[thick] (\i,\j) -- (\i+1,\j);
				}
				\foreach \i/\j in {0/0, 0/-1, 0/-2, 0/-3, 0/-4, 0/-5, 1/0, 1/-1, 1/-2, 1/-3, 1/-4, 1/-5, 2/0, 2/-1, 2/-2, 2/-3, 3/0, 3/-1, 3/-2, 4/0, 4/-1, 4/-2, 5/0, 5/-1, 6/0, 6/-1, 7/0}
				{
					\draw[thick] (\i,\j) -- (\i,\j-1);
				}
				\foreach \i/\j in {}
				{
					\draw[thick] (\i,\j) -- (\i-1,\j-1);
				}
				\foreach \x/\y in {6/0, 5/1, 3/2, 1/3, 0/5}
				{
					\draw[fill=teal] (\x,-\y) -- (\x+1,-\y) -- (\x+1,-\y-1) -- (\x,-\y-1) -- cycle;
				}
				\node at (4,-10) {(c) \ $\text{Rem}(\lambda$)};
			\end{tikzpicture}
		\end{array}
	\end{equation}
	
	Action of higher $\hat{e}_{k},\hat{f}_{k}$ is a direct consequence of \eqref{higher e and f via psi}:
	\begin{align}
		\begin{aligned}
			\hat{e}_k \, S_{\lambda} = \sum_{\begin{tikzpicture}[scale=0.15]
					\foreach \i/\j in {0/0, 0/-1}
					{
						\draw[thick] (\i,\j) -- (\i+1,\j);
					}
					\foreach \i/\j in {0/0, 1/0}
					{
						\draw[thick] (\i,\j) -- (\i,\j-1);
					}
					\foreach \i/\j in {}
					{
						\draw[thick] (\i,\j) -- (\i-1,\j-1);
					}
					\foreach \x/\y in {0/0}
					{
						\draw[fill=pink] (\x,-\y) -- (\x+1,-\y) -- (\x+1,-\y-1) -- (\x,-\y-1) -- cycle;
					}
				\end{tikzpicture} \in \text{Add}(\lambda)} \ (\omega_{\begin{tikzpicture}[scale=0.15]
					\foreach \i/\j in {0/0, 0/-1}
					{
						\draw[thick] (\i,\j) -- (\i+1,\j);
					}
					\foreach \i/\j in {0/0, 1/0}
					{
						\draw[thick] (\i,\j) -- (\i,\j-1);
					}
					\foreach \i/\j in {}
					{
						\draw[thick] (\i,\j) -- (\i-1,\j-1);
					}
					\foreach \x/\y in {0/0}
					{
						\draw[fill=pink] (\x,-\y) -- (\x+1,-\y) -- (\x+1,-\y-1) -- (\x,-\y-1) -- cycle;
					} 
			\end{tikzpicture}})^{k} \,  S_{\lambda + \begin{tikzpicture}[scale=0.15]
					\foreach \i/\j in {0/0, 0/-1}
					{
						\draw[thick] (\i,\j) -- (\i+1,\j);
					}
					\foreach \i/\j in {0/0, 1/0}
					{
						\draw[thick] (\i,\j) -- (\i,\j-1);
					}
					\foreach \i/\j in {}
					{
						\draw[thick] (\i,\j) -- (\i-1,\j-1);
					}
					\foreach \x/\y in {0/0}
					{
						\draw[fill=pink] (\x,-\y) -- (\x+1,-\y) -- (\x+1,-\y-1) -- (\x,-\y-1) -- cycle;
					} 
			\end{tikzpicture}}  \\
			\hat{f}_k \, S_{\lambda} = \sum_{\begin{tikzpicture}[scale=0.15]
					\foreach \i/\j in {0/0, 0/-1}
					{
						\draw[thick] (\i,\j) -- (\i+1,\j);
					}
					\foreach \i/\j in {0/0, 1/0}
					{
						\draw[thick] (\i,\j) -- (\i,\j-1);
					}
					\foreach \i/\j in {}
					{
						\draw[thick] (\i,\j) -- (\i-1,\j-1);
					}
					\foreach \x/\y in {0/0}
					{
						\draw[fill=teal] (\x,-\y) -- (\x+1,-\y) -- (\x+1,-\y-1) -- (\x,-\y-1) -- cycle;
					}
				\end{tikzpicture} \in \text{Rem}(\lambda)} \ (\omega_{\begin{tikzpicture}[scale=0.15]
					\foreach \i/\j in {0/0, 0/-1}
					{
						\draw[thick] (\i,\j) -- (\i+1,\j);
					}
					\foreach \i/\j in {0/0, 1/0}
					{
						\draw[thick] (\i,\j) -- (\i,\j-1);
					}
					\foreach \i/\j in {}
					{
						\draw[thick] (\i,\j) -- (\i-1,\j-1);
					}
					\foreach \x/\y in {0/0}
					{
						\draw[fill=teal] (\x,-\y) -- (\x+1,-\y) -- (\x+1,-\y-1) -- (\x,-\y-1) -- cycle;
					}
			\end{tikzpicture}})^{k} \, S_{\lambda - \begin{tikzpicture}[scale=0.15]
					\foreach \i/\j in {0/0, 0/-1}
					{
						\draw[thick] (\i,\j) -- (\i+1,\j);
					}
					\foreach \i/\j in {0/0, 1/0}
					{
						\draw[thick] (\i,\j) -- (\i,\j-1);
					}
					\foreach \i/\j in {}
					{
						\draw[thick] (\i,\j) -- (\i-1,\j-1);
					}
					\foreach \x/\y in {0/0}
					{
						\draw[fill=teal] (\x,-\y) -- (\x+1,-\y) -- (\x+1,-\y-1) -- (\x,-\y-1) -- cycle;
					}
			\end{tikzpicture}}
		\end{aligned}
	\end{align}
	
	$\hat{W}$-operator of Gaussian Hermitian model is expressed via $\hat{e}_k$ by the following simple formula:
	\begin{equation}
		\hat{W} = \frac{1}{2} \Big[ \hat{e}_2, \hat{e}_1 \Big] 
	\end{equation}
	
	Action of $\hat{W}$ operator on Schur polynomials \eqref{W action} follows from the above formulas by direct computation. Therefore superintegrable formula \eqref{SI formula for GHM} acquires natural interpretation in terms of $W_{1 + \infty}$ algebra.
	
	Remarkable fact is that all components of the above story \textit{simultaneously} admit one parametric deformation widely known as $\beta$-deformation. Under this deformation the matrix integral becomes eigenvalue $\beta$-ensemble \cite{Mishnyakov:2022bkg, Wang:2022lzj, Bawane:2022cpd}; superintegrable property persist with one parametric deformation of Schur polynomials $S_{\lambda}$ that bear the name of Jack polynomials $J_{\lambda}^{\beta}$; the underlying algebra becomes affine Yangian $Y(\hat{\mathfrak{gl}}_1)$ \cite{Tsymbaliuk2017, 2012arXiv1211.1287M, Prochazka:2015deb, Morozov:2023vra}. This story certainly deserves attention, however, it is under the scope of this paper.

	Motivated by the tight relation of superintegrable property of Gaussian Hermitian model and $W_{1 + \infty}$ algebra we propose the conjecture: superintegrability is a manifestation of underlying symmetry algebra. However, it seems a difficult question to tell the algebra for a given matrix model. Basing on the above mentioned example we propose a set of assumptions:
	\begin{itemize}
		\item \textit{Diagrams}.
		
		In GH model superintegrable basis was enumerated by Young diagrams, in particular, number of graded polynomials $P(n)$ on the level $n$ of time-variables $p_a$ matches the number of Young diagrams with $n$ boxes. 
		
		We suppose that set of time-variables $T_{a}$ is known for a given matrix model. Then we analyze the number of graded polynomials of corresponding time-variables of this model and choose the proper set of diagrams $\Lambda$ by counting argument. This step of the algorithm is rather a clever guess. For now we do not have any additional general arguments to systematically extract the shape of diagrams from the given matrix model. However, in each particular case there are may be more arguments specific for a given model.
		
		Superintegrable basis $\text{SIB}_{\Lambda}(T_a)$ then is enumerated by diagrams $\Lambda$.
		
		\item \textit{Box adding/removing rule}.
		
		Usually the first (the smallest) time-variable, say $T_1$, correspond to first superintegrable polynomial corresponding to just $\Box$ diagram. Therefore we suppose that the first time-variable $T_1$ and its derivative add and remove boxes:
		\begin{align}
			\begin{aligned}
				T_1 \cdot \text{SIB}_{\Lambda} = \sum_{\Box \in \text{Add}(\Lambda)} C_{\Lambda, \Lambda + \Box} \, \text{SIB}_{\Lambda + \Box} \\
				\frac{\partial}{\partial T_1} \, \text{SIB}_{\Lambda} = \sum_{\Box \in \text{Rem}(\Lambda)} \tilde{C}_{\Lambda, \Lambda - \Box} \, \text{SIB}_{\Lambda - \Box} \\
			\end{aligned}
		\end{align}
		We do not impose any relations on coefficients $C_{\Lambda, \Lambda + \Box},\tilde{C}_{\Lambda, \Lambda - \Box}$ and treat them as unknown constants. The main idea here is that the sum on r.h.s. runs only over limited subsets $\text{Add}(\Lambda)$,$\text{Rem}(\Lambda)$ of diagrams of size $|\Lambda| + 1$ and $|\Lambda|-1$ respectively. These assumptions are rather strong constraints on the form of $\text{SIB}_{\Lambda}$ polynomials. 
		
		Higher time-variables and Virasoro operators of given matrix model may act on superintegrable polynomials by adding and removing rule. However, for a non minimal grading there are in general several different adding/removing rules.
		
		Remarkably, if a set of diagrams $\Lambda$ contain linear (single-row or single-column) diagrams then this set is closed under operation of removing boxes. We use this fact to define linear system of equation on single-row polynomials.
		
		\item \textit{Hamiltonian for} $\text{SIB}_{\Lambda}$ \textit{polynomials}.
		
		Box adding/removing rules impose strong constraints on $\text{SIB}_{\Lambda}$ polynomials, however, they can not fix the form of these polynomials completely. Motivated by the $W_{1 + \infty}$ example we postulate that $\text{SIB}_{\Lambda}$ polynomials should be eigenfunctions of some Hamiltonian operator $\hat{H}$:
		\begin{equation}
			\hat{H} \, \text{SIB}_{\Lambda} = \kappa_{\Lambda} \cdot \text{SIB}_{\Lambda}
		\end{equation}
		with unknown eigenvalues $\kappa_{\Lambda}$. The only assumption for the Hamiltonian operator is \textit{finite spin constraint}. Where by \textit{spin} we mean the number of time variables $T_a$ and derivatives $\frac{\partial}{\partial T_a}$ that enter each term (in normal ordered form) of the Hamiltonian. For example operator \eqref{Ham for Schurs} has spin 3. We will consider Hamiltonians of minimal possible spin in our presentation.
		
	\end{itemize}
	
	We argue that these two main assumptions: \textit{box adding/removing rule} and existence of \textit{Hamiltonian} may allow one to fully determine the form of superintegrable polynomials in some cases. 
	\\
	
	This paper is organized as follows. In Section \ref{sec::GH model} we demonstrate in details our method of construction of superintegrable basis. First, we analyze the diagrams. Second, we consider single-row family of diagrams and construct naively overdetermined system of equations for corresponding superintegrable polynomials, that has unique solution. Third, we introduce ansatz for Hamiltonian of finite (minimal) spin with unknown coefficients and partially fix these coefficients from single-row superintegrable polynomials. Finally, using box adding and removing rule we complete the basis and find the form of Hamiltonian. In Section \ref{sec::Kontsevich model} we apply the above mentioned algorithm to cubic Kontsevich matrix model and step by step restore its superintegrable basis -- Q-Schur polynomials. In Section \ref{sec::conclusion} we briefly summarize our results.
	\section{Gaussian Hermitian model}
	\label{sec::GH model}
	In this section we perform steps of our algorithm without references to Schur polynomials themselves, i.e. pretending that we do not know form of Schur polynomials.
	Our method starts from partition function of Gaussian Hermitian model that is given by the following integral over $N \times N$ Hermitian matrices $H$:
	\begin{equation}
		\label{partition func Hermitian model 1}
		Z_{GH} \left( p_k \right) = \frac{1}{Z} \int dH \exp\left( - \frac{1}{2} \tr H^2 + \sum_{k = 1}^{\infty} \frac{p_k}{k} \tr H^k \right)
	\end{equation}
	Partition function as a function of infinite family of time-variables $p_k$ obeys infinitely many linear equations (Ward identities) - Virasoro constraints \cite{David:1990ge,Ambjorn:1990ji,Mironov:1990im, Itoyama:1990mf}:
	\begin{equation}
		\hat{L}_n \, Z_{GH} = (n+2)\frac{\partial Z_{GH}}{\partial p_{n+2}}, \hspace{20mm} n = -1, 0, 1, 2, \ldots
	\end{equation}
	
	Explicit formulas for Virasoro operators depend on the model and in this particular case they have the following form:
	\begin{equation}
		\label{Virasoro GH}
		\hat{L}_n = \sum_{a=1}^{n-1} a(n-a) \, \frac{\partial}{\partial p_a} \frac{\partial}{\partial p_{n-a}} + \sum_{k=1}^{\infty} (n+k) \, p_k \frac{\partial}{\partial p_{n+k}} + 2 N n \, \frac{\partial}{\partial p_n} + N^2 \, \delta_{n,0} + N \, p_1 \, \delta_{n+1,0}
	\end{equation}
	Virasoro operators $L_n$ obey commutation relations of Virasoro algebra:
	\begin{equation}
		\Big[ \hat{L}_n, \hat{L}_m \Big] = (n - m) \hat{L}_{n+m} \hspace{20mm} n,m \geqslant -1
	\end{equation}

	\subsection{Step 1. Diagrams}
	Time-variables $p_k$ have a parameter - grading $[p_k]$. The gradings are encoded in $L_0$ Virasoro operator:
	\begin{equation}
		\hat{L}_0 = \sum_{k=1}^{\infty} \ k \cdot \, p_k \frac{\partial}{\partial p_{k}} + N^2 
	\end{equation}
	\begin{equation}
		[p_k] = k
	\end{equation}
	Usually it is possible to identify polynomial basis in time-variables with the set of diagrams. In case of time-variables $p_a$ with gradings $[p_a] = a$ the set of diagrams is given by Young diagrams.
	\begin{table}[h!]
		\centering
		$\begin{array}{|c|c|c|}
			\hline
			\text{Level} & \text{Young diagrams} & \text{Basis polynomials} \\
			\hline
			1 & \begin{array}{c}
				\begin{tikzpicture}
					\node(a1) at (0,0) {$\begin{array}{c}
							\begin{tikzpicture}[scale=0.3]
								\foreach \i/\j in {0/0, 0/-1}
								{
									\draw[thick] (\i,\j) -- (\i+1,\j);
								}
								\foreach \i/\j in {0/0, 1/0}
								{
									\draw[thick] (\i,\j) -- (\i,\j-1);
								}
								\foreach \i/\j in {}
								{
									\draw[thick] (\i,\j) -- (\i-1,\j-1);
								}
							\end{tikzpicture}
						\end{array}$};
				\end{tikzpicture}
			\end{array} &
			\begin{array}{c}
				p_1
			\end{array} \\
			\hline
			2 &  \begin{array}{c}
				\begin{tikzpicture}
					\node(a1) at (0,0) {$\begin{array}{c}
							\begin{tikzpicture}[scale=0.3]
								\foreach \i/\j in {0/0, 0/-1, 1/0, 1/-1}
								{
									\draw[thick] (\i,\j) -- (\i+1,\j);
								}
								\foreach \i/\j in {0/0, 1/0, 2/0}
								{
									\draw[thick] (\i,\j) -- (\i,\j-1);
								}
								\foreach \i/\j in {}
								{
									\draw[thick] (\i,\j) -- (\i-1,\j-1);
								}
							\end{tikzpicture}
						\end{array}$};
					\node(a2) at (0.8,0) {$\begin{array}{c}
							\begin{tikzpicture}[scale=0.3]
								\foreach \i/\j in {0/0, 0/-1, 0/-2}
								{
									\draw[thick] (\i,\j) -- (\i+1,\j);
								}
								\foreach \i/\j in {0/0, 1/0, 0/-1, 1/-1}
								{
									\draw[thick] (\i,\j) -- (\i,\j-1);
								}
								\foreach \i/\j in {}
								{
									\draw[thick] (\i,\j) -- (\i-1,\j-1);
								}
							\end{tikzpicture}
						\end{array}$};
				\end{tikzpicture}
			\end{array} &
			\begin{array}{c}
				p_2, p_1^2
			\end{array} \\
			\hline
			3 &  \begin{array}{c}
				\begin{tikzpicture}
					\node(a1) at (0,0) {$\begin{array}{c}
							\begin{tikzpicture}[scale=0.3]
								\foreach \i/\j in {0/0, 0/-1, 1/0, 1/-1, 2/0, 2/-1}
								{
									\draw[thick] (\i,\j) -- (\i+1,\j);
								}
								\foreach \i/\j in {0/0, 1/0, 2/0, 3/0}
								{
									\draw[thick] (\i,\j) -- (\i,\j-1);
								}
								\foreach \i/\j in {}
								{
									\draw[thick] (\i,\j) -- (\i-1,\j-1);
								}
							\end{tikzpicture}
						\end{array}$};
					\node(a2) at (1.15,0) {$\begin{array}{c}
							\begin{tikzpicture}[scale=0.3]
								\foreach \i/\j in {0/0, 0/-1, 0/-2, 1/0, 1/-1}
								{
									\draw[thick] (\i,\j) -- (\i+1,\j);
								}
								\foreach \i/\j in {0/0, 1/0, 0/-1, 1/-1, 2/0}
								{
									\draw[thick] (\i,\j) -- (\i,\j-1);
								}
								\foreach \i/\j in {}
								{
									\draw[thick] (\i,\j) -- (\i-1,\j-1);
								}
							\end{tikzpicture}
						\end{array}$};
					\node(a3) at (2,0) {$\begin{array}{c}
							\begin{tikzpicture}[scale=0.3]
								\foreach \i/\j in {0/0, 0/-1, 0/-2, 0/-3}
								{
									\draw[thick] (\i,\j) -- (\i+1,\j);
								}
								\foreach \i/\j in {0/0, 1/0, 0/-1, 1/-1, 0/-2, 1/-2}
								{
									\draw[thick] (\i,\j) -- (\i,\j-1);
								}
								\foreach \i/\j in {}
								{
									\draw[thick] (\i,\j) -- (\i-1,\j-1);
								}
							\end{tikzpicture}
						\end{array}$};
				\end{tikzpicture}
			\end{array} &
			\begin{array}{c}
				p_3, p_2 p_1, p_1^3
			\end{array} \\
			\hline
			4 &
			\begin{array}{c}
				\begin{tikzpicture}
					\node(a1) at (0,0) {$\begin{array}{c}
							\begin{tikzpicture}[scale=0.3]
								\foreach \i/\j in {0/0, 0/-1, 1/0, 1/-1, 2/0, 2/-1, 3/0, 3/-1}
								{
									\draw[thick] (\i,\j) -- (\i+1,\j);
								}
								\foreach \i/\j in {0/0, 1/0, 2/0, 3/0, 4/0}
								{
									\draw[thick] (\i,\j) -- (\i,\j-1);
								}
								\foreach \i/\j in {}
								{
									\draw[thick] (\i,\j) -- (\i-1,\j-1);
								}
							\end{tikzpicture}
						\end{array}$};
					\node(a2) at (1.4,0) {$\begin{array}{c}
							\begin{tikzpicture}[scale=0.3]
								\foreach \i/\j in {0/0, 0/-1, 0/-2, 1/0, 1/-1, 2/0, 2/-1}
								{
									\draw[thick] (\i,\j) -- (\i+1,\j);
								}
								\foreach \i/\j in {0/0, 1/0, 0/-1, 1/-1, 2/0, 3/0}
								{
									\draw[thick] (\i,\j) -- (\i,\j-1);
								}
								\foreach \i/\j in {}
								{
									\draw[thick] (\i,\j) -- (\i-1,\j-1);
								}
							\end{tikzpicture}
						\end{array}$};
					\node(a3) at (2.45,0) {$\begin{array}{c}
							\begin{tikzpicture}[scale=0.3]
								\foreach \i/\j in {0/0, 0/-1, 0/-2, 1/0, 1/-1, 1/-2}
								{
									\draw[thick] (\i,\j) -- (\i+1,\j);
								}
								\foreach \i/\j in {0/0, 1/0, 0/-1, 1/-1, 2/0, 2/-1}
								{
									\draw[thick] (\i,\j) -- (\i,\j-1);
								}
								\foreach \i/\j in {}
								{
									\draw[thick] (\i,\j) -- (\i-1,\j-1);
								}
							\end{tikzpicture}
						\end{array}$};
					\node(a4) at (3.4,0) {$\begin{array}{c}
							\begin{tikzpicture}[scale=0.3]
								\foreach \i/\j in {0/0, 0/-1, 0/-2, 0/-3, 1/0, 1/-1}
								{
									\draw[thick] (\i,\j) -- (\i+1,\j);
								}
								\foreach \i/\j in {0/0, 1/0, 0/-1, 1/-1, 0/-2, 1/-2, 2/0}
								{
									\draw[thick] (\i,\j) -- (\i,\j-1);
								}
								\foreach \i/\j in {}
								{
									\draw[thick] (\i,\j) -- (\i-1,\j-1);
								}
							\end{tikzpicture}
						\end{array}$};
					\node(a5) at (4.2,0) {$\begin{array}{c}
							\begin{tikzpicture}[scale=0.3]
								\foreach \i/\j in {0/0, 0/-1, 0/-2, 0/-3, 0/-4}
								{
									\draw[thick] (\i,\j) -- (\i+1,\j);
								}
								\foreach \i/\j in {0/0, 1/0, 0/-1, 1/-1, 0/-2, 1/-2, 0/-3, 1/-3}
								{
									\draw[thick] (\i,\j) -- (\i,\j-1);
								}
								\foreach \i/\j in {}
								{
									\draw[thick] (\i,\j) -- (\i-1,\j-1);
								}
							\end{tikzpicture}
						\end{array}$};
				\end{tikzpicture}
			\end{array} &
			\begin{array}{c}
				p_4, p_3 p_1, p_2^2, p_2 p_1^2, p_1^4
			\end{array} \\
			\hline
			5 &
			\begin{array}{c}
				\begin{tikzpicture}
					\node(a1) at (0,0) {$\begin{array}{c}
							\begin{tikzpicture}[scale=0.3]
								\foreach \i/\j in {0/0, 0/-1, 1/0, 1/-1, 2/0, 2/-1, 3/0, 3/-1, 4/0, 4/-1}
								{
									\draw[thick] (\i,\j) -- (\i+1,\j);
								}
								\foreach \i/\j in {0/0, 1/0, 2/0, 3/0, 4/0, 5/0}
								{
									\draw[thick] (\i,\j) -- (\i,\j-1);
								}
								\foreach \i/\j in {}
								{
									\draw[thick] (\i,\j) -- (\i-1,\j-1);
								}
							\end{tikzpicture}
						\end{array}$};
					\node(a2) at (1.7,0) {$\begin{array}{c}
							\begin{tikzpicture}[scale=0.3]
								\foreach \i/\j in {0/0, 0/-1, 0/-2, 1/0, 1/-1, 2/0, 2/-1, 3/0, 3/-1}
								{
									\draw[thick] (\i,\j) -- (\i+1,\j);
								}
								\foreach \i/\j in {0/0, 1/0, 0/-1, 1/-1, 2/0, 3/0, 4/0}
								{
									\draw[thick] (\i,\j) -- (\i,\j-1);
								}
								\foreach \i/\j in {}
								{
									\draw[thick] (\i,\j) -- (\i-1,\j-1);
								}
							\end{tikzpicture}
						\end{array}$};
					\node(a3) at (3.1,0) {$\begin{array}{c}
							\begin{tikzpicture}[scale=0.3]
								\foreach \i/\j in {0/0, 0/-1, 0/-2, 1/0, 1/-1, 1/-2, 2/0, 2/-1}
								{
									\draw[thick] (\i,\j) -- (\i+1,\j);
								}
								\foreach \i/\j in {0/0, 1/0, 0/-1, 1/-1, 2/0, 2/-1, 3/0}
								{
									\draw[thick] (\i,\j) -- (\i,\j-1);
								}
								\foreach \i/\j in {}
								{
									\draw[thick] (\i,\j) -- (\i-1,\j-1);
								}
							\end{tikzpicture}
						\end{array}$};
					\node(a2) at (4.35,0) {$\begin{array}{c}
							\begin{tikzpicture}[scale=0.3]
								\foreach \i/\j in {0/0, 0/-1, 0/-2, 1/0, 1/-1, 2/0, 2/-1, 0/-3}
								{
									\draw[thick] (\i,\j) -- (\i+1,\j);
								}
								\foreach \i/\j in {0/0, 1/0, 0/-1, 1/-1, 2/0, 3/0, 0/-2, 1/-2}
								{
									\draw[thick] (\i,\j) -- (\i,\j-1);
								}
								\foreach \i/\j in {}
								{
									\draw[thick] (\i,\j) -- (\i-1,\j-1);
								}
							\end{tikzpicture}
						\end{array}$};
					\node(a4) at (5.4,0) {$\begin{array}{c}
							\begin{tikzpicture}[scale=0.3]
								\foreach \i/\j in {0/0, 0/-1, 0/-2, 1/0, 1/-1, 1/-2, 0/-3}
								{
									\draw[thick] (\i,\j) -- (\i+1,\j);
								}
								\foreach \i/\j in {0/0, 1/0, 0/-1, 1/-1, 2/0, 2/-1, 0/-2, 1/-2}
								{
									\draw[thick] (\i,\j) -- (\i,\j-1);
								}
								\foreach \i/\j in {}
								{
									\draw[thick] (\i,\j) -- (\i-1,\j-1);
								}
							\end{tikzpicture}
						\end{array}$};
					\node(a4) at (6.3,0) {$\begin{array}{c}
							\begin{tikzpicture}[scale=0.3]
								\foreach \i/\j in {0/0, 0/-1, 0/-2, 0/-3, 1/0, 1/-1, 0/-4}
								{
									\draw[thick] (\i,\j) -- (\i+1,\j);
								}
								\foreach \i/\j in {0/0, 1/0, 0/-1, 1/-1, 0/-2, 1/-2, 2/0, 0/-3, 1/-3}
								{
									\draw[thick] (\i,\j) -- (\i,\j-1);
								}
								\foreach \i/\j in {}
								{
									\draw[thick] (\i,\j) -- (\i-1,\j-1);
								}
							\end{tikzpicture}
						\end{array}$};
					\node(a5) at (7,0) {$\begin{array}{c}
							\begin{tikzpicture}[scale=0.3]
								\foreach \i/\j in {0/0, 0/-1, 0/-2, 0/-3, 0/-4, 0/-5}
								{
									\draw[thick] (\i,\j) -- (\i+1,\j);
								}
								\foreach \i/\j in {0/0, 1/0, 0/-1, 1/-1, 0/-2, 1/-2, 0/-3, 1/-3, 0/-4, 1/-4}
								{
									\draw[thick] (\i,\j) -- (\i,\j-1);
								}
								\foreach \i/\j in {}
								{
									\draw[thick] (\i,\j) -- (\i-1,\j-1);
								}
							\end{tikzpicture}
						\end{array}$};
				\end{tikzpicture}
			\end{array} &
			\begin{array}{c}
				p_5, p_4 p_1, p_3 p_2, p_3 p_1^2, p_2^2 p_1, p_2 p_1^3, p_1^5
			\end{array} \\
			\hline
		\end{array}$
		\caption{\small{Young diagrams and graded polynomials on small levels 1-5. Young diagrams are in one-to-one correspondence to the graded polynomials in time-variables $p_a$.}}
		\label{Boson table}
	\end{table}
	
	Generating function of number of Young diagrams $P^{YD}(n)$ on level $n$ have the following form:
	\begin{equation}
		\prod_{k = 1} \frac{1}{1 - q^k} = \sum_{n = 0} q^n P^{YD}(n) = 1 + q + 2 q^2 + 3q^3 + 5 q^4 + 7 q^5 + \ldots
	\end{equation}

	\subsection{Step 2. System of equations for single-row polynomials}
	
	Virasoro operators $L_n$ and time-variables $p_k, \frac{\partial}{\partial p_k}$ have the following gradings:
	\begin{table}[h!]
		\centering
		$\begin{array}{|c|c|c|c|c|c|c|c|c|c|}
			\hline
			\text{grading} & \ldots & 3 & 2 & 1 & 0 & -1 & -2 & -3 & \ldots \\
			\hline
			\text{time-variables} & \ldots & p_3 & p_2 & p_1 & 1 & \frac{\partial}{\partial p_1} & \frac{\partial}{\partial p_2} & \frac{\partial}{\partial p_3} & \ldots \\
			\hline
			\text{Virasoro operators} &  &  &  & L_{-1} & L_{0} & L_{1} & L_{2} & L_{3} & \ldots \\
			\hline
		\end{array}$
	\end{table} 
	
	According to the list of assumptions (listed in the end of Section \ref{sec::intro}) we observe the following property of single-row subset of Young diagrams. If we act on a SIB (i.e. SuperIntegrable Basis) polynomial $\text{SIB}_{[n]}$ labeled by single-row diagram $[n]$ of length $n$ with operator $\mathcal{O}_m$ of negative grading $\left[ \mathcal{O}_m \right] = -m$ we get $\text{SIB}_{[n-m]}$ without any option, because there is only  \textit{one way} to remove $m$ boxes from single-row diagram. Pictorially this statement looks as follows in a particular example $n = 8, m = 3$:
	\begin{equation}
		\hat{\mathcal{O}}_{3} \, \text{SIB}_{\begin{ytableau}
				\ & \ & \ & \ & \ & *(teal) & *(teal) & *(teal) 
		\end{ytableau}} \sim \text{SIB}_{\begin{ytableau}
				\ & \ & \ & \ & \  
		\end{ytableau}}
	\end{equation}
	Here we colored the boxes that are removed after the action of operator. Our approach is based on removing boxes, because adding boxes immediately takes us out from single-row subset. For example, operator of positive grading 1 add a box into second row:
	\begin{equation}
		\hat{\mathcal{O}}_{-1} \, \text{SIB}_{\begin{ytableau}
				\ & \ & \ & \ & \ 
		\end{ytableau}} \sim \text{SIB}_{\begin{ytableau}
				\ & \ & \ & \ & \ & *(pink) \\
		\end{ytableau}} + \text{SIB}_{\begin{ytableau}
				\ & \ & \ & \ & \ \\
				*(pink)
		\end{ytableau}}
	\end{equation}
	
	We consider the following system of equations:
	\begin{tcolorbox}
		\begin{align}
			\label{single row system}
			\frac{\partial \, \text{SIB}_{[n]}}{\partial p_k} = u_{n,k} \, \text{SIB}_{[n-k]} \\
			\hat{L}_k \, \text{SIB}_{[n]} = v_{n,k} \, \text{SIB}_{[n-k]}
		\end{align}
	\end{tcolorbox}
	This system naively is heavily overdetermined; however it is self-consistent and has a unique solution. We consider the following general form of SIB polynomials:
	
	\begin{align}
		\begin{aligned}
			\text{SIB}_{[0]} &= 1 \\
			\text{SIB}_{[1]} &= p_1 \\
			\text{SIB}_{[2]} &= \frac{p_1^2}{2} + a_{2,1} \, p_2 \\
			\text{SIB}_{[3]} &= \frac{p_1^3}{6} + a_{3,1} \, p_2p_1 + a_{3,2} \, p_3 \\
			\text{SIB}_{[4]} &= \frac{p_1^4}{24} + a_{4,1} \, p_2p_1^2 + a_{4,2} \, p_2^2 + a_{4,3} \, p_{3} p_1 + a_{4,4} \, p_4 \\
			\text{SIB}_{[5]} &= \frac{p_1^5}{120} + a_{5,1} \, p_2p_1^3 + a_{5,2} \, p_2^2p_1 + a_{5,3} \, p_{3} p_1^2 + a_{5,4} \, p_4p_1 + a_{5,5} \, p_3p_2 + a_{5,6} \, p_5 \\
			\ldots
		\end{aligned}
	\end{align}
	
	The overall normalization of polynomials is fixed by the first coefficient that we choose to be $1/n!$. This choice makes the coefficients $u_{n,1} = 1$ trivial. The solution of this system is unique and parametrized by one parameter $x$:
	\begin{align}
		\label{SIB HGM}
		\begin{aligned}
			\text{SIB}_{[0]} &= 1 \\
			\text{SIB}_{[1]} &= p_1 \\
			\text{SIB}_{[2]} &= \frac{p_1^2}{2}+x \, \frac{p_2}{2} \\
			\text{SIB}_{[3]} &= \frac{p_1^3}{6}+x \, \frac{p_1 p_2}{2}+x^2 \, \frac{p_3}{3} \\
			\text{SIB}_{[4]} &= \frac{p_1^4}{24}+  x \, \frac{p_1^2 p_2}{4}+ x^2 \, \frac{p_2^2}{8}+x^2 \, \frac{p_1 p_3 }{3}+x^3 \, \frac{p_4}{4} \\
			\text{SIB}_{[5]} &= \frac{p_1^5}{120}+ x \, \frac{p_1^3 p_2}{12}+ x^2 \, \frac{p_1 p_2^2}{8}+ x^2 \, \frac{p_1^2 p_3}{6}+x^3  \, \frac{p_2 p_3}{6}+x^3 \, \frac{ p_1 p_4}{4}+x^4 \, \frac{p_5}{5} \\
			\ldots
		\end{aligned}
	\end{align}
	Formulas for coefficients $u_{n,k},v_{n,k}$ have the following form:
	\begin{align}
		\label{u v coefs}
		u_{n,k}=x^{k-1}\\
		v_{n,k}= x^{k-2}\left(k-1+2x N+(n-k)x^2\right)
	\end{align}
	
	
	Note, that free parameter $x$ is nothing but rescaling of time-variables $p_k \to x^{k-1} \, p_k$. However, Virasoro operators \eqref{Virasoro GH} have fixed normalization of time-variables. We observe, that correct normalization is restored when coefficients $u_{n,k}, v_{n,k}$ do not have power dependence on $k$. In this case it corresponds to value $\boxed{x = 1}$.
	
	\subsection{Step 3. Hamiltonian for SIB polynomials}
	To complete the polynomials $\text{SIB}_{[n]}$ to complete basis we use additional assumption. The first of them is a finite-spin Hamiltonian $\hat{H}$ that has SIB as the set of eigenfunctions:
	\begin{equation}
		\label{Ham Sch}
		\hat{H} \, \text{SIB}_{\lambda} = \kappa_{\lambda} \, \text{SIB}_{\lambda}
	\end{equation}
	Here we use assumption that $\hat{H}$ has the smallest possible spin = number of time-variable and derivatives. In this particular case it is possible to construct non-trivial zero grading operator of spin 3:
	\begin{equation}
		\label{Hamiltonian HGM}
		\hat{H} = \sum_{a,b = 1}^{\infty} A_{a,b} \cdot p_a p_b \frac{\partial}{\partial p_{a+b}} + B_{a,b} \cdot p_{a+b} \frac{\partial}{\partial p_{a}} \frac{\partial}{\partial p_{b}}
	\end{equation}
	where $A_{a,b} = A_{b,a}$ and $B_{a,b} = B_{b,a}$ are some unknown coefficients. Similar formulas were considered in \cite{Morozov:2022ndt}. Solving equation \eqref{Ham Sch} for the already found functions \eqref{SIB HGM}:
	\begin{equation}
		\label{single row eigen eq}
		\hat{H} \, \text{SIB}_{[n]} = \kappa_{[n]} \, \text{SIB}_{[n]}
	\end{equation}
	we can fix the form of coefficients $A_{a,b}, B_{a,b}$. We choose overall normalization of Hamiltonian by setting $B_{1,1} = 1$. Then eigenvalues are restored immediately:
	
	\begin{table}[h!]
		\centering
		$\begin{array}{|c|c|c|c|c|c|c|c|c|c|c|}
			\hline
			n & 0 & 1 & 2 & 3 & 4 & 5 & 6 & 7 & 8 & \ldots \\
			\hline
			\kappa_{[n]} & 0 & 0 & 2 & 6 & 12 & 20 & 30 & 42 & 56 & \ldots \\
			\hline
		\end{array}$
	\end{table}
	
	The resulting formula for eigenvalues reads:
	\begin{equation}
		\kappa_{\lambda} = n(n-1)
	\end{equation}
	
	Coefficients $A_{a,b}$ are \textit{fully} restored from \eqref{single row eigen eq}:
	
	\begin{table}[h!]
		\centering
		$\begin{array}{|c||c|c|c|c|c|c|c|c|}
			\hline
			A_{a,b} & 1 & 2 & 3 & 4 & 5 & 6 & \ldots \\
			\hline
			\hline
			1 & 2 & 3 & 4 & 5 & 6 & 7 & \ldots \\
			\hline
			2 & 3 & 4 & 5 & 6 & 7 & 8 & \ldots \\
			\hline
			3 & 4 & 5 & 6 & 7 & 8 & 9 & \ldots \\
			\hline
			4 & 5 & 6 & 7 & 8 & 9 & 10 & \ldots \\
			\hline
			5 & 6 & 7 & 8 & 9 & 10 & 11 & \ldots \\
			\hline
			6 & 7 & 8 & 9 & 10 & 11 & 12 & \ldots \\
			\hline
			\ldots & \ldots & \ldots & \ldots & \ldots & \ldots & \ldots & \ldots \\
			\hline
		\end{array}$
	\end{table}
	
	The resulting formula reads:
	\begin{equation}
		\boxed{
			A_{a,b} = a + b 
		}
	\end{equation}
	
	The situation with coefficients $B_{a,b}$ is more involved. They can not be fully determined from the above reasoning, however, there are linear equations on them. One equation for each level:
	\begin{align}
		B_{1,1} = 1  \\
		\frac{1}{2}\, B_{1,2} + \frac{1}{2} \, B_{2,1} = 2 \\
		\frac{1}{3}\, B_{1,3} + \frac{1}{4} \, B_{2,2} + \frac{1}{3}\,B_{3,1} = 3  \\
		\frac{1}{4}\, B_{1,4} + \frac{1}{6}\, B_{2,3} + \frac{1}{6}\, B_{3,2} + \frac{1}{4}\, B_{4,1} = 4  \\
		\frac{1}{5}\, B_{1,5} + \frac{1}{8}\, B_{2,4} + \frac{1}{9}\, B_{3,3} + \frac{1}{8}\,  B_{4,2} + \frac{1}{5}\,  B_{5,1} = 5 \\	
		\ldots	
	\end{align}
	We intensionally do not use property $B_{a,b} = B_{b,a}$ in order to represent this system in a symmetric form. Generally, these equations read:
	\begin{equation}
		\label{eq on B coef}
		\sum_{a = 1}^{n-1} \frac{1}{a(n-a)} \cdot B_{a,n-a} = n
	\end{equation}
	\subsection{Step 4. Box adding and removing rule}
	\subsubsection{Small levels}
	For now we have considered two objects:
	\begin{itemize}
		\item Subset of single-row $\text{SIB}_{[n]}$ polynomials \eqref{SIB HGM}
		\item Spin-3 ansatz for simplest Hamiltonian \eqref{Hamiltonian HGM}
	\end{itemize}
	that obviously can not determine $\text{SIB}$ completely, however, they allow us to \textit{fully} determine complete SIB on 2 and 3 level, because the corresponding coefficients $B_{1,1} = 1$ and $B_{1,2}=B_{2,1}=2$ are known. \\
	
	2 level:
	\begin{align}
		\text{SIB}_{\begin{ytableau}
				\ & \ 
		\end{ytableau}} &= \frac{p_1^2}{2}+\frac{p_2}{2} &\hspace{20 mm} \kappa_{\begin{ytableau}
				\ & \ 
		\end{ytableau}} &= 2 \\
		\text{SIB}_{\begin{ytableau}
				\ \\
				\
		\end{ytableau}} &= \frac{p_1^2}{2}-\frac{p_2}{2}  &\hspace{20 mm} 
		\kappa_{\begin{ytableau}
				\ \\
				\
		\end{ytableau}} &= -2
	\end{align}
	
	3 level:
	\begin{align}
		\text{SIB}_{\begin{ytableau}
				\ & \ & \
		\end{ytableau}} &= \frac{p_1^3}{6}+\frac{p_1 p_2}{2}+\frac{p_3}{3}  & \hspace{20 mm} \kappa_{\begin{ytableau}
				\ & \ & \
		\end{ytableau}} &= 6 \\
		\text{SIB}_{\begin{ytableau}
				\ & \ \\
				\
		\end{ytableau}} &= \frac{p_1^3}{3}-\frac{p_3}{3} & \hspace{20mm} \kappa_{\begin{ytableau}
				\ & \ \\
				\
		\end{ytableau}} &= 0 \\
		\text{SIB}_{\begin{ytableau}
				\ \\
				\ \\
				\
		\end{ytableau}} &= \frac{p_1^3}{6}-\frac{p_1 p_2}{2}+\frac{p_3}{3} &\hspace{20 mm}  \kappa_{\begin{ytableau}
				\ \\
				\ \\
				\
		\end{ytableau}} &= -6
	\end{align}
	These results are in full agreement with box adding and removing rule. For example:
	\begin{equation}
		p_1 \cdot \text{SIB}_{\begin{ytableau}
				\ & \ 
		\end{ytableau}} = \text{SIB}_{\begin{ytableau}
				\ & \ & *(pink)
		\end{ytableau}} + \text{SIB}_{\begin{ytableau}
				\ & \ \\
				*(pink)
		\end{ytableau}}
	\end{equation}
	\begin{equation}
		\frac{\partial}{\partial p_1}\text{SIB}_{\begin{ytableau}
				\ & \ \\
				\
		\end{ytableau}} = \text{SIB}_{\begin{ytableau}
				\ & \ 
		\end{ytableau}} + \text{SIB}_{\begin{ytableau}
				\ \\
				\
		\end{ytableau}}
	\end{equation}
	\subsubsection{Higher levels}
	One can proceed further in computing SIB polynomials via box adding rules. Multiplication by $p_1$ increases the grading by one, therefore we write the following ansatz for adding a box:
	\begin{equation}
		p_1 \cdot \text{SIB}_{\begin{ytableau}
				\ & \ & \
		\end{ytableau}} = \alpha_1 \cdot \text{SIB}_{\begin{ytableau}
				\ & \ & \ & *(pink)
		\end{ytableau}} + \alpha_2 \cdot \text{SIB}_{\begin{ytableau}
				\ & \ & \ \\
				*(pink)
		\end{ytableau}}
	\end{equation}
	with some constants $\alpha_1, \alpha_2$ and the following general form of new polynomial:
	\begin{equation}
		\text{SIB}_{\begin{ytableau}
				\ & \ & \ \\
				\
		\end{ytableau}} = \frac{p_1^4}{8} + b_1 \cdot p_1^2 p_2 + b_2 \cdot p_2^2 + b_3 \cdot p_1 p_3 + b_4 \cdot p_4
	\end{equation}
	
	The solution to the above equation is the following:
	\begin{equation}
		\alpha_1 = 4-3\alpha_2 \hspace{10mm} b_1 = \frac{3}{4}-\frac{1}{2\alpha_2} \hspace{10mm} b_2 = \frac{3}{8}-\frac{1}{2\alpha_2} \hspace{10mm} b_3 = 1-\frac{1}{\alpha_2 } \hspace{10mm} b_4 = \frac{3}{4}-\frac{1}{\alpha_2}
	\end{equation}
	where only one unknown parameter $\alpha_2$ remained. We can find it via the Hamiltonian:
	\begin{equation}
		\hat{H} \, \text{SIB}_{\begin{ytableau}
				\ & \ & \ \\
				\
		\end{ytableau}} = \kappa_{\begin{ytableau}
				\ & \ & \ \\
				\
		\end{ytableau}} \, \text{SIB}_{\begin{ytableau}
				\ & \ & \ \\
				\
		\end{ytableau}}
	\end{equation}
	The remaining parameter $\alpha_2$ is fixed from the above equation, moreover the unknown coefficient $B_{1,3}$ of the Hamiltonian is fixed too:
	\begin{equation}
		\alpha_2 = 1 \hspace{20mm} B_{1,3} = 3
	\end{equation}
	Now all the coefficients of Hamiltonian of the 4-th level are fixed and the basis of 4-th level is known in full generality:
	\begin{align}
		\text{SIB}_{\begin{ytableau}
				\ & \ & \ & \
		\end{ytableau}} &= \frac{p_1^4}{24}+\frac{p_1^2 p_2}{4}+\frac{p_2^2}{8} +\frac{p_1 p_3}{3}+ \frac{p_4}{4}  & \hspace{20 mm} \kappa_{\begin{ytableau}
				\ & \ & \ & \
		\end{ytableau}} &= 12 \\
		\text{SIB}_{\begin{ytableau}
				\ & \ & \ \\
				\
		\end{ytableau}} &= \frac{p_1^4}{8}+\frac{p_1^2 p_2}{4}-\frac{p_2^2}{8} - \frac{p_4}{4}  & \hspace{20 mm} \kappa_{\begin{ytableau}
				\ & \ & \ \\
				\
		\end{ytableau}} &= 4 \\
		\text{SIB}_{\begin{ytableau}
				\ & \  \\
				\ & \
		\end{ytableau}} &= \frac{p_1^4}{12}+\frac{p_2^2}{4} -\frac{p_1 p_3}{3}  & \hspace{20 mm} \kappa_{\begin{ytableau}
				\ & \ \\
				\ & \
		\end{ytableau}} &= 0 \\
		\text{SIB}_{\begin{ytableau}
				\ & \  \\
				\ \\
				\
		\end{ytableau}} &= \frac{p_1^4}{8}-\frac{p_1^2 p_2}{4}-\frac{p_2^2}{8} +\frac{p_4}{4}  & \hspace{20 mm} \kappa_{\begin{ytableau}
				\ & \ \\
				\ \\
				\
		\end{ytableau}} &= -4 \\
		\text{SIB}_{\begin{ytableau}
				\ \\
				\ \\
				\ \\
				\
		\end{ytableau}} &= \frac{p_1^4}{24}-\frac{p_1^2 p_2}{4}+\frac{p_2^2}{8} +\frac{p_1 p_3}{3}- \frac{p_4}{4}  & \hspace{20 mm} \kappa_{\begin{ytableau}
				\ \\
				\ \\
				\ \\
				\
		\end{ytableau}} &= -12 \\
	\end{align}
	One can proceed further using box adding rule with Hamiltonian and level by level compute $\text{SIB}$ polynomials. In particular, one finds remaining coefficients of the Hamiltonian:
	\begin{align}
		\begin{aligned}
			B_{1,1} = 1\\
			\frac{1}{2}B_{1,2}=\frac{1}{2}B_{2,1} = 1\\
			\frac{1}{3}B_{1,3} = \frac{1}{4}B_{2,2} = \frac{1}{3} B_{3,1}=1 \\
			\frac{1}{4}B_{1,4} = \frac{1}{6}B_{2,3} =\frac{1}{6}B_{3,2} = \frac{1}{4} B_{4,1}=1 \\
			\ldots
		\end{aligned}
	\end{align}
	One can guess the solution for system of equations \eqref{eq on B coef}:
	\begin{equation}
		\boxed{B_{a,b} = a \cdot b}
	\end{equation}
	On this step we fully determine the Hamiltonian:
	\begin{equation}
		\label{Schur Hamiltonian}
		\hat{H} = \sum_{a,b = 1}^{\infty} (a+b) \cdot p_a p_b \frac{\partial}{\partial p_{a+b}} + a b \cdot p_{a+b} \frac{\partial}{\partial p_{a}} \frac{\partial}{\partial p_{b}}
	\end{equation}
	which has eigenvalues:
	\begin{equation}
		\label{Ham eigvals}
		\kappa_{\lambda} = 2 \cdot \sum_{\Box \in \lambda} j_{\Box} - i_{\Box}
	\end{equation}
	Remarkably, the eigenvalues are integer numbers that is a rather nontrivial property of Hamiltonian. Even if we know the full form of Hamiltonian, it does not determine SIB polynomials completely, because the spectrum is degenerate: eigenvalues start to coincide from 6-th level. However eigenfunction property equipped with box adding/removing rules \textit{determines} $\text{SIB}_{\lambda}$ \textit{completely}.
	
	\subsection{Action of Virasoro operators on $\text{SIB}_{\lambda}$ polynomials}
	
	Notably, the overall coefficients of $\text{SIB}_{\lambda}$ polynomials can be chosen in a way that the following equations have unit coefficients in front of polynomials in the r.h.s (Pierri rules):
	\begin{align}
		\label{SIB add remove box}
		p_1 \cdot \text{SIB}_{\lambda} &= \sum_{\Box \in \text{Add}(\lambda)} \text{SIB}_{\lambda + \Box} \\
		\frac{\partial}{\partial p_1} \, \text{SIB}_{\lambda} &= \sum_{\Box \in \text{Rem}(\lambda)}  \text{SIB}_{\lambda - \Box}
	\end{align}
	
	Another intriguing property of resulting $\text{SIB}_{\lambda}$ polynomials is that Virasoro operators add and remove boxes. We demonstrate this fact by several examples:
	\begin{equation}
		\hat{L}_{-1} \, \text{SIB}_{\lambda} = \sum_{\Box \in \text{Add}(\lambda)} (N + j_{\Box} - i_{\Box}) \cdot \text{SIB}_{\lambda + \Box}
	\end{equation}
	
	\begin{equation}
		\hat{L}_{1} \, \text{SIB}_{\lambda} = \sum_{\Box \in \text{Rem}(\lambda)} (2N + j_{\Box} - i_{\Box}) \cdot \text{SIB}_{\lambda - \Box}
	\end{equation}
	Operators $\hat{L}_{-1}$, $\hat{L}_1$ have gradings $1$ and $-1$ respectively, therefore they add and remove only one box. The situation is more involved for higher Virasoro operators and higher time-derivatives, they remove skew diagrams that are called \textit{border strips}. It is instructive to consider similar formulas for higher time-variables:
	\begin{equation}
		\label{GH model time derivatives}
		k \frac{\partial}{\partial p_k} \text{SIB}_{\lambda} = \sum_{\substack{ \text{bs} \in \text{BS}(\lambda) \\ |\text{bs}| = k }} (-)^{h(\text{bs})+1} \, \text{SIB}_{\lambda - \text{bs}}
	\end{equation} 
	We denote as $\text{BS}(\lambda)$ a set of border strips of diagram $\lambda$. A border strip (also called ribbon or rim-hook) is a skew diagram that is connected, has unit width. An example of border strip $\text{bs} \in \text{BS}([7,6,4,2,1,1])$ of size $|\text{bs}| = 7$ and height $h(\text{bs}) = 3$:
	
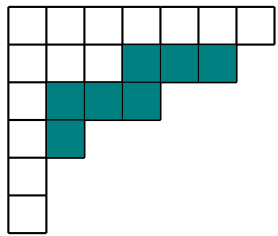
\begin{equation}
		\begin{tikzpicture}
			\node(B) at (6,0) {$\begin{array}{c}
					\begin{tikzpicture}[scale=0.5]
						\foreach \i/\j in {0/0, 1/0, 2/0, 3/0, 4/0, 5/0, 6/0, 0/-1, 1/-1, 2/-1, 3/-1, 4/-1, 5/-1, 6/-1, 0/-2, 1/-2, 2/-2, 3/-2, 4/-2, 5/-2, 0/-3, 1/-3, 2/-3, 3/-3, 0/-4, 1/-4, 0/-5, 0/-6}
						{
							\draw[thick] (\i,\j) -- (\i+1,\j);
						}
						\foreach \i/\j in {0/0, 0/-1, 0/-2, 0/-3, 0/-4, 0/-5, 1/0, 1/-1, 1/-2, 1/-3, 1/-4, 1/-5, 2/0, 2/-1, 2/-2, 2/-3, 3/0, 3/-1, 3/-2, 4/0, 4/-1, 4/-2, 5/0, 5/-1, 6/0, 6/-1, 7/0}
						{
							\draw[thick] (\i,\j) -- (\i,\j-1);
						}
						\foreach \i/\j in {}
						{
							\draw[thick] (\i,\j) -- (\i-1,\j-1);
						}
						\foreach \x/\y in {5/1, 4/1, 3/1, 3/2, 2/2, 1/2, 1/3}
						{
							\draw[fill=teal] (\x,-\y) -- (\x+1,-\y) -- (\x+1,-\y-1) -- (\x,-\y-1) -- cycle;
						}
					\end{tikzpicture}
				\end{array}$};
		\end{tikzpicture}
	\end{equation}
	Then formula for $L_2$ Virasoro operator reads:
	\begin{align}
		\begin{aligned}
			\hat{L}_2 \, \text{SIB}_{\lambda} = \sum_{\substack{ \text{bs} \in \text{BS}(\lambda) \\ |\text{bs}| = 2 }} (-)^{h(\text{bs})+1} \, \left( 2N + j_{\begin{tikzpicture}[scale=0.2]
					\foreach \x/\y in {0/0}
					{
						\draw[fill=red] (\x,-\y) -- (\x+1,-\y) -- (\x+1,-\y-1) -- (\x,-\y-1) -- cycle;
					}
				\end{tikzpicture}_{\, \text{bs}}} - i_{\begin{tikzpicture}[scale=0.2]
					\foreach \x/\y in {0/0}
					{
						\draw[fill=red] (\x,-\y) -- (\x+1,-\y) -- (\x+1,-\y-1) -- (\x,-\y-1) -- cycle;
					}
				\end{tikzpicture}_{\, \text{bs}}} \right) \text{SIB}_{\lambda - \text{bs}} +\\
			+\sum_{\substack{ \text{bs}, \text{bs}^{\prime} \in \text{BS}(\lambda) \\ |\text{bs}| + |\text{bs}^{\prime}| = 2 }} (-)^{h(\text{bs}) + h(\text{bs}^{\prime}) } \, \text{SIB}_{\lambda - \text{bs} - \text{bs}^{\prime}}
		\end{aligned}
	\end{align}
	In the above formula there are two sums, the first sum over border strips of size 2 while the second sum runs over pairs of (non-empty) border strips of overall size 2. In this particular case it means the second sum runs over pairs of (non-adjacent) boxes from set $\text{Rem}(\lambda)$. 
	
	In the first sum of the above formula content $j_{\Box} - i_{\Box}$ is evaluated in the special point $\begin{tikzpicture}[scale=0.2]
		\foreach \x/\y in {0/0}
		{
			\draw[fill=red] (\x,-\y) -- (\x+1,-\y) -- (\x+1,-\y-1) -- (\x,-\y-1) -- cycle;
		}
	\end{tikzpicture}_{\, \text{bs}}$ that depends on border strip. There are only $2^{k-1}$ possible border strips of size $k$ and we provide them explicitly for $k=2$:
	\begin{equation}
		\begin{array}{c}
			\begin{tikzpicture}[scale=0.3]
				\foreach \i/\j in {0/0, 0/-1, 1/0, 1/-1}
				{
					\draw[thick] (\i,\j) -- (\i+1,\j);
				}
				\foreach \i/\j in {0/0, 1/0, 2/0}
				{
					\draw[thick] (\i,\j) -- (\i,\j-1);
				}
				\foreach \i/\j in {}
				{
					\draw[thick] (\i,\j) -- (\i-1,\j-1);
				}
				\foreach \x/\y in {1/0}
				{
					\draw[fill=red] (\x,-\y) -- (\x+1,-\y) -- (\x+1,-\y-1) -- (\x,-\y-1) -- cycle;
				}
			\end{tikzpicture}
		\end{array}
		\hspace{10mm}
		\begin{array}{c}
			\begin{tikzpicture}[scale=0.3]
				\foreach \i/\j in {0/0, 0/-1, 0/-2}
				{
					\draw[thick] (\i,\j) -- (\i+1,\j);
				}
				\foreach \i/\j in {0/0, 1/0, 0/-1, 1/-1}
				{
					\draw[thick] (\i,\j) -- (\i,\j-1);
				}
				\foreach \i/\j in {}
				{
					\draw[thick] (\i,\j) -- (\i-1,\j-1);
				}
				\foreach \x/\y in {0/1}
				{
					\draw[fill=red] (\x,-\y) -- (\x+1,-\y) -- (\x+1,-\y-1) -- (\x,-\y-1) -- cycle;
				}
			\end{tikzpicture}
		\end{array}
	\end{equation}
	
	Details of the above formula are clear from explicit example:
	\begin{equation}
		\hat{L}_2 \, \text{SIB}_{\begin{ytableau}
				\ & \ & \ \\
				\ \\
				\ 
		\end{ytableau}} = (-)(2N - 2)\text{SIB}_{\begin{ytableau}
				\ & \ & \
		\end{ytableau}} + (2N + 2)\text{SIB}_{\begin{ytableau}
				\ \\
				\ \\
				\ 
		\end{ytableau}} + \text{SIB}_{\begin{ytableau}
				\ & \ \\
				\ 
		\end{ytableau}}
	\end{equation} 
	
	General formula for $\hat{L}_k$ operator action has the following form:
	\begin{tcolorbox}
		\begin{align}
			\label{GH model Virasoro action}
			\begin{aligned}
				\hat{L}_k \, \text{SIB}_{\lambda} = \sum_{\substack{ \text{bs} \in \text{BS}(\lambda) \\ |\text{bs}| = k }} (-)^{h(\text{bs})+1} \, \left( 2N + j_{\begin{tikzpicture}[scale=0.2]
						\foreach \x/\y in {0/0}
						{
							\draw[fill=red] (\x,-\y) -- (\x+1,-\y) -- (\x+1,-\y-1) -- (\x,-\y-1) -- cycle;
						}
					\end{tikzpicture}_{\, \text{bs}}} - i_{\begin{tikzpicture}[scale=0.2]
						\foreach \x/\y in {0/0}
						{
							\draw[fill=red] (\x,-\y) -- (\x+1,-\y) -- (\x+1,-\y-1) -- (\x,-\y-1) -- cycle;
						}
					\end{tikzpicture}_{\, \text{bs}}} \right) \text{SIB}_{\lambda - \text{bs}} +\\
				+\sum_{\substack{ \text{bs}, \text{bs}^{\prime} \in \text{BS}(\lambda) \\ |\text{bs}| + |\text{bs}^{\prime}| = k }} (-)^{h(\text{bs}) + h(\text{bs}^{\prime}) } \, \text{SIB}_{\lambda - \text{bs} - \text{bs}^{\prime}}
			\end{aligned}
		\end{align}
	\end{tcolorbox}
	
	The remaining element of the general formula that does not have yet any description is a position of the special box $\begin{tikzpicture}[scale=0.2]
		\foreach \x/\y in {0/0}
		{
			\draw[fill=red] (\x,-\y) -- (\x+1,-\y) -- (\x+1,-\y-1) -- (\x,-\y-1) -- cycle;
		}
	\end{tikzpicture}_{\, \text{bs}}$. We show all border strips of size 3 and 4 explicitly:
	\begin{equation}
		\begin{array}{c}
			\begin{tikzpicture}[scale=0.3]
				\foreach \i/\j in {0/0, 0/-1, 1/0, 1/-1, 2/0, 2/-1}
				{
					\draw[thick] (\i,\j) -- (\i+1,\j);
				}
				\foreach \i/\j in {0/0, 1/0, 2/0, 3/0}
				{
					\draw[thick] (\i,\j) -- (\i,\j-1);
				}
				\foreach \x/\y in {2/0}
				{
					\draw[fill=red] (\x,-\y) -- (\x+1,-\y) -- (\x+1,-\y-1) -- (\x,-\y-1) -- cycle;
				}
			\end{tikzpicture}
		\end{array} 
		\hspace{10mm}
		\begin{array}{c}
			\begin{tikzpicture}[scale=0.3]
				\foreach \i/\j in {0/0, 0/-1, 0/-2, 0/-3}
				{
					\draw[thick] (\i,\j) -- (\i+1,\j);
				}
				\foreach \i/\j in {0/0, 1/0, 0/-1, 1/-1, 0/-2, 1/-2}
				{
					\draw[thick] (\i,\j) -- (\i,\j-1);
				}
				\foreach \x/\y in {0/2}
				{
					\draw[fill=red] (\x,-\y) -- (\x+1,-\y) -- (\x+1,-\y-1) -- (\x,-\y-1) -- cycle;
				}
			\end{tikzpicture}
		\end{array}
		\hspace{10mm}
		\begin{array}{c}
			\begin{tikzpicture}[scale=0.3]
				\foreach \i/\j in {0/0, 0/-1, 0/-2, 1/-1, 1/0}
				{
					\draw[thick] (\i,\j) -- (\i+1,\j);
				}
				\foreach \i/\j in {0/0, 1/0, 0/-1, 1/-1, 2/0}
				{
					\draw[thick] (\i,\j) -- (\i,\j-1);
				}
				\foreach \x/\y in {0/0}
				{
					\draw[fill=red] (\x,-\y) -- (\x+1,-\y) -- (\x+1,-\y-1) -- (\x,-\y-1) -- cycle;
				}
			\end{tikzpicture}
		\end{array}
		\hspace{10mm}
		\begin{array}{c}
			\begin{tikzpicture}[scale=0.3]
				\foreach \i/\j in { 0/-1, 0/-2, 1/-1, 1/0, 1/-2}
				{
					\draw[thick] (\i,\j) -- (\i+1,\j);
				}
				\foreach \i/\j in { 1/0, 0/-1, 1/-1, 2/0, 2/-1}
				{
					\draw[thick] (\i,\j) -- (\i,\j-1);
				}
				\foreach \x/\y in {1/1}
				{
					\draw[fill=red] (\x,-\y) -- (\x+1,-\y) -- (\x+1,-\y-1) -- (\x,-\y-1) -- cycle;
				}
			\end{tikzpicture}
		\end{array}
	\end{equation}
	
	\begin{equation}
		\begin{array}{c}
			\begin{tikzpicture}[scale=0.3]
				\foreach \i/\j in {0/0, 0/-1, 1/0, 1/-1, 2/0, 2/-1, 3/0, 3/-1}
				{
					\draw[thick] (\i,\j) -- (\i+1,\j);
				}
				\foreach \i/\j in {0/0, 1/0, 2/0, 3/0, 4/0}
				{
					\draw[thick] (\i,\j) -- (\i,\j-1);
				}
				\foreach \x/\y in {3/0}
				{
					\draw[fill=red] (\x,-\y) -- (\x+1,-\y) -- (\x+1,-\y-1) -- (\x,-\y-1) -- cycle;
				}
			\end{tikzpicture}
		\end{array} 
		\hspace{5mm}
		\begin{array}{c}
			\begin{tikzpicture}[scale=0.3]
				\foreach \i/\j in {0/0, 0/-1, 0/-2, 0/-3, 0/-4}
				{
					\draw[thick] (\i,\j) -- (\i+1,\j);
				}
				\foreach \i/\j in {0/0, 1/0, 0/-1, 1/-1, 0/-2, 1/-2, 0/-3, 1/-3}
				{
					\draw[thick] (\i,\j) -- (\i,\j-1);
				}
				\foreach \x/\y in {0/3}
				{
					\draw[fill=red] (\x,-\y) -- (\x+1,-\y) -- (\x+1,-\y-1) -- (\x,-\y-1) -- cycle;
				}
			\end{tikzpicture}
		\end{array}
		\hspace{5mm}
		\begin{array}{c}
			\begin{tikzpicture}[scale=0.3]
				\foreach \i/\j in {0/0, 0/-1, 0/-2, 1/-1, 1/0, 2/-1, 2/0}
				{
					\draw[thick] (\i,\j) -- (\i+1,\j);
				}
				\foreach \i/\j in {0/0, 1/0, 0/-1, 1/-1, 2/0, 3/0}
				{
					\draw[thick] (\i,\j) -- (\i,\j-1);
				}
				\foreach \x/\y in {1/0}
				{
					\draw[fill=red] (\x,-\y) -- (\x+1,-\y) -- (\x+1,-\y-1) -- (\x,-\y-1) -- cycle;
				}
			\end{tikzpicture}
		\end{array}
		\hspace{5mm}
		\begin{array}{c}
			\begin{tikzpicture}[scale=0.3]
				\foreach \i/\j in {0/0, 0/-1, 0/-2, 1/-1, 1/0, 0/-3}
				{
					\draw[thick] (\i,\j) -- (\i+1,\j);
				}
				\foreach \i/\j in {0/0, 1/0, 0/-1, 1/-1, 2/0, 0/-2, 1/-2}
				{
					\draw[thick] (\i,\j) -- (\i,\j-1);
				}
				\foreach \x/\y in {0/1}
				{
					\draw[fill=red] (\x,-\y) -- (\x+1,-\y) -- (\x+1,-\y-1) -- (\x,-\y-1) -- cycle;
				}
			\end{tikzpicture}
		\end{array}
		\hspace{5mm}
		\begin{array}{c}
			\begin{tikzpicture}[scale=0.3]
				\foreach \i/\j in {0/0, 0/-1, 0/-2, 1/-1, 1/0, 1/1}
				{
					\draw[thick] (\i,\j) -- (\i+1,\j);
				}
				\foreach \i/\j in {0/0, 1/0, 0/-1, 1/-1, 2/0, 1/1, 2/1}
				{
					\draw[thick] (\i,\j) -- (\i,\j-1);
				}
				\foreach \x/\y in {0/0}
				{
					\draw[fill=red] (\x,-\y) -- (\x+1,-\y) -- (\x+1,-\y-1) -- (\x,-\y-1) -- cycle;
				}
			\end{tikzpicture}
		\end{array}
		\hspace{5mm}
		\begin{array}{c}
			\begin{tikzpicture}[scale=0.3]
				\foreach \i/\j in {0/0, 0/-1, 0/-2, 1/-1, 1/0, -1/-1, -1/-2}
				{
					\draw[thick] (\i,\j) -- (\i+1,\j);
				}
				\foreach \i/\j in {0/0, 1/0, 0/-1, 1/-1, 2/0, -1/-1}
				{
					\draw[thick] (\i,\j) -- (\i,\j-1);
				}
				\foreach \x/\y in {0/0}
				{
					\draw[fill=red] (\x,-\y) -- (\x+1,-\y) -- (\x+1,-\y-1) -- (\x,-\y-1) -- cycle;
				}
			\end{tikzpicture}
		\end{array}
		\hspace{5mm}
		\begin{array}{c}
			\begin{tikzpicture}[scale=0.3]
				\foreach \i/\j in { 0/-1, 0/-2, 1/-1, 1/0, 1/-2, 1/1}
				{
					\draw[thick] (\i,\j) -- (\i+1,\j);
				}
				\foreach \i/\j in { 1/0, 0/-1, 1/-1, 2/0, 2/-1, 1/1, 2/1}
				{
					\draw[thick] (\i,\j) -- (\i,\j-1);
				}
				\foreach \x/\y in {1/1}
				{
					\draw[fill=red] (\x,-\y) -- (\x+1,-\y) -- (\x+1,-\y-1) -- (\x,-\y-1) -- cycle;
				}
			\end{tikzpicture}
		\end{array}
		\hspace{5mm}
		\begin{array}{c}
			\begin{tikzpicture}[scale=0.3]
				\foreach \i/\j in { 0/-1, 0/-2, 1/-1, 1/0, 1/-2, -1/-1, -1/-2}
				{
					\draw[thick] (\i,\j) -- (\i+1,\j);
				}
				\foreach \i/\j in { 1/0, 0/-1, 1/-1, 2/0, 2/-1, -1/-1}
				{
					\draw[thick] (\i,\j) -- (\i,\j-1);
				}
				\foreach \x/\y in {1/1}
				{
					\draw[fill=red] (\x,-\y) -- (\x+1,-\y) -- (\x+1,-\y-1) -- (\x,-\y-1) -- cycle;
				}
			\end{tikzpicture}
		\end{array}
	\end{equation}
	However, this general formula demonstrates important property: Virasoro operators act on superintegrable polynomials by removing boxes from corresponding diagrams. Similar formulas was considered in \cite{Liu:2022oej}

	\section{Cubic Kontsevich model}
	
	\label{sec::Kontsevich model}
	
	In this section we will apply our method and construct superintegrable basis for Kontsevich matrix model \cite{Kontsevich:1992ti} in Kontsevich phase. According to the recent progress \cite{Mironov:2020tjf} the superintegrable basis for this model is given by Q-Schur polynomials \cite{Macdonald}, that are tightly connected to BKP integrability of this model \cite{Alexandrov:2020yxf, Drachov:2023xyz}. Our presentation will be parallel to the previous section.

	First, we define the partition function for Kontsevich model and write down Virasoro operators that realize the Ward identities. Next we discuss the time-variables of the model and analyze possible families of diagrams, that are combinatorial objects indexing basis graded polynomials.
	
	We start construction of superintegrable basis $\text{SIB}_{\lambda}$ \footnote{We denote superintegrable basis of Kontsevich model by the same name $\text{SIB}_{\lambda}$, however they are different from superintegrable basis of Gaussian Hermitian model. } from the single-row sector, which is closed with respect to the action of the box removing operators. Next we consider minimal spin ansatz for the Hamiltonian (in this case minimal spin is 4) for which SIB polynomials are eigenfunctions. To complete single-row subset to the full basis we impose box adding/removing rule for the minimal time-variable. These assumptions allow us to completely fix the form of SIB polynomials. Remarkably, the resulting Hamiltonian has only integer eigenvalues and higher Virasoro operators act on SIB polynomials by removing boxes.
	
	Partition function of Kontsevich matrix integral in Kontsevich phase is defined by the following integral over Hermitian $N \times N$ matrices $H$ with external matrix  $\Phi$:
	\begin{equation}
		\mathcal{Z}_K \left( \Phi \right)=\frac{1}{\mathcal{Z}(\Phi)}\int dH \exp \left(-\frac{1}{3!} \tr H^3-\frac{1}{2} \tr H^2 \Phi \right)
	\end{equation}
	where normalization factor $\mathcal{Z}(\Phi)$ is defined as follows:
	\begin{equation}
		\mathcal{Z}(\Phi) = \int dH \exp \left(-\frac{1}{2} \tr H^2 \Phi \right)
	\end{equation}
	Partition function is a function of an external matrix $\Phi$ and in Kontsevich phase it becomes a power series in variables $\tr \Phi^{-k}$ \cite{Kharchev:1991cu, Kharchev:1991cy, DiFrancesco:1992sm}. More importantly it depends only on odd negative powers of external matrix that we treat (in the limit $N \to \infty$) as independent time-variables $\tau_k$:
	\begin{gather}
		\tau_k=\frac{\tr \Phi^{-2k-1}}{2k+1}  \hspace{20mm} k = 0,1,2,3, \ldots
	\end{gather}
	As a function of time-variables partition function $\mathcal{Z}_K(\tau_k)$ obeys infinitely many linear equations - Virasoro constraints \cite{Marshakov:1991fu,Alexandrov:2008yq}:
	\begin{gather}
		\hat{\mathcal{L}}_n \, \mathcal{Z}_K=\frac{\partial \mathcal{Z}_K}{\partial\tau_{n+1}} \hspace{20mm}  n = -1, 0, 1, 2, \ldots
	\end{gather}
	Virasoro operators of this model has the following explicit form \cite{Alexandrov:2010bn}:
	\begin{equation}
		\hat{\mathcal{L}}_n = \frac{1}{8} \sum_{k=0}^{n-1} \frac{\partial}{\partial \tau_k} \frac{\partial}{\partial \tau_{n-k-1}} + \sum_{k=0}^{\infty} \left( k + \frac{1}{2} \right) \tau_k \frac{\partial}{\partial \tau_{n+k}} + \frac{\delta_{n,0}}{16} + \delta_{n+1,0} \, \frac{\tau_0^2}{2}
	\end{equation}
	and obey relations of Virasoro algebra:
	\begin{gather}
		\Big[ \hat{\mathcal{L}}_n , \hat{\mathcal{L}}_m \Big]=(n-m)\hat{\mathcal{L}}_{n+m} \hspace{15mm} n,m \geqslant -1
	\end{gather}
	\subsection{Step 1. Diagrams}
	
	\begin{table}[h!]
		\centering
		$\begin{array}{|c|c|c|c|}
			\hline
			\text{Lvl.} & \text{Odd partitions (OP)} & \boxed{\text{Distinct partitions (DP)}} & \text{Basis polynomials} \\
			\hline
			1 & 
			\begin{array}{c}
				\begin{tikzpicture}
					\node(a1) at (0,0) {$\begin{array}{c}
							\begin{tikzpicture}[scale=0.3]
								\foreach \i/\j in {0/0, 0/-1}
								{
									\draw[thick] (\i,\j) -- (\i+1,\j);
								}
								\foreach \i/\j in {0/0, 1/0}
								{
									\draw[thick] (\i,\j) -- (\i,\j-1);
								}
								\foreach \i/\j in {}
								{
									\draw[thick] (\i,\j) -- (\i-1,\j-1);
								}
							\end{tikzpicture}
						\end{array}$};
				\end{tikzpicture}
			\end{array} &
			\begin{array}{c}
				\begin{tikzpicture}
					\node(a1) at (0,0) {$\begin{array}{c}
							\begin{tikzpicture}[scale=0.3]
								\foreach \i/\j in {0/0, 0/-1}
								{
									\draw[thick] (\i,\j) -- (\i+1,\j);
								}
								\foreach \i/\j in {0/0, 1/0}
								{
									\draw[thick] (\i,\j) -- (\i,\j-1);
								}
								\foreach \i/\j in {}
								{
									\draw[thick] (\i,\j) -- (\i-1,\j-1);
								}
							\end{tikzpicture}
						\end{array}$};
				\end{tikzpicture}
			\end{array} &
			\begin{array}{c}
				\tau_0
			\end{array} \\
			\hline
			2 &  
			\begin{array}{c}
				\begin{tikzpicture}
					\node(a2) at (0.8,0) {$\begin{array}{c}
							\begin{tikzpicture}[scale=0.3]
								\foreach \i/\j in {0/0, 0/-1, 0/-2}
								{
									\draw[thick] (\i,\j) -- (\i+1,\j);
								}
								\foreach \i/\j in {0/0, 1/0, 0/-1, 1/-1}
								{
									\draw[thick] (\i,\j) -- (\i,\j-1);
								}
								\foreach \i/\j in {}
								{
									\draw[thick] (\i,\j) -- (\i-1,\j-1);
								}
							\end{tikzpicture}
						\end{array}$};
				\end{tikzpicture}
			\end{array}
			&
			\begin{array}{c}
				\begin{tikzpicture}
					\node(a1) at (0,0) {$\begin{array}{c}
							\begin{tikzpicture}[scale=0.3]
								\foreach \i/\j in {0/0, 0/-1, 1/0, 1/-1}
								{
									\draw[thick] (\i,\j) -- (\i+1,\j);
								}
								\foreach \i/\j in {0/0, 1/0, 2/0}
								{
									\draw[thick] (\i,\j) -- (\i,\j-1);
								}
								\foreach \i/\j in {}
								{
									\draw[thick] (\i,\j) -- (\i-1,\j-1);
								}
							\end{tikzpicture}
						\end{array}$};
				\end{tikzpicture} 
			\end{array}
			&
			\begin{array}{c}
				\tau_0^2
			\end{array} \\
			\hline
			3 &  \begin{array}{c}
				\begin{tikzpicture}
					\node(a1) at (0,0) {$\begin{array}{c}
							\begin{tikzpicture}[scale=0.3]
								\foreach \i/\j in {0/0, 0/-1, 1/0, 1/-1, 2/0, 2/-1}
								{
									\draw[thick] (\i,\j) -- (\i+1,\j);
								}
								\foreach \i/\j in {0/0, 1/0, 2/0, 3/0}
								{
									\draw[thick] (\i,\j) -- (\i,\j-1);
								}
								\foreach \i/\j in {}
								{
									\draw[thick] (\i,\j) -- (\i-1,\j-1);
								}
							\end{tikzpicture}
						\end{array}$};
					\node(a3) at (1,0) {$\begin{array}{c}
							\begin{tikzpicture}[scale=0.3]
								\foreach \i/\j in {0/0, 0/-1, 0/-2, 0/-3}
								{
									\draw[thick] (\i,\j) -- (\i+1,\j);
								}
								\foreach \i/\j in {0/0, 1/0, 0/-1, 1/-1, 0/-2, 1/-2}
								{
									\draw[thick] (\i,\j) -- (\i,\j-1);
								}
								\foreach \i/\j in {}
								{
									\draw[thick] (\i,\j) -- (\i-1,\j-1);
								}
							\end{tikzpicture}
						\end{array}$};
				\end{tikzpicture}
			\end{array} 
			&
			\begin{array}{c}
				\begin{tikzpicture}
					\node(a1) at (0,0) {$\begin{array}{c}
							\begin{tikzpicture}[scale=0.3]
								\foreach \i/\j in {0/0, 0/-1, 1/0, 1/-1, 2/0, 2/-1}
								{
									\draw[thick] (\i,\j) -- (\i+1,\j);
								}
								\foreach \i/\j in {0/0, 1/0, 2/0, 3/0}
								{
									\draw[thick] (\i,\j) -- (\i,\j-1);
								}
								\foreach \i/\j in {}
								{
									\draw[thick] (\i,\j) -- (\i-1,\j-1);
								}
							\end{tikzpicture}
						\end{array}$};
					\node(a2) at (1.15,0) {$\begin{array}{c}
							\begin{tikzpicture}[scale=0.3]
								\foreach \i/\j in {0/0, 0/-1, 0/-2, 1/0, 1/-1}
								{
									\draw[thick] (\i,\j) -- (\i+1,\j);
								}
								\foreach \i/\j in {0/0, 1/0, 0/-1, 1/-1, 2/0}
								{
									\draw[thick] (\i,\j) -- (\i,\j-1);
								}
								\foreach \i/\j in {}
								{
									\draw[thick] (\i,\j) -- (\i-1,\j-1);
								}
							\end{tikzpicture}
						\end{array}$};
				\end{tikzpicture}
			\end{array}
			&
			\begin{array}{c}
				\tau_1, \tau_0^3
			\end{array} \\
			\hline
			4 &
			\begin{array}{c}
				\begin{tikzpicture}
					\node(a2) at (0,0) {$\begin{array}{c}
							\begin{tikzpicture}[scale=0.3]
								\foreach \i/\j in {0/0, 0/-1, 0/-2, 1/0, 1/-1, 2/0, 2/-1}
								{
									\draw[thick] (\i,\j) -- (\i+1,\j);
								}
								\foreach \i/\j in {0/0, 1/0, 0/-1, 1/-1, 2/0, 3/0}
								{
									\draw[thick] (\i,\j) -- (\i,\j-1);
								}
								\foreach \i/\j in {}
								{
									\draw[thick] (\i,\j) -- (\i-1,\j-1);
								}
							\end{tikzpicture}
						\end{array}$};
					\node(a5) at (1,0) {$\begin{array}{c}
							\begin{tikzpicture}[scale=0.3]
								\foreach \i/\j in {0/0, 0/-1, 0/-2, 0/-3, 0/-4}
								{
									\draw[thick] (\i,\j) -- (\i+1,\j);
								}
								\foreach \i/\j in {0/0, 1/0, 0/-1, 1/-1, 0/-2, 1/-2, 0/-3, 1/-3}
								{
									\draw[thick] (\i,\j) -- (\i,\j-1);
								}
								\foreach \i/\j in {}
								{
									\draw[thick] (\i,\j) -- (\i-1,\j-1);
								}
							\end{tikzpicture}
						\end{array}$};
				\end{tikzpicture}
			\end{array} 
			&
			\begin{array}{c}
				\begin{tikzpicture}
					\node(a1) at (0,0) {$\begin{array}{c}
							\begin{tikzpicture}[scale=0.3]
								\foreach \i/\j in {0/0, 0/-1, 1/0, 1/-1, 2/0, 2/-1, 3/0, 3/-1}
								{
									\draw[thick] (\i,\j) -- (\i+1,\j);
								}
								\foreach \i/\j in {0/0, 1/0, 2/0, 3/0, 4/0}
								{
									\draw[thick] (\i,\j) -- (\i,\j-1);
								}
								\foreach \i/\j in {}
								{
									\draw[thick] (\i,\j) -- (\i-1,\j-1);
								}
							\end{tikzpicture}
						\end{array}$};
					\node(a2) at (1.4,0) {$\begin{array}{c}
							\begin{tikzpicture}[scale=0.3]
								\foreach \i/\j in {0/0, 0/-1, 0/-2, 1/0, 1/-1, 2/0, 2/-1}
								{
									\draw[thick] (\i,\j) -- (\i+1,\j);
								}
								\foreach \i/\j in {0/0, 1/0, 0/-1, 1/-1, 2/0, 3/0}
								{
									\draw[thick] (\i,\j) -- (\i,\j-1);
								}
								\foreach \i/\j in {}
								{
									\draw[thick] (\i,\j) -- (\i-1,\j-1);
								}
							\end{tikzpicture}
						\end{array}$};
				\end{tikzpicture}
			\end{array}
			&
			\begin{array}{c}
				\tau_1 \tau_0, \tau_0^4
			\end{array} \\
			\hline
			5 &
			\begin{array}{c}
				\begin{tikzpicture}
					\node(a1) at (0,0) {$\begin{array}{c}
							\begin{tikzpicture}[scale=0.3]
								\foreach \i/\j in {0/0, 0/-1, 1/0, 1/-1, 2/0, 2/-1, 3/0, 3/-1, 4/0, 4/-1}
								{
									\draw[thick] (\i,\j) -- (\i+1,\j);
								}
								\foreach \i/\j in {0/0, 1/0, 2/0, 3/0, 4/0, 5/0}
								{
									\draw[thick] (\i,\j) -- (\i,\j-1);
								}
								\foreach \i/\j in {}
								{
									\draw[thick] (\i,\j) -- (\i-1,\j-1);
								}
							\end{tikzpicture}
						\end{array}$};
					\node(a2) at (1.5,0) {$\begin{array}{c}
							\begin{tikzpicture}[scale=0.3]
								\foreach \i/\j in {0/0, 0/-1, 0/-2, 1/0, 1/-1, 2/0, 2/-1, 0/-3}
								{
									\draw[thick] (\i,\j) -- (\i+1,\j);
								}
								\foreach \i/\j in {0/0, 1/0, 0/-1, 1/-1, 2/0, 3/0, 0/-2, 1/-2}
								{
									\draw[thick] (\i,\j) -- (\i,\j-1);
								}
								\foreach \i/\j in {}
								{
									\draw[thick] (\i,\j) -- (\i-1,\j-1);
								}
							\end{tikzpicture}
						\end{array}$};
					\node(a5) at (2.4,0) {$\begin{array}{c}
							\begin{tikzpicture}[scale=0.3]
								\foreach \i/\j in {0/0, 0/-1, 0/-2, 0/-3, 0/-4, 0/-5}
								{
									\draw[thick] (\i,\j) -- (\i+1,\j);
								}
								\foreach \i/\j in {0/0, 1/0, 0/-1, 1/-1, 0/-2, 1/-2, 0/-3, 1/-3, 0/-4, 1/-4}
								{
									\draw[thick] (\i,\j) -- (\i,\j-1);
								}
								\foreach \i/\j in {}
								{
									\draw[thick] (\i,\j) -- (\i-1,\j-1);
								}
							\end{tikzpicture}
						\end{array}$};
				\end{tikzpicture}
			\end{array} 
			&
			\begin{array}{c}
				\begin{tikzpicture}
					\node(a1) at (0,0) {$\begin{array}{c}
							\begin{tikzpicture}[scale=0.3]
								\foreach \i/\j in {0/0, 0/-1, 1/0, 1/-1, 2/0, 2/-1, 3/0, 3/-1, 4/0, 4/-1}
								{
									\draw[thick] (\i,\j) -- (\i+1,\j);
								}
								\foreach \i/\j in {0/0, 1/0, 2/0, 3/0, 4/0, 5/0}
								{
									\draw[thick] (\i,\j) -- (\i,\j-1);
								}
								\foreach \i/\j in {}
								{
									\draw[thick] (\i,\j) -- (\i-1,\j-1);
								}
							\end{tikzpicture}
						\end{array}$};
					\node(a2) at (1.65,0) {$\begin{array}{c}
							\begin{tikzpicture}[scale=0.3]
								\foreach \i/\j in {0/0, 0/-1, 0/-2, 1/0, 1/-1, 2/0, 2/-1, 3/0, 3/-1}
								{
									\draw[thick] (\i,\j) -- (\i+1,\j);
								}
								\foreach \i/\j in {0/0, 1/0, 0/-1, 1/-1, 2/0, 3/0, 4/0}
								{
									\draw[thick] (\i,\j) -- (\i,\j-1);
								}
								\foreach \i/\j in {}
								{
									\draw[thick] (\i,\j) -- (\i-1,\j-1);
								}
							\end{tikzpicture}
						\end{array}$};
					\node(a3) at (3,0) {$\begin{array}{c}
							\begin{tikzpicture}[scale=0.3]
								\foreach \i/\j in {0/0, 0/-1, 0/-2, 1/0, 1/-1, 1/-2, 2/0, 2/-1}
								{
									\draw[thick] (\i,\j) -- (\i+1,\j);
								}
								\foreach \i/\j in {0/0, 1/0, 0/-1, 1/-1, 2/0, 2/-1, 3/0}
								{
									\draw[thick] (\i,\j) -- (\i,\j-1);
								}
								\foreach \i/\j in {}
								{
									\draw[thick] (\i,\j) -- (\i-1,\j-1);
								}
							\end{tikzpicture}
						\end{array}$};
				\end{tikzpicture}
			\end{array}
			&
			\begin{array}{c}
				\tau_2, \tau_1 \tau_0^2, \tau_0^5
			\end{array} \\
			\hline
			6 & 
			\begin{array}{c}
				\begin{tikzpicture}
					\node(a1) at (0,0) {$\begin{array}{c}
							\begin{tikzpicture}[scale=0.3]
								\foreach \i/\j in {0/0, 0/-1, 1/0, 1/-1, 2/0, 2/-1, 3/0, 3/-1, 4/0, 4/-1, 0/-2}
								{
									\draw[thick] (\i,\j) -- (\i+1,\j);
								}
								\foreach \i/\j in {0/0, 1/0, 2/0, 3/0, 4/0, 5/0, 0/-1, 1/-1}
								{
									\draw[thick] (\i,\j) -- (\i,\j-1);
								}
								\foreach \i/\j in {}
								{
									\draw[thick] (\i,\j) -- (\i-1,\j-1);
								}
							\end{tikzpicture}
						\end{array}$};
					\node(a3) at (1.4,0) {$\begin{array}{c}
							\begin{tikzpicture}[scale=0.3]
								\foreach \i/\j in {0/0, 0/-1, 0/-2, 1/0, 1/-1, 1/-2, 2/0, 2/-1, 2/-2}
								{
									\draw[thick] (\i,\j) -- (\i+1,\j);
								}
								\foreach \i/\j in {0/0, 1/0, 0/-1, 1/-1, 2/0, 2/-1, 3/0, 3/-1}
								{
									\draw[thick] (\i,\j) -- (\i,\j-1);
								}
								\foreach \i/\j in {}
								{
									\draw[thick] (\i,\j) -- (\i-1,\j-1);
								}
							\end{tikzpicture}
						\end{array}$};
					\node(a2) at (2.5,0) {$\begin{array}{c}
							\begin{tikzpicture}[scale=0.3]
								\foreach \i/\j in {0/0, 0/-1, 0/-2, 1/0, 1/-1, 2/0, 2/-1, 0/-3, 0/-4}
								{
									\draw[thick] (\i,\j) -- (\i+1,\j);
								}
								\foreach \i/\j in {0/0, 1/0, 0/-1, 1/-1, 2/0, 3/0, 0/-2, 1/-2, 0/-3, 1/-3}
								{
									\draw[thick] (\i,\j) -- (\i,\j-1);
								}
								\foreach \i/\j in {}
								{
									\draw[thick] (\i,\j) -- (\i-1,\j-1);
								}
							\end{tikzpicture}
						\end{array}$};
					\node(a5) at (3.3,0) {$\begin{array}{c}
							\begin{tikzpicture}[scale=0.3]
								\foreach \i/\j in {0/0, 0/-1, 0/-2, 0/-3, 0/-4, 0/-5}
								{
									\draw[thick] (\i,\j) -- (\i+1,\j);
								}
								\foreach \i/\j in {0/0, 1/0, 0/-1, 1/-1, 0/-2, 1/-2, 0/-3, 1/-3, 0/-4, 1/-4}
								{
									\draw[thick] (\i,\j) -- (\i,\j-1);
								}
								\foreach \i/\j in {}
								{
									\draw[thick] (\i,\j) -- (\i-1,\j-1);
								}
							\end{tikzpicture}
						\end{array}$};
				\end{tikzpicture}
			\end{array} 
			&
			\begin{array}{c}
				\begin{tikzpicture}
					\node(a1) at (0,0) {$\begin{array}{c}
							\begin{tikzpicture}[scale=0.3]
								\foreach \i/\j in {0/0, 0/-1, 1/0, 1/-1, 2/0, 2/-1, 3/0, 3/-1, 4/0, 4/-1, 5/0, 5/-1}
								{
									\draw[thick] (\i,\j) -- (\i+1,\j);
								}
								\foreach \i/\j in {0/0, 1/0, 2/0, 3/0, 4/0, 5/0, 6/0}
								{
									\draw[thick] (\i,\j) -- (\i,\j-1);
								}
								\foreach \i/\j in {}
								{
									\draw[thick] (\i,\j) -- (\i-1,\j-1);
								}
							\end{tikzpicture}
						\end{array}$};
					\node(a2) at (1.85,0) {$\begin{array}{c}
							\begin{tikzpicture}[scale=0.3]
								\foreach \i/\j in {0/0, 0/-1, 0/-2, 1/0, 1/-1, 2/0, 2/-1, 3/0, 3/-1, 4/0, 4/-1}
								{
									\draw[thick] (\i,\j) -- (\i+1,\j);
								}
								\foreach \i/\j in {0/0, 1/0, 0/-1, 1/-1, 2/0, 3/0, 4/0, 5/0}
								{
									\draw[thick] (\i,\j) -- (\i,\j-1);
								}
								\foreach \i/\j in {}
								{
									\draw[thick] (\i,\j) -- (\i-1,\j-1);
								}
							\end{tikzpicture}
						\end{array}$};
					\node(a3) at (3.4,0) {$\begin{array}{c}
							\begin{tikzpicture}[scale=0.3]
								\foreach \i/\j in {0/0, 0/-1, 0/-2, 1/0, 1/-1, 1/-2, 2/0, 2/-1, 3/0, 3/-1}
								{
									\draw[thick] (\i,\j) -- (\i+1,\j);
								}
								\foreach \i/\j in {0/0, 1/0, 0/-1, 1/-1, 2/0, 2/-1, 3/0, 4/0}
								{
									\draw[thick] (\i,\j) -- (\i,\j-1);
								}
								\foreach \i/\j in {}
								{
									\draw[thick] (\i,\j) -- (\i-1,\j-1);
								}
							\end{tikzpicture}
						\end{array}$};
					\node(a2) at (4.65,0) {$\begin{array}{c}
							\begin{tikzpicture}[scale=0.3]
								\foreach \i/\j in {0/0, 0/-1, 0/-2, 1/0, 1/-1, 2/0, 2/-1, 0/-3, 1/-2}
								{
									\draw[thick] (\i,\j) -- (\i+1,\j);
								}
								\foreach \i/\j in {0/0, 1/0, 0/-1, 1/-1, 2/0, 3/0, 0/-2, 1/-2, 2/-1}
								{
									\draw[thick] (\i,\j) -- (\i,\j-1);
								}
								\foreach \i/\j in {}
								{
									\draw[thick] (\i,\j) -- (\i-1,\j-1);
								}
							\end{tikzpicture}
						\end{array}$};
				\end{tikzpicture}
			\end{array}
			&
			\begin{array}{c}
				\tau_2 \tau_0, \tau_1^2, \tau_1 \tau_0^3, \tau_0^6
			\end{array} \\
			\hline
		\end{array}$
		\caption{\small{Odd partitions, distinct (strict) partition and graded polynomials of time-variables $\tau_k$ for small levels 1-6. Odd partitions (OP) have obvious mapping to graded polynomials in time-variables $\tau_k$. Distinct partitions (DP) do not have a clear connection to basis monomials, however their dimensions indeed coincide on each level.}}
		\label{Kontsevich time variables}
	\end{table}
	
	Time-variables $\tau_k$ have natural gradings that are encoded in zero Virasoro operator:
	\begin{align}
		\hat{\mathcal{L}}_0=\sum_{k=0}^{\infty}\left(k+\frac{1}{2} \right)\tau_k\frac{\partial}{\partial \tau_k} + \frac{1}{16}\\
		[\tau_k]=2\left(k+\frac{1}{2} \right)= 2k + 1
	\end{align}
	where for simplicity we rescaled all the gradings by 2.
	
	Similar set of time-variables can be obtained by reduction $p_{2k} \to 0$ of time-variables $p_a$ of GH model from the previous section.
	The set of $\tau_k$ time-variables can be described by diagrams in at least two ways. First option is Young diagrams with odd row lengths -- \textit{odd partitions} (OP). Second option is Young diagrams with different row lengths -- \textit{distinct partitions} (DP). The last diagrams are also called strict partitions in the literature. Examples of these diagrams are collected in Table \ref{Kontsevich time variables}. 
	
	Generating function of number of odd and distinct partitions $P^{OP}(n) = P^{DP}(n)$ \footnote{The idea of proof $P^{OP}(n) = P^{DP}(n)$ is that every natural number can be expressed uniquely as a power of 2 multiplied by an odd number.} on level $n$ has the following form:
	\begin{equation}
		\prod_{k = 1} \frac{1}{1 - q^{2k-1}} = \sum_{n = 0} q^n P^{DP}(n) = 1 + q + q^2 + 2q^3 + 2q^4 + 3 q^5 + 4 q^6 + \ldots
	\end{equation}
	
	Odd partitions are \textit{not suitable} in our method and we provide the following arguments below. Odd partitions have linear one-parametric subset of one-column diagrams $[1^n]$. According to our assumptions (see Section \ref{sec::intro}) the first time-variable $\tau_0$ and its derivative $\frac{\partial}{\partial \tau_0}$ acts by adding and removing a box to corresponding diagrams. In particular, subset of one-column diagrams of OP is closed under operation of removing a box, therefore the following equation holds for corresponding polynomials :
	\begin{equation}
		\frac{\partial}{\partial \tau_0} \text{SIB}^{OP}_{[1^{n}]} = c_{n} \, \text{SIB}^{OP}_{[1^{n-1}]}
	\end{equation}
	with some coefficients $c_n$. On the other hand, the box adding operation also does not output from the subset of one-column diagrams:
	\begin{equation}
		\tau_0 \, \text{SIB}^{OP}_{[1^{n}]} = c^{\prime}_{n} \, \text{SIB}^{OP}_{[1^{n+1}]}
	\end{equation}
	because the only other possible diagram $[2,1^{n-1}]$ contains even number and therefore is not included in odd partition family. Therefore for OP family there is only trivial solution:
	\begin{equation}
		\text{SIB}_{[1^n]}^{OP} \sim \tau_0^n
	\end{equation}
	We do not consider OP diagrams in this paper and focus on distinct partitions (DP).
	\subsection{Step 2. System of equations on single-row polynomials}
	Virasoro operators $\hat{\mathcal{L}}_n$ and variables $\tau_k, \frac{\partial}{\partial \tau_k}$ have the following gradings:
	\begin{table}[h!]
		\centering
		$\begin{array}{|c|c|c|c|c|c|c|c|c|c|}
			\hline
			\text{grading} & \ldots & 3 & 2 & 1 & 0 & -1 & -2 & -3 & \ldots \\
			\hline
			\text{time-variables} & \ldots & \tau_1 & - & \tau_0 & 1 & \frac{\partial}{\partial \tau_0} & - & \frac{\partial}{\partial \tau_1} & \ldots \\
			\hline
			\text{Virasoro operators} &  &  & \hat{\mathcal{L}}_{-1} & - & \hat{\mathcal{L}}_0 & - & \hat{\mathcal{L}}_1 & - & \ldots \\
			\hline
		\end{array}$
	\end{table}  
	
	We assume that single-row subspace $\text{SIB}_{[n]}$ is closed under action of Virasoro operators and derivatives and consider the following system of equations (very similar to \eqref{single row system}):
	\begin{tcolorbox}
		\begin{align}
			\frac{\partial \, \text{SIB}_{[n]}}{\partial \tau_k}  &=  u_{n,k} \, \text{SIB}_{[n-2k-1]} \\
			\hat{\mathcal{L}}_k \,\text{SIB}_{[n]} &=  v_{n,k} \, \text{SIB}_{[n-2k]}  
		\end{align}
	\end{tcolorbox}
	
	This system is naively overdetermined, however it is self-consistent and has a unique solution. We consider general form of SIB polynomials:
	\begin{align}
		\begin{aligned}
			\text{SIB}_{[0]}&=1\\
			\text{SIB}_{[1]}&=\tau_0\\
			\text{SIB}_{[2]}&=\tau_0^2\\
			\text{SIB}_{[3]}&=\frac{2\tau_0^3}{3}+a_{3,1} \, \tau_1\\
			\text{SIB}_{[4]}&=\frac{\tau_0^4}{3}+a_{4,1} \, \tau_1\tau_0\\
			\text{SIB}_{[5]}&=\frac{2\tau_0^5}{15}+a_{5,1} \, \tau_1\tau_0^2+a_{5,2} \, \tau_2 \\
			\text{SIB}_{[6]}&=\frac{2\tau _0^6}{45}+a_{6,1} \, \tau _0^3 \tau _1+ a_{6,2} \, \tau _1^2+ a_{6,3} \, \tau _0 \tau _2 \\
			...
		\end{aligned}
	\end{align}
	We put first coefficient $2^{n-1}/n!$ to fix the overall normalization for future convenience. Solution of the above system is unique and parametrized by one parameter $x$:
	\begin{align}
		\begin{aligned}
			\text{SIB}_{[0]}&=1\\
			\text{SIB}_{[1]}&=\tau_0\\
			\text{SIB}_{[2]}&=\tau_0^2\\
			\text{SIB}_{[3]}&=\frac{2\tau_0^3}{3}+4x \,\tau_1\\
			\text{SIB}_{[4]}&=\frac{\tau_0^4}{3}+8x \, \tau_1\tau_0\\
			\text{SIB}_{[5]}&=\frac{2\tau _0^5}{15}+8x \,  \tau _0^2 \tau _1+16x^2 \, \tau _2\\
			\text{SIB}_{[6]}&=\frac{2\tau _0^6}{45}+\frac{16}{3} x \, \tau _0^3 \tau _1+16 x^2 \, \tau _1^2+ 32x^2 \, \tau _0 \tau _2 \\
			...
		\end{aligned}
	\end{align}
	From the above formulas one can see that $x$ parameter is a trivial rescaling of time-variables $\tau_k \to x^k \tau_k$. The explicit formula for coefficients $u_{n,k}, v_{n,k}$:
	\begin{align}
		u_{n,k} = 2 \cdot (4x)^k \\
		v_{n,k} = \frac{1}{2} \cdot (4x)^{k-1}\left(k+4x(n-2k)\right)
	\end{align}
	According to our observation after formula \eqref{u v coefs} the true normalization of time-variables is restored when power dependence on $k$ is eliminated from $u_{n,k}, v_{n,k}$ coefficient. In this particular case there only one solution $ \boxed{x = \frac{1}{4}} $.
	
	\subsection{Step 3. Hamiltonian for SIB polynomials}
	
	In order to complete the basis we consider the simplest Hamiltonian $\hat{\mathcal{H}}$ for $\text{SIB}_{\lambda}$ polynomials:
	\begin{equation}
		\label{Kontsevich Ham}
		\hat{\mathcal{H}} \, \text{SIB}_{\lambda} = \varepsilon_{\lambda} \, \text{SIB}_{\lambda}
	\end{equation} 
	
	For the time-variable $\tau_k$ the simplest non-trivial Hamiltonian has spin 4: 
	\begin{align}
		\label{Hamiltonian Konts}
		\begin{aligned}
			\hat{\mathcal{H}}=\sum_{a,b,c=0}^{\infty}A_{a,b,c} \, \tau_{a+b+c+1}\frac{\partial^3}{\partial\tau_a\partial\tau_b\partial\tau_c}+B_{a,b,c} \, \tau_a\tau_b\tau_c\frac{\partial}{\partial\tau_{a+b+c+1}}+\\ +\sum_{a+b=c+d}C_{a,b|c,d} \, \tau_a\tau_b\frac{\partial^2}{\partial\tau_c\partial\tau_d}+\sum_{a=0}^{\infty}G_a \, \tau_a\frac{\partial}{\partial\tau_a}
		\end{aligned}
	\end{align}
	
	Coefficients $A_{a,b,c}$ and $B_{a,b,c}$ are absolutely symmetric with respect to permutations of indices $a,b,c$, while coefficients $C_{a,b|c,d}$ have symmetry $a \leftrightarrow b$, $c \leftrightarrow d$. 
	
	It is instructive to consider grading operator:
	\begin{equation}
		\hat{D} = \sum_{k = 0}^{\infty} (2 k + 1) \tau_k \frac{\partial}{\partial \tau_k}
	\end{equation}
	then there are two trivial Hamiltonians $\hat{D}$ and $\hat{D}^2$ among operators of zero grading and spin at most 4. Eigenvalues of these trivial Hamiltonians on single-row polynomials:
	\begin{align}
		\label{gradings D and D2}
		\hat{D} \, \text{SIB}_{[n]} &= n \cdot \text{SIB}_{[n]} \\
		\hat{D}^2 \, \text{SIB}_{[n]} &= n^2 \cdot \text{SIB}_{[n]}
	\end{align} 
	
	We impose eigenfunction condition for single-row polynomials:
	\begin{equation}
		\hat{\mathcal{H}} \, \text{SIB}_{[n]} = \varepsilon_{[n]} \cdot \text{SIB}_{[n]}
	\end{equation}
	and solve corresponding equations. As an output we get eigenvalues $\varepsilon_{[n]}$ and represent them as a functions of $\varepsilon_{[1]}, \varepsilon_{[2]}, \varepsilon_{[3]}$:
	\begin{align}
		\begin{aligned}
			\varepsilon_{[4]}&=4\varepsilon_{[1]}-6\varepsilon_{[2]}+4\varepsilon_{[3]}
			\\ 
			\varepsilon_{[5]}&=15\varepsilon_{[1]}-20\varepsilon_{[2]}+10\varepsilon_{[3]}
			\\ 
			\varepsilon_{[6]}&=36\varepsilon_{[1]}-45\varepsilon_{[2]}+20\varepsilon_{[3]}
			\\ 
			\varepsilon_{[7]}&=70\varepsilon_{[1]}-84\varepsilon_{[2]}+35\varepsilon_{[3]}
			\\
			\varepsilon_{[8]}&=120\varepsilon_{[1]}-140\varepsilon_{[2]}+56\varepsilon_{[3]}
			\\ 
			\varepsilon_{[9]}&=189\varepsilon_{[1]}-216\varepsilon_{[2]}+84\varepsilon_{[3]}
			\\ 
			\varepsilon_{[10]}&=280\varepsilon_{[1]}-315\varepsilon_{[2]}+120\varepsilon_{[3]}
			\\
			\ldots
		\end{aligned}
	\end{align}
	\begin{equation}
		\varepsilon_{[n]}=\frac{n(n-2)(n-3)}{2}\varepsilon_{[1]}-\frac{n(n-1)(n-3)}{2}\varepsilon_{[2]}+\frac{n(n-1)(n-2)}{6}\varepsilon_{[3]},
	\end{equation}
	Three parameters $\varepsilon_{[1]},\varepsilon_{[3]},\varepsilon_{[3]}$ allow one to vary the form of single-row eigenvalues $\varepsilon_{[n]}$ and we argue that there is one distinguished choice of the parameters. Motivated by eigenvalues \eqref{gradings D and D2} we impose:
	\begin{equation}
		\boxed{
			\varepsilon_{[n]} = n^3
		}
	\end{equation}
	Notably, this condition is satisfied when $\varepsilon_{[1]} = 1, \varepsilon_{[2]} = 2^3, \varepsilon_{[3]} = 3^3$. From the above equations on single-row polynomials we find $B_{a,b,c}$ coefficients:
	\begin{align}
		\begin{aligned}
			B_{0,0,0} &= 4 \\
			B_{0,0,1} &= 4 \cdot 3 \\
			B_{0,0,2} &= 4 \cdot 5 \\
			B_{0,1,1} &= 4 \cdot 3^2 \\
			B_{0,0,3} &= 4 \cdot 7 \\
			B_{0,1,2} &= 4 \cdot 3 \cdot 5 \\
			B_{1,1,1} &= 4 \cdot 3^3 \\
			\ldots
		\end{aligned}
	\end{align}
	therefore
	\begin{equation}
		B_{a,b,c} = 4 \cdot (2a + 1)(2b + 1)(2c + 1)
	\end{equation}
	We do not provide explicitely all the other constraints on $A_{a,b,c}$, $C_{a,b|c,d}$, $G_{a}$ coefficients but several examples:
	\begin{align}
		\begin{aligned}
			C_{0,0|0,0} &= 3 \\
			C_{0,1|0,1} &= 9 \\
			2C_{1,1|0,2} + C_{1,1|1,1} &= 81 \\
			C_{0,2|1,1} + 2 C_{0,2|0,2} &= 45 \\
			\ldots \\
		\end{aligned}
		\hspace{10mm}
		\begin{aligned}
			G_{0} &= 1 \\
			A_{0,0,0} &= \frac{1}{4} (27 - G_1) \\
			A_{0,0,1} &= \frac{1}{12}(125 - G_2) \\
			A_{0,0,2} + A_{0,1,1} &= \frac{1}{12}(343 - G_3) \\
			\ldots
		\end{aligned}
	\end{align}
	\subsection{Step 4. Box adding and removing rule}
	
	On the next step we impose box adding/removing rules with eigenfunction property \eqref{Kontsevich Ham} for SIB polynomials up to 5 level. Box adding rules read:
	\begin{align}
		\label{eq 5 level add}
		\begin{aligned}
			\tau_0 \cdot \text{SIB}_{\begin{ytableau}
					\ & \ 
			\end{ytableau}} &= c_{1} \, \text{SIB}_{\begin{ytableau}
					\ & \ & *(pink)
			\end{ytableau}} + c_{2} \, \text{SIB}_{\begin{ytableau}
					\ & \ \\
					*(pink)
			\end{ytableau}} \\
			\tau_0 \cdot \text{SIB}_{\begin{ytableau}
					\ & \ & \
			\end{ytableau}} &= c_{3} \, \text{SIB}_{\begin{ytableau}
					\ & \ & \ & *(pink)
			\end{ytableau}} + c_{4} \, \text{SIB}_{\begin{ytableau}
					\ & \ & \ \\
					*(pink)
			\end{ytableau}} \\
			\tau_0 \cdot \text{SIB}_{\begin{ytableau}
					\ & \ \\
					\
			\end{ytableau}} &= c_{5} \, \text{SIB}_{\begin{ytableau}
					\ & \ & *(pink) \\
					\
			\end{ytableau}} \\
			\tau_0 \cdot \text{SIB}_{\begin{ytableau}
					\ & \ & \ & \
			\end{ytableau}} &= c_{6} \, \text{SIB}_{\begin{ytableau}
					\ & \ & \ & \ & *(pink)
			\end{ytableau}} + c_{7} \, \text{SIB}_{\begin{ytableau}
					\ & \ & \ & \ \\
					*(pink)
			\end{ytableau}} \\
			\tau_0 \cdot \text{SIB}_{\begin{ytableau}
					\ & \ & \ \\
					\
			\end{ytableau}} &= c_{8} \, \text{SIB}_{\begin{ytableau}
					\ & \ & \ & *(pink) \\
					\
			\end{ytableau}} + c_{9} \, \text{SIB}_{\begin{ytableau}
					\ & \ & \  \\
					\ & *(pink)
			\end{ytableau}} \\
		\end{aligned}
	\end{align}
	Box removing rules (we do not mention here single-row polynomials, because they were analyzed earlier):
	\begin{align}
		\label{eq 5 level remove}
		\begin{aligned}
			\frac{\partial}{\partial \tau_0} \, \text{SIB}_{\begin{ytableau}
					\ & \ \\
					\
			\end{ytableau}} &= d_{1} \, \text{SIB}_{\begin{ytableau}
					\ & \ 
			\end{ytableau}} \\
			\frac{\partial}{\partial \tau_0} \, \text{SIB}_{\begin{ytableau}
					\ & \ & \ \\
					\
			\end{ytableau}} &= d_{2} \, \text{SIB}_{\begin{ytableau}
					\ & \ & \  
			\end{ytableau}} + d_{3} \, \text{SIB}_{\begin{ytableau}
					\ & \  \\
					\ 
			\end{ytableau}} \\
			\frac{\partial}{\partial \tau_0} \, \text{SIB}_{\begin{ytableau}
					\ & \ & \ & \ \\
					\
			\end{ytableau}} &= d_{4} \, \text{SIB}_{\begin{ytableau}
					\ & \ & \ & \
			\end{ytableau}} + d_{5} \, \text{SIB}_{\begin{ytableau}
					\ & \ & \ \\
					\ 
			\end{ytableau}} \\
			\frac{\partial}{\partial \tau_0} \, \text{SIB}_{\begin{ytableau}
					\ & \ & \ \\
					\ & \
			\end{ytableau}} &= d_{6} \, \text{SIB}_{\begin{ytableau}
					\ & \ & \ \\ 
					\
			\end{ytableau}}
		\end{aligned}
	\end{align}
	Here $c_i, d_i$ are some constants. Solving the above equations and eigenfunction equations \eqref{Kontsevich Ham} simultaneously with respect to polynomial coeficients we obtain the following solution:
	\begin{align}
		\label{eq 6 level}
		\begin{aligned}
			\text{SIB}_{\begin{ytableau}
					\ & \ \\
					\
			\end{ytableau}} = \frac{\tau _0^3}{3}+ (\gamma-27) \frac{\tau _1}{12} \\
			\text{SIB}_{\begin{ytableau}
					\ & \ & \ \\
					\
			\end{ytableau}} = \frac{\tau _0^4}{3} +(\gamma-27)\frac{\tau _0 \tau _1}{12} \\
			\text{SIB}_{\begin{ytableau}
					\ & \ & \ & \ \\
					\
			\end{ytableau}} = \frac{\tau _0^5}{5}  +(\gamma-15) \frac{\tau _0^2 \tau _1}{20}+(\gamma-39) \frac{\tau _2}{24} \\
			\text{SIB}_{\begin{ytableau}
					\ & \ & \  \\
					\ & \
			\end{ytableau}} = \frac{2 \tau _0^5}{15}+ (\gamma-27) \frac{\tau _0^2 \tau _1}{12}+ (\gamma-39) (\gamma-27) \frac{\tau _2}{288}
		\end{aligned}
	\end{align} 
	where $\gamma := G_1$ is a free parameter that is not fixed from the above mentioned equations. However, if we consider equations from the next level 6:
	\begin{align}
		\begin{aligned}
			\tau_0 \cdot \text{SIB}_{\begin{ytableau}
					\ & \ & \ & \ & \
			\end{ytableau}} &= c_{10} \, \text{SIB}_{\begin{ytableau}
					\ & \ & \ & \ & \ & *(pink)
			\end{ytableau}} + c_{11} \, \text{SIB}_{\begin{ytableau}
					\ & \ & \ & \ & \ \\
					*(pink)
			\end{ytableau}} \\
			\tau_0 \cdot \text{SIB}_{\begin{ytableau}
					\ & \ & \ & \ \\
					\
			\end{ytableau}} &= c_{12} \, \text{SIB}_{\begin{ytableau}
					\ & \ & \ & \ & *(pink) \\
					\
			\end{ytableau}} + c_{13} \, \text{SIB}_{\begin{ytableau}
					\ & \ & \ & \ \\
					\ & *(pink)
			\end{ytableau}} \\
			\tau_0 \cdot \text{SIB}_{\begin{ytableau}
					\ & \ & \  \\
					\ & \
			\end{ytableau}} &= c_{13} \, \text{SIB}_{\begin{ytableau}
					\ & \ & \ & *(pink) \\
					\ & \
			\end{ytableau}} + c_{14} \, \text{SIB}_{\begin{ytableau}
					\ & \ & \ \\
					\ & \ \\
					*(pink)
			\end{ytableau}} \\
			\frac{\partial}{\partial \tau_0} \, \text{SIB}_{\begin{ytableau}
					\ & \ & \ & \ & \ \\
					\
			\end{ytableau}} &= d_{7} \, \text{SIB}_{\begin{ytableau}
					\ & \ & \ & \ & \
			\end{ytableau}} + d_{8} \, \text{SIB}_{\begin{ytableau}
					\ & \ & \ & \ \\
					\ 
			\end{ytableau}} \\
			\frac{\partial}{\partial \tau_0} \, \text{SIB}_{\begin{ytableau}
					\ & \ & \ & \  \\
					\ & \
			\end{ytableau}} &= d_{9} \, \text{SIB}_{\begin{ytableau}
					\ & \ & \ & \ \\
					\
			\end{ytableau}} + d_{10} \, \text{SIB}_{\begin{ytableau}
					\ & \ & \  \\
					\ & \
			\end{ytableau}} \\
			\frac{\partial}{\partial \tau_0} \, \text{SIB}_{\begin{ytableau}
					\ & \ & \  \\
					\ & \ \\
					\
			\end{ytableau}} &= d_{11} \, \text{SIB}_{\begin{ytableau}
					\ & \ & \ \\
					\ & \
			\end{ytableau}} 
		\end{aligned}
	\end{align}
	this parameter is fixed uniqely $\boxed{\gamma = 15}$. We provide expicit formulas for SIB polynomials un to 6 level:
	\begin{align}
		\begin{aligned}
			\text{SIB}_{\begin{ytableau}
					\ 
			\end{ytableau}} &= \tau _0 &\hspace{10mm} \varepsilon_{\begin{ytableau}
					\ 
			\end{ytableau}} &= 1\\
			\text{SIB}_{\begin{ytableau}
					\ & \
			\end{ytableau}} &= \tau _0^2 &\hspace{10mm} \varepsilon_{\begin{ytableau}
					\ & \
			\end{ytableau}} &= 8\\
			\text{SIB}_{\begin{ytableau}
					\ & \ & \
			\end{ytableau}} &= \frac{2 \tau _0^3}{3}+\tau _1 &\hspace{10mm} \varepsilon_{\begin{ytableau}
					\ & \ & \
			\end{ytableau}} &= 27\\
			\text{SIB}_{\begin{ytableau}
					\ & \ \\
					\
			\end{ytableau}} &= \frac{\tau _0^3}{3}-\tau _1 &\hspace{10mm} \varepsilon_{\begin{ytableau}
					\ & \ \\
					\
			\end{ytableau}} &= 9\\
			\text{SIB}_{\begin{ytableau}
					\ & \ & \ & \
			\end{ytableau}} &= \frac{\tau _0^4}{3}+2 \tau _0 \tau _1 &\hspace{10mm} \varepsilon_{\begin{ytableau}
					\ & \ & \ & \
			\end{ytableau}} &= 64\\
			\text{SIB}_{\begin{ytableau}
					\ & \ & \ \\
					\
			\end{ytableau}} &= \frac{\tau _0^4}{3}-\tau _0 \tau _1 &\hspace{10mm} \varepsilon_{\begin{ytableau}
					\ & \ & \ \\
					\
			\end{ytableau}} &= 28\\
			\text{SIB}_{\begin{ytableau}
					\ & \ & \ & \ & \
			\end{ytableau}} &= \frac{2 \tau _0^5}{15}+2 \tau _0^2 \tau _1+\tau _2 &\hspace{10mm} \varepsilon_{\begin{ytableau}
					\ & \ & \ & \ & \
			\end{ytableau}} &= 125\\
			\text{SIB}_{\begin{ytableau}
					\ & \ & \ & \ \\
					\
			\end{ytableau}} &= \frac{\tau _0^5}{5}-\tau _2 &\hspace{10mm} \varepsilon_{\begin{ytableau}
					\ & \ & \ & \ \\
					\
			\end{ytableau}} &= 64 \\
			\text{SIB}_{\begin{ytableau}
					\ & \ & \ \\
					\ & \
			\end{ytableau}} &= \frac{2 \tau _0^5}{15}-\tau _0^2 \tau _1+\tau _2 &\hspace{10mm} \varepsilon_{\begin{ytableau}
					\ & \ & \ \\
					\ & \
			\end{ytableau}} &= 35\\
			\text{SIB}_{\begin{ytableau}
					\ & \ & \ & \ & \ & \
			\end{ytableau}} &= \frac{2 \tau _0^6}{45}+\frac{4}{3} \tau _0^3 \tau _1+\tau _1^2+2 \tau _0 \tau _2 &\hspace{10mm} \varepsilon_{\begin{ytableau}
					\ & \ & \ & \ & \ & \
			\end{ytableau}} &= 216 \\
			\text{SIB}_{\begin{ytableau}
					\ & \ & \ & \ & \ \\
					\
			\end{ytableau}} &= \frac{4 \tau _0^6}{45}+\frac{2}{3} \tau _0^3 \tau _1-\tau _1^2-\tau _0 \tau _2 &\hspace{10mm} \varepsilon_{\begin{ytableau}
					\ & \ & \ & \ & \ \\
					\
			\end{ytableau}} &= 126 \\
			\text{SIB}_{\begin{ytableau}
					\ & \ & \ & \ \\
					\ & \
			\end{ytableau}} &= \frac{\tau _0^6}{9}-\frac{2}{3} \tau _0^3 \tau _1+\tau _1^2 &\hspace{10mm} \varepsilon_{\begin{ytableau}
					\ & \ & \ & \ \\
					\ & \
			\end{ytableau}} &= 72 \\
			\text{SIB}_{\begin{ytableau}
					\ & \ & \ \\
					\ & \ \\
					\
			\end{ytableau}} &= \frac{\tau _0^6}{45}-\frac{1}{3} \tau _0^3 \tau _1-\tau _1^2+\tau _0 \tau _2 &\hspace{10mm} \varepsilon_{\begin{ytableau}
					\ & \ & \ \\
					\ & \ \\
					\
			\end{ytableau}} &= 36\\
		\end{aligned} 
	\end{align}
	Now one can see that SIB polynomials are nothing but well-known Q-Schur polynomials \cite{Macdonald,Mironov:2020tjf}.
	On this limited set of examples we observe remarkable "sum over boxes" property of eigenvalues:
	\begin{equation}
		\label{eigval Kontsevich}
		\varepsilon_{\lambda} = \sum_{\Box \in \lambda} j_{\Box}^3 -(j_{\Box}-1)^3 = \sum_{i = 1}^{l(\lambda)} \lambda_i^3 
	\end{equation}
	Here we denote by $j_{\Box}$ the horizontal coordinate of a box $\Box$ and the first box has coordinate 1. Solving equations like \eqref{eq 5 level add}, \eqref{eq 5 level remove}, \eqref{eq 6 level} for box adding/removing rules and equations \eqref{Kontsevich Ham} one can level by level compute SIB polynomials. From the other hand, eigenfunction conditions \eqref{Kontsevich Ham} fix the form of Hamiltonian coefficients:
	\begin{align}
		\begin{aligned}
			A_{0,0,0} & = 3 \\
			A_{0,0,1} & = 5 \\
			A_{0,0,2} & = 7 \\
			A_{0,1,1} & = 7 \\
			A_{0,0,3} & = 9 \\
			A_{0,1,2} & = 9 \\
			A_{1,1,1} & = 9 \\
			\ldots
		\end{aligned}
		\hspace{10mm}
		\begin{aligned}
			C_{0,0|0,0} & = 3 \\
			C_{0,1|0,1} & = 3 \cdot 3 \\
			C_{0,2|0,2} & = 3 \cdot 5 \\
			C_{0,2|1,1} & = 3 \cdot 5 \\
			C_{1,1|0,2} & = 3 \cdot 3^2 \\
			C_{1,1|1,1} & = 3 \cdot 3^2 
		\end{aligned}
		\hspace{10mm}
		\begin{aligned}
			C_{0,3|0,3} & = 3 \cdot 7 \\
			C_{0,3|1,2} & = 3 \cdot 7 \\
			C_{1,2|0,3} & = 3 \cdot 3 \cdot 5 \\
			C_{1,2|1,2} & = 3 \cdot 3 \cdot 5 \\
			\ldots
		\end{aligned}
		\hspace{10mm}
		\begin{aligned}
			G_0 &= 1 \\
			G_1 &= 15 \\
			G_2 &= 65 \\
			G_3 &= 175 \\
			G_4 &= 369 \\
			G_5 &= 671 \\
			\ldots
		\end{aligned}
	\end{align}
	These values of coefficients are enough to reconstruct the full dependence:
	\begin{align}
		\begin{aligned}
			A_{a,b,c} = 2(a+b+c) + 3 \\
			C_{a,b|c,d} = 3 (2a+1)(2b+1) \\
			G_{a} = (2a+1)(2a^2 + 2a + 1)	
		\end{aligned}
	\end{align}
	Now we are able to write the full Hamiltonian for SIB polynomials:
	\begin{tcolorbox}
		\begin{align}
			\label{Kontsevich Hamiltonian}
			\begin{aligned}
				\hat{\mathcal{H}}=\sum_{a,b,c=0}^{\infty}(2(a+b+c)+3)\tau_{a+b+c+1}\frac{\partial^3}{\partial\tau_a\partial\tau_b\partial\tau_c}+ \sum_{a+b=c+d}3(2a+1)(2b+1)\tau_a\tau_b\frac{\partial^2}{\partial\tau_c\partial\tau_d} \\
				\sum_{a,b,c=0}^{\infty}4(2a+1)(2b+1)(2c+1)\tau_a\tau_b\tau_c\frac{\partial}{\partial\tau_{a+b+c+1}}+ \sum_{a=0}^{\infty}(2 a+1) \left(2 a^2+2 a+1\right)\tau_a\frac{\partial}{\partial\tau_a}
			\end{aligned}
		\end{align}
	\end{tcolorbox}
	One can check that eigenvalues of the above operator are positive integers and obey formula \eqref{eigval Kontsevich}. This Hamiltonian itself does not define SIB basis due to degenerate spectrum, for example $\varepsilon_{[8,7]} = 8^3 + 7^3 = 855 = 9^3 + 5^3 + 1^3 = \varepsilon_{[9,5,1]}$. However, with adding/removing rules the basis is fixed completely.
	
	\subsection{Action of Virasoro operators on $\text{SIB}_{\lambda}$ polynomials}
	Motivated by the action of time-variables and Virasoro operators on SIB polynomials that we discuss in this section it is instructive to slightly change notation of diagrams DP - distinct partitions. For this purpose we represent a Young diagram (turned upside down) as a way of tight packing of square boxes in a corner of a room. In this setting distinct partitions (DP) are in one-to-one correspondence with possible ways of tight packing of square boxes under a ladder. We demonstrate examples of these diagrams, left picture correspond to usual Young diagram, right one correspond to distinct partition:
	\begin{equation}
		\begin{tikzpicture}[scale=0.35]
			\fill[pattern=north east lines] (0,0) rectangle (16,-0.5);
			\fill[pattern=north east lines] (-0.5,-0.5) rectangle (0, 16);
			\draw[line width=0.5pt](0,0)-- (16,0); 
			\draw[line width=0.5pt](0,0)-- (0,16);
			\foreach \i/\j in {0/0, 0/1, 0/2, 0/3, 0/4, 0/5, 0/6, 0/7, 0/8, 0/9, 0/10, 0/11, 0/12, 0/13, 1/0, 1/1, 1/2, 1/3, 1/4, 1/5, 1/6, 1/7, 1/8, 1/9, 1/10, 2/0, 2/1, 2/2, 2/3, 2/4, 2/5, 2/6, 2/7, 2/8, 2/9, 3/0, 3/1, 3/2, 3/3, 3/4, 3/5, 3/6, 3/7, 4/0, 4/1, 4/2, 4/3, 4/4, 4/5, 4/6, 4/7, 5/0, 5/1, 5/2, 5/3, 5/4, 5/5, 5/6, 6/0, 6/1, 6/2, 6/3, 7/0, 7/1, 7/2, 7/3, 8/0, 8/1, 8/2, 9/0, 9/1, 10/0, 10/1, 11/0, 12/0, 13/0}
			{
				\draw[thick] (\i,\j) -- (\i+1,\j) -- (\i+1,\j+1) -- (\i,\j+1) -- (\i,\j);
			}
			\node at (7,-3) {(a) \ $\text{Usual partition}$};
		\end{tikzpicture}
		\hspace{25mm}
		\begin{tikzpicture}[scale=0.35]
			\foreach \i/\j in {0/0, 1/0, 2/0, 3/0, 4/0, 5/0, 6/0, 7/0, 8/0, 9/0, 10/0, 11/0, 12/0, 13/0, 14/0, 1/1, 2/1, 3/1, 4/1, 5/1, 6/1, 7/1, 8/1, 9/1, 10/1, 11/1, 12/1, 2/2, 3/2, 4/2, 5/2, 6/2, 7/2, 8/2, 9/2, 10/2, 11/2, 3/3, 4/3, 5/3, 6/3, 7/3, 8/3, 9/3, 10/3, 11/3, 4/4, 5/4, 6/4, 7/4, 8/4, 5/5, 6/5, 7/5, 6/6, 7/6}
			{
				\draw[thick] (\i,\j) -- (\i+1,\j) -- (\i+1,\j+1) -- (\i,\j+1) -- (\i,\j);
			}
			\fill[pattern=north east lines] (0,0) rectangle (16,-0.5);
			\fill[pattern=north east lines] (-0.5,-0.5) rectangle (0, 1.0);
			\fill[pattern=north east lines] (0.5, 1.0) rectangle (1.0, 2.0);
			\fill[pattern=north east lines] (1.5, 2.0) rectangle (2.0, 3.0);
			\fill[pattern=north east lines] (2.5, 3.0) rectangle (3.0, 4.0);
			\fill[pattern=north east lines] (3.5, 4.0) rectangle (4.0, 5.0);
			\fill[pattern=north east lines] (4.5, 5.0) rectangle (5.0, 6.0);
			\fill[pattern=north east lines] (5.5, 6.0) rectangle (6.0, 7.0);
			\fill[pattern=north east lines] (6.5, 7.0) rectangle (7.0, 8.0);
			\fill[pattern=north east lines] (7.5, 8.0) rectangle (8.0, 9.0);
			\fill[pattern=north east lines] (8.5, 9.0) rectangle (9.0, 10.0);
			\fill[pattern=north east lines] (9.5, 10.0) rectangle (10.0, 11.0);
			\fill[pattern=north east lines] (-0.5, 1.0) rectangle (0.5, 1.5);
			\fill[pattern=north east lines] (0.5, 2.0) rectangle (1.5, 2.5);
			\fill[pattern=north east lines] (1.5, 3.0) rectangle (2.5, 3.5);
			\fill[pattern=north east lines] (2.5, 4.0) rectangle (3.5, 4.5);
			\fill[pattern=north east lines] (3.5, 5.0) rectangle (4.5, 5.5);
			\fill[pattern=north east lines] (4.5, 6.0) rectangle (5.5, 6.5);
			\fill[pattern=north east lines] (5.5, 7.0) rectangle (6.5, 7.5);
			\fill[pattern=north east lines] (6.5, 8.0) rectangle (7.5, 8.5);
			\fill[pattern=north east lines] (7.5, 9.0) rectangle (8.5, 9.5);
			\fill[pattern=north east lines] (8.5, 10.0) rectangle (9.5, 10.5);
			\draw[line width=0.5pt](0,0)-- (16,0); 
			\draw[line width=0.5pt](0,0)-- (0,1);
			\draw[line width=0.5pt](1,1)-- (1,2);
			\draw[line width=0.5pt](2,2)-- (2,3);
			\draw[line width=0.5pt](3,3)-- (3,4);
			\draw[line width=0.5pt](4,4)-- (4,5);
			\draw[line width=0.5pt](5,5)-- (5,6);
			\draw[line width=0.5pt](6,6)-- (6,7);
			\draw[line width=0.5pt](7,7)-- (7,8);
			\draw[line width=0.5pt](8,8)-- (8,9);
			\draw[line width=0.5pt](9,9)-- (9,10);
			\draw[line width=0.5pt](10,10)-- (10,11);
			\draw[line width=0.5pt](0,1)-- (1,1);
			\draw[line width=0.5pt](1,2)-- (2,2);
			\draw[line width=0.5pt](2,3)-- (3,3);
			\draw[line width=0.5pt](3,4)-- (4,4);
			\draw[line width=0.5pt](4,5)-- (5,5);
			\draw[line width=0.5pt](5,6)-- (6,6);
			\draw[line width=0.5pt](6,7)-- (7,7);
			\draw[line width=0.5pt](7,8)-- (8,8);
			\draw[line width=0.5pt](8,9)-- (9,9);
			\draw[line width=0.5pt](9,10)-- (10,10);
			\node at (7,-3) {(a) \ $\text{Distinct partition}$};
		\end{tikzpicture}
		\label{DP example}
	\end{equation}
	
	Form now on we represent distinct partitions as Young diagrams under the ladder. With these notations the action of $\tau_0$ and $\frac{\partial }{\partial \tau_0}$ have the following from:
	\begin{equation}
		\tau_0 \cdot \text{SIB}_{\lambda} = \sum_{\Box \in \text{AddDP}(\lambda)} \text{SIB}_{\lambda + \Box}
	\end{equation}
	\begin{equation}
		\frac{\partial}{\partial \tau_0} \, \text{SIB}_{\lambda} = \sum_{\Box \in \text{RemDP}(\lambda)} (2 - \delta_{i_{\Box}, j_{\Box}}) \cdot \text{SIB}_{\lambda - \Box}
	\end{equation}
	Note, that coefficients in the r.h.s. almost always are units except the diagonal $i_{\Box} = j_{\Box}$. The sets $\text{AddDP}(\lambda)$ and $\text{RemDP}(\lambda)$ are defined in the same way as in the previous section \eqref{Add Rem definition Schur}. We provide explicit example of $\text{AddDP}(\lambda)$ and $\text{RemDP}(\lambda)$:
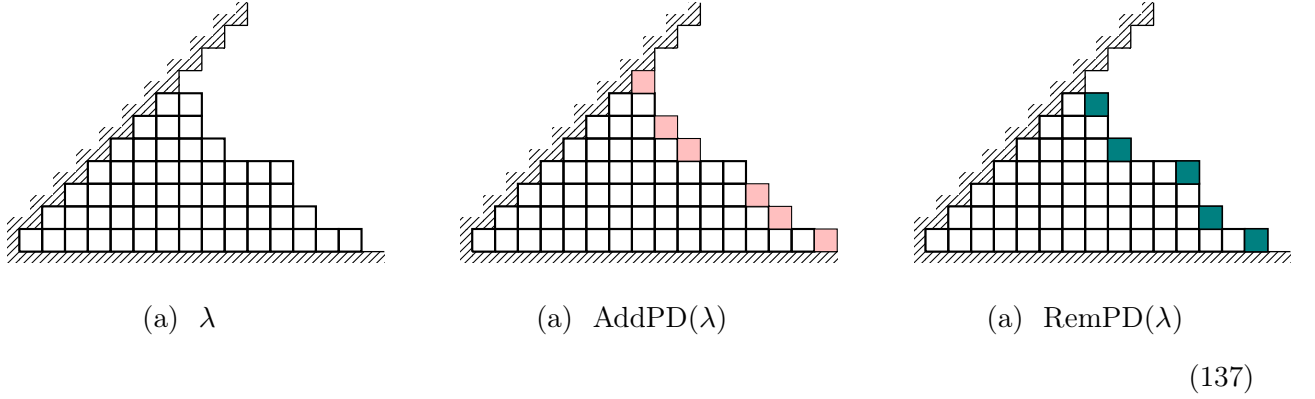
	\begin{equation}
		\begin{tikzpicture}[scale=0.3]
			\foreach \i/\j in {0/0, 1/0, 2/0, 3/0, 4/0, 5/0, 6/0, 7/0, 8/0, 9/0, 10/0, 11/0, 12/0, 13/0, 14/0, 1/1, 2/1, 3/1, 4/1, 5/1, 6/1, 7/1, 8/1, 9/1, 10/1, 11/1, 12/1, 2/2, 3/2, 4/2, 5/2, 6/2, 7/2, 8/2, 9/2, 10/2, 11/2, 3/3, 4/3, 5/3, 6/3, 7/3, 8/3, 9/3, 10/3, 11/3, 4/4, 5/4, 6/4, 7/4, 8/4, 5/5, 6/5, 7/5, 6/6, 7/6}
			{
				\draw[thick] (\i,\j) -- (\i+1,\j) -- (\i+1,\j+1) -- (\i,\j+1) -- (\i,\j);
			}
			\fill[pattern=north east lines] (0,0) rectangle (16,-0.5);
			\fill[pattern=north east lines] (-0.5,-0.5) rectangle (0, 1.0);
			\fill[pattern=north east lines] (0.5, 1.0) rectangle (1.0, 2.0);
			\fill[pattern=north east lines] (1.5, 2.0) rectangle (2.0, 3.0);
			\fill[pattern=north east lines] (2.5, 3.0) rectangle (3.0, 4.0);
			\fill[pattern=north east lines] (3.5, 4.0) rectangle (4.0, 5.0);
			\fill[pattern=north east lines] (4.5, 5.0) rectangle (5.0, 6.0);
			\fill[pattern=north east lines] (5.5, 6.0) rectangle (6.0, 7.0);
			\fill[pattern=north east lines] (6.5, 7.0) rectangle (7.0, 8.0);
			\fill[pattern=north east lines] (7.5, 8.0) rectangle (8.0, 9.0);
			\fill[pattern=north east lines] (8.5, 9.0) rectangle (9.0, 10.0);
			\fill[pattern=north east lines] (9.5, 10.0) rectangle (10.0, 11.0);
			\fill[pattern=north east lines] (-0.5, 1.0) rectangle (0.5, 1.5);
			\fill[pattern=north east lines] (0.5, 2.0) rectangle (1.5, 2.5);
			\fill[pattern=north east lines] (1.5, 3.0) rectangle (2.5, 3.5);
			\fill[pattern=north east lines] (2.5, 4.0) rectangle (3.5, 4.5);
			\fill[pattern=north east lines] (3.5, 5.0) rectangle (4.5, 5.5);
			\fill[pattern=north east lines] (4.5, 6.0) rectangle (5.5, 6.5);
			\fill[pattern=north east lines] (5.5, 7.0) rectangle (6.5, 7.5);
			\fill[pattern=north east lines] (6.5, 8.0) rectangle (7.5, 8.5);
			\fill[pattern=north east lines] (7.5, 9.0) rectangle (8.5, 9.5);
			\fill[pattern=north east lines] (8.5, 10.0) rectangle (9.5, 10.5);
			\draw[line width=0.5pt](0,0)-- (16,0); 
			\draw[line width=0.5pt](0,0)-- (0,1);
			\draw[line width=0.5pt](1,1)-- (1,2);
			\draw[line width=0.5pt](2,2)-- (2,3);
			\draw[line width=0.5pt](3,3)-- (3,4);
			\draw[line width=0.5pt](4,4)-- (4,5);
			\draw[line width=0.5pt](5,5)-- (5,6);
			\draw[line width=0.5pt](6,6)-- (6,7);
			\draw[line width=0.5pt](7,7)-- (7,8);
			\draw[line width=0.5pt](8,8)-- (8,9);
			\draw[line width=0.5pt](9,9)-- (9,10);
			\draw[line width=0.5pt](10,10)-- (10,11);
			\draw[line width=0.5pt](0,1)-- (1,1);
			\draw[line width=0.5pt](1,2)-- (2,2);
			\draw[line width=0.5pt](2,3)-- (3,3);
			\draw[line width=0.5pt](3,4)-- (4,4);
			\draw[line width=0.5pt](4,5)-- (5,5);
			\draw[line width=0.5pt](5,6)-- (6,6);
			\draw[line width=0.5pt](6,7)-- (7,7);
			\draw[line width=0.5pt](7,8)-- (8,8);
			\draw[line width=0.5pt](8,9)-- (9,9);
			\draw[line width=0.5pt](9,10)-- (10,10);
			\node at (7,-3) {(a) \ $\lambda$};
		\end{tikzpicture}
		\hspace{10mm}
		\begin{tikzpicture}[scale=0.3]
			\foreach \i/\j in {0/0, 1/0, 2/0, 3/0, 4/0, 5/0, 6/0, 7/0, 8/0, 9/0, 10/0, 11/0, 12/0, 13/0, 14/0, 1/1, 2/1, 3/1, 4/1, 5/1, 6/1, 7/1, 8/1, 9/1, 10/1, 11/1, 12/1, 2/2, 3/2, 4/2, 5/2, 6/2, 7/2, 8/2, 9/2, 10/2, 11/2, 3/3, 4/3, 5/3, 6/3, 7/3, 8/3, 9/3, 10/3, 11/3, 4/4, 5/4, 6/4, 7/4, 8/4, 5/5, 6/5, 7/5, 6/6, 7/6}
			{
				\draw[thick] (\i,\j) -- (\i+1,\j) -- (\i+1,\j+1) -- (\i,\j+1) -- (\i,\j);
			}
			\fill[pattern=north east lines] (0,0) rectangle (16,-0.5);
			\fill[pattern=north east lines] (-0.5,-0.5) rectangle (0, 1.0);
			\fill[pattern=north east lines] (0.5, 1.0) rectangle (1.0, 2.0);
			\fill[pattern=north east lines] (1.5, 2.0) rectangle (2.0, 3.0);
			\fill[pattern=north east lines] (2.5, 3.0) rectangle (3.0, 4.0);
			\fill[pattern=north east lines] (3.5, 4.0) rectangle (4.0, 5.0);
			\fill[pattern=north east lines] (4.5, 5.0) rectangle (5.0, 6.0);
			\fill[pattern=north east lines] (5.5, 6.0) rectangle (6.0, 7.0);
			\fill[pattern=north east lines] (6.5, 7.0) rectangle (7.0, 8.0);
			\fill[pattern=north east lines] (7.5, 8.0) rectangle (8.0, 9.0);
			\fill[pattern=north east lines] (8.5, 9.0) rectangle (9.0, 10.0);
			\fill[pattern=north east lines] (9.5, 10.0) rectangle (10.0, 11.0);
			\fill[pattern=north east lines] (-0.5, 1.0) rectangle (0.5, 1.5);
			\fill[pattern=north east lines] (0.5, 2.0) rectangle (1.5, 2.5);
			\fill[pattern=north east lines] (1.5, 3.0) rectangle (2.5, 3.5);
			\fill[pattern=north east lines] (2.5, 4.0) rectangle (3.5, 4.5);
			\fill[pattern=north east lines] (3.5, 5.0) rectangle (4.5, 5.5);
			\fill[pattern=north east lines] (4.5, 6.0) rectangle (5.5, 6.5);
			\fill[pattern=north east lines] (5.5, 7.0) rectangle (6.5, 7.5);
			\fill[pattern=north east lines] (6.5, 8.0) rectangle (7.5, 8.5);
			\fill[pattern=north east lines] (7.5, 9.0) rectangle (8.5, 9.5);
			\fill[pattern=north east lines] (8.5, 10.0) rectangle (9.5, 10.5);
			\draw[line width=0.5pt](0,0)-- (16,0); 
			\draw[line width=0.5pt](0,0)-- (0,1);
			\draw[line width=0.5pt](1,1)-- (1,2);
			\draw[line width=0.5pt](2,2)-- (2,3);
			\draw[line width=0.5pt](3,3)-- (3,4);
			\draw[line width=0.5pt](4,4)-- (4,5);
			\draw[line width=0.5pt](5,5)-- (5,6);
			\draw[line width=0.5pt](6,6)-- (6,7);
			\draw[line width=0.5pt](7,7)-- (7,8);
			\draw[line width=0.5pt](8,8)-- (8,9);
			\draw[line width=0.5pt](9,9)-- (9,10);
			\draw[line width=0.5pt](10,10)-- (10,11);
			\draw[line width=0.5pt](0,1)-- (1,1);
			\draw[line width=0.5pt](1,2)-- (2,2);
			\draw[line width=0.5pt](2,3)-- (3,3);
			\draw[line width=0.5pt](3,4)-- (4,4);
			\draw[line width=0.5pt](4,5)-- (5,5);
			\draw[line width=0.5pt](5,6)-- (6,6);
			\draw[line width=0.5pt](6,7)-- (7,7);
			\draw[line width=0.5pt](7,8)-- (8,8);
			\draw[line width=0.5pt](8,9)-- (9,9);
			\draw[line width=0.5pt](9,10)-- (10,10);
			\foreach \x/\y in {7/-8, 8/-6, 9/-5, 12/-3, 13/-2, 15/-1}
			{
				\draw[fill=pink] (\x,-\y) -- (\x+1,-\y) -- (\x+1,-\y-1) -- (\x,-\y-1) -- cycle;
			}
			\node at (7,-3) {(a) \ $\text{AddPD}(\lambda)$};
		\end{tikzpicture}
		\hspace{10mm}
		\begin{tikzpicture}[scale=0.3]
			\foreach \i/\j in {0/0, 1/0, 2/0, 3/0, 4/0, 5/0, 6/0, 7/0, 8/0, 9/0, 10/0, 11/0, 12/0, 13/0, 14/0, 1/1, 2/1, 3/1, 4/1, 5/1, 6/1, 7/1, 8/1, 9/1, 10/1, 11/1, 12/1, 2/2, 3/2, 4/2, 5/2, 6/2, 7/2, 8/2, 9/2, 10/2, 11/2, 3/3, 4/3, 5/3, 6/3, 7/3, 8/3, 9/3, 10/3, 11/3, 4/4, 5/4, 6/4, 7/4, 8/4, 5/5, 6/5, 7/5, 6/6, 7/6}
			{
				\draw[thick] (\i,\j) -- (\i+1,\j) -- (\i+1,\j+1) -- (\i,\j+1) -- (\i,\j);
			}
			\fill[pattern=north east lines] (0,0) rectangle (16,-0.5);
			\fill[pattern=north east lines] (-0.5,-0.5) rectangle (0, 1.0);
			\fill[pattern=north east lines] (0.5, 1.0) rectangle (1.0, 2.0);
			\fill[pattern=north east lines] (1.5, 2.0) rectangle (2.0, 3.0);
			\fill[pattern=north east lines] (2.5, 3.0) rectangle (3.0, 4.0);
			\fill[pattern=north east lines] (3.5, 4.0) rectangle (4.0, 5.0);
			\fill[pattern=north east lines] (4.5, 5.0) rectangle (5.0, 6.0);
			\fill[pattern=north east lines] (5.5, 6.0) rectangle (6.0, 7.0);
			\fill[pattern=north east lines] (6.5, 7.0) rectangle (7.0, 8.0);
			\fill[pattern=north east lines] (7.5, 8.0) rectangle (8.0, 9.0);
			\fill[pattern=north east lines] (8.5, 9.0) rectangle (9.0, 10.0);
			\fill[pattern=north east lines] (9.5, 10.0) rectangle (10.0, 11.0);
			\fill[pattern=north east lines] (-0.5, 1.0) rectangle (0.5, 1.5);
			\fill[pattern=north east lines] (0.5, 2.0) rectangle (1.5, 2.5);
			\fill[pattern=north east lines] (1.5, 3.0) rectangle (2.5, 3.5);
			\fill[pattern=north east lines] (2.5, 4.0) rectangle (3.5, 4.5);
			\fill[pattern=north east lines] (3.5, 5.0) rectangle (4.5, 5.5);
			\fill[pattern=north east lines] (4.5, 6.0) rectangle (5.5, 6.5);
			\fill[pattern=north east lines] (5.5, 7.0) rectangle (6.5, 7.5);
			\fill[pattern=north east lines] (6.5, 8.0) rectangle (7.5, 8.5);
			\fill[pattern=north east lines] (7.5, 9.0) rectangle (8.5, 9.5);
			\fill[pattern=north east lines] (8.5, 10.0) rectangle (9.5, 10.5);
			\draw[line width=0.5pt](0,0)-- (16,0); 
			\draw[line width=0.5pt](0,0)-- (0,1);
			\draw[line width=0.5pt](1,1)-- (1,2);
			\draw[line width=0.5pt](2,2)-- (2,3);
			\draw[line width=0.5pt](3,3)-- (3,4);
			\draw[line width=0.5pt](4,4)-- (4,5);
			\draw[line width=0.5pt](5,5)-- (5,6);
			\draw[line width=0.5pt](6,6)-- (6,7);
			\draw[line width=0.5pt](7,7)-- (7,8);
			\draw[line width=0.5pt](8,8)-- (8,9);
			\draw[line width=0.5pt](9,9)-- (9,10);
			\draw[line width=0.5pt](10,10)-- (10,11);
			\draw[line width=0.5pt](0,1)-- (1,1);
			\draw[line width=0.5pt](1,2)-- (2,2);
			\draw[line width=0.5pt](2,3)-- (3,3);
			\draw[line width=0.5pt](3,4)-- (4,4);
			\draw[line width=0.5pt](4,5)-- (5,5);
			\draw[line width=0.5pt](5,6)-- (6,6);
			\draw[line width=0.5pt](6,7)-- (7,7);
			\draw[line width=0.5pt](7,8)-- (8,8);
			\draw[line width=0.5pt](8,9)-- (9,9);
			\draw[line width=0.5pt](9,10)-- (10,10);
			\foreach \x/\y in {7/-7, 8/-5, 11/-4, 12/-2, 14/-1}
			{
				\draw[fill=teal] (\x,-\y) -- (\x+1,-\y) -- (\x+1,-\y-1) -- (\x,-\y-1) -- cycle;
			}
			\node at (7,-3) {(a) \ $\text{RemPD}(\lambda)$};
		\end{tikzpicture}
	\end{equation} 
	
	Derivatives with respect to higher time-variables $\tau_k$ act on superintegrable polynomials via the following rules:
	\begin{equation}
		\label{Konts time derivative}
		\frac{\partial}{\partial \tau_k} \text{SIB}_{\lambda} = \sum_{\substack{ \text{bs} \in \text{BS}^{*}(\lambda) \\ |\text{bs}| = 2k+1}} (-1)^{\frac{|\text{bs}^{(2)}|}{2} + h(\text{bs}^{(1)}) + 1} \cdot (2 - \delta_{i_{\Box_{\text{bs}}}, j_{\Box_{\text{bs}}}}) \cdot \text{SIB}_{\lambda - \text{bs}}
	\end{equation}
	Where $\text{BS}^{*}(\lambda)$ is a set of \textit{double strips} \cite{Macdonald}. A double strip is a skew diagram (i.e. it can be represented as $\lambda/\mu$, where $\mu \subset \lambda$, $\lambda, \mu$ are distinct partitions), that is connected and has width one or two. Subdiagram of width two always touches on the ladder-border. A double strip can be a usual border strip if it has width one everywhere. We demostrate two examples of double strips of the same distinct partitions:
	
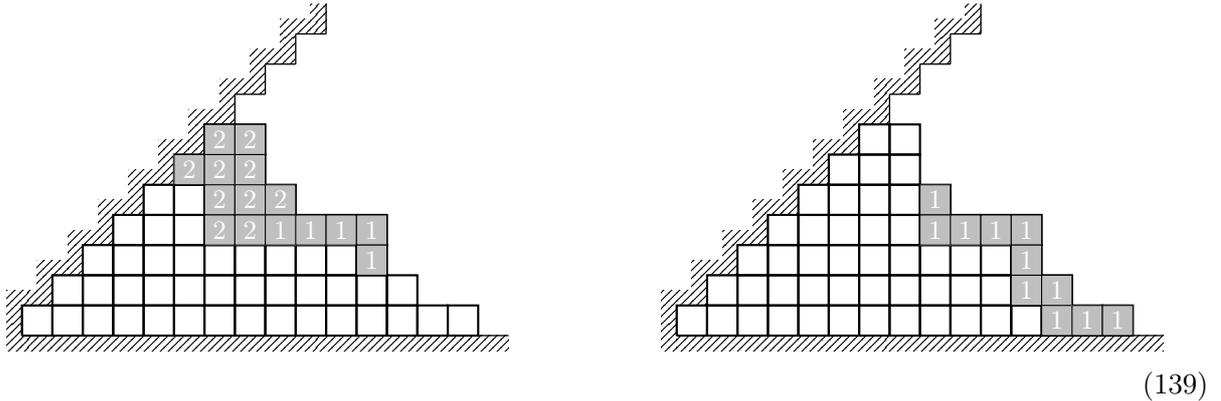
\begin{equation}
		\begin{tikzpicture}[scale=0.4]
			\foreach \i/\j in {0/0, 1/0, 2/0, 3/0, 4/0, 5/0, 6/0, 7/0, 8/0, 9/0, 10/0, 11/0, 12/0, 13/0, 14/0, 1/1, 2/1, 3/1, 4/1, 5/1, 6/1, 7/1, 8/1, 9/1, 10/1, 11/1, 12/1, 2/2, 3/2, 4/2, 5/2, 6/2, 7/2, 8/2, 9/2, 10/2, 11/2, 3/3, 4/3, 5/3, 6/3, 7/3, 8/3, 9/3, 10/3, 11/3, 4/4, 5/4, 6/4, 7/4, 8/4, 5/5, 6/5, 7/5, 6/6, 7/6}
			{
				\draw[thick] (\i,\j) -- (\i+1,\j) -- (\i+1,\j+1) -- (\i,\j+1) -- (\i,\j);
			}
			\fill[pattern=north east lines] (0,0) rectangle (16,-0.5);
			\fill[pattern=north east lines] (-0.5,-0.5) rectangle (0, 1.0);
			\fill[pattern=north east lines] (0.5, 1.0) rectangle (1.0, 2.0);
			\fill[pattern=north east lines] (1.5, 2.0) rectangle (2.0, 3.0);
			\fill[pattern=north east lines] (2.5, 3.0) rectangle (3.0, 4.0);
			\fill[pattern=north east lines] (3.5, 4.0) rectangle (4.0, 5.0);
			\fill[pattern=north east lines] (4.5, 5.0) rectangle (5.0, 6.0);
			\fill[pattern=north east lines] (5.5, 6.0) rectangle (6.0, 7.0);
			\fill[pattern=north east lines] (6.5, 7.0) rectangle (7.0, 8.0);
			\fill[pattern=north east lines] (7.5, 8.0) rectangle (8.0, 9.0);
			\fill[pattern=north east lines] (8.5, 9.0) rectangle (9.0, 10.0);
			\fill[pattern=north east lines] (9.5, 10.0) rectangle (10.0, 11.0);
			\fill[pattern=north east lines] (-0.5, 1.0) rectangle (0.5, 1.5);
			\fill[pattern=north east lines] (0.5, 2.0) rectangle (1.5, 2.5);
			\fill[pattern=north east lines] (1.5, 3.0) rectangle (2.5, 3.5);
			\fill[pattern=north east lines] (2.5, 4.0) rectangle (3.5, 4.5);
			\fill[pattern=north east lines] (3.5, 5.0) rectangle (4.5, 5.5);
			\fill[pattern=north east lines] (4.5, 6.0) rectangle (5.5, 6.5);
			\fill[pattern=north east lines] (5.5, 7.0) rectangle (6.5, 7.5);
			\fill[pattern=north east lines] (6.5, 8.0) rectangle (7.5, 8.5);
			\fill[pattern=north east lines] (7.5, 9.0) rectangle (8.5, 9.5);
			\fill[pattern=north east lines] (8.5, 10.0) rectangle (9.5, 10.5);
			\draw[line width=0.5pt](0,0)-- (16,0); 
			\draw[line width=0.5pt](0,0)-- (0,1);
			\draw[line width=0.5pt](1,1)-- (1,2);
			\draw[line width=0.5pt](2,2)-- (2,3);
			\draw[line width=0.5pt](3,3)-- (3,4);
			\draw[line width=0.5pt](4,4)-- (4,5);
			\draw[line width=0.5pt](5,5)-- (5,6);
			\draw[line width=0.5pt](6,6)-- (6,7);
			\draw[line width=0.5pt](7,7)-- (7,8);
			\draw[line width=0.5pt](8,8)-- (8,9);
			\draw[line width=0.5pt](9,9)-- (9,10);
			\draw[line width=0.5pt](10,10)-- (10,11);
			\draw[line width=0.5pt](0,1)-- (1,1);
			\draw[line width=0.5pt](1,2)-- (2,2);
			\draw[line width=0.5pt](2,3)-- (3,3);
			\draw[line width=0.5pt](3,4)-- (4,4);
			\draw[line width=0.5pt](4,5)-- (5,5);
			\draw[line width=0.5pt](5,6)-- (6,6);
			\draw[line width=0.5pt](6,7)-- (7,7);
			\draw[line width=0.5pt](7,8)-- (8,8);
			\draw[line width=0.5pt](8,9)-- (9,9);
			\draw[line width=0.5pt](9,10)-- (10,10);
			\foreach \x/\y in {5/-6, 6/-7, 6/-6, 7/-7, 6/-5, 7/-6, 6/-4, 7/-5, 7/-4, 8/-5, 8/-4, 9/-4, 10/-4, 11/-4, 11/-3}
			{
				\draw[fill=lightgray] (\x,-\y) -- (\x+1,-\y) -- (\x+1,-\y-1) -- (\x,-\y-1) -- cycle;
			}
			\foreach \x/\y in {8/-4, 9/-4, 10/-4, 11/-4, 11/-3}
			{
				\node at (\x+0.5,-\y-0.5) { \textcolor{white}{\small{1}}};
			}
			\foreach \x/\y in {5/-6, 6/-7, 6/-6, 7/-7, 6/-5, 7/-6, 6/-4, 7/-5, 7/-4, 8/-5}
			{
				\node at (\x+0.5,-\y-0.5) { \textcolor{white}{\small{2}}};
			}
		\end{tikzpicture}
		\hspace{20mm}
		\begin{tikzpicture}[scale=0.4]
			\foreach \i/\j in {0/0, 1/0, 2/0, 3/0, 4/0, 5/0, 6/0, 7/0, 8/0, 9/0, 10/0, 11/0, 12/0, 13/0, 14/0, 1/1, 2/1, 3/1, 4/1, 5/1, 6/1, 7/1, 8/1, 9/1, 10/1, 11/1, 12/1, 2/2, 3/2, 4/2, 5/2, 6/2, 7/2, 8/2, 9/2, 10/2, 11/2, 3/3, 4/3, 5/3, 6/3, 7/3, 8/3, 9/3, 10/3, 11/3, 4/4, 5/4, 6/4, 7/4, 8/4, 5/5, 6/5, 7/5, 6/6, 7/6}
			{
				\draw[thick] (\i,\j) -- (\i+1,\j) -- (\i+1,\j+1) -- (\i,\j+1) -- (\i,\j);
			}
			\fill[pattern=north east lines] (0,0) rectangle (16,-0.5);
			\fill[pattern=north east lines] (-0.5,-0.5) rectangle (0, 1.0);
			\fill[pattern=north east lines] (0.5, 1.0) rectangle (1.0, 2.0);
			\fill[pattern=north east lines] (1.5, 2.0) rectangle (2.0, 3.0);
			\fill[pattern=north east lines] (2.5, 3.0) rectangle (3.0, 4.0);
			\fill[pattern=north east lines] (3.5, 4.0) rectangle (4.0, 5.0);
			\fill[pattern=north east lines] (4.5, 5.0) rectangle (5.0, 6.0);
			\fill[pattern=north east lines] (5.5, 6.0) rectangle (6.0, 7.0);
			\fill[pattern=north east lines] (6.5, 7.0) rectangle (7.0, 8.0);
			\fill[pattern=north east lines] (7.5, 8.0) rectangle (8.0, 9.0);
			\fill[pattern=north east lines] (8.5, 9.0) rectangle (9.0, 10.0);
			\fill[pattern=north east lines] (9.5, 10.0) rectangle (10.0, 11.0);
			\fill[pattern=north east lines] (-0.5, 1.0) rectangle (0.5, 1.5);
			\fill[pattern=north east lines] (0.5, 2.0) rectangle (1.5, 2.5);
			\fill[pattern=north east lines] (1.5, 3.0) rectangle (2.5, 3.5);
			\fill[pattern=north east lines] (2.5, 4.0) rectangle (3.5, 4.5);
			\fill[pattern=north east lines] (3.5, 5.0) rectangle (4.5, 5.5);
			\fill[pattern=north east lines] (4.5, 6.0) rectangle (5.5, 6.5);
			\fill[pattern=north east lines] (5.5, 7.0) rectangle (6.5, 7.5);
			\fill[pattern=north east lines] (6.5, 8.0) rectangle (7.5, 8.5);
			\fill[pattern=north east lines] (7.5, 9.0) rectangle (8.5, 9.5);
			\fill[pattern=north east lines] (8.5, 10.0) rectangle (9.5, 10.5);
			\draw[line width=0.5pt](0,0)-- (16,0); 
			\draw[line width=0.5pt](0,0)-- (0,1);
			\draw[line width=0.5pt](1,1)-- (1,2);
			\draw[line width=0.5pt](2,2)-- (2,3);
			\draw[line width=0.5pt](3,3)-- (3,4);
			\draw[line width=0.5pt](4,4)-- (4,5);
			\draw[line width=0.5pt](5,5)-- (5,6);
			\draw[line width=0.5pt](6,6)-- (6,7);
			\draw[line width=0.5pt](7,7)-- (7,8);
			\draw[line width=0.5pt](8,8)-- (8,9);
			\draw[line width=0.5pt](9,9)-- (9,10);
			\draw[line width=0.5pt](10,10)-- (10,11);
			\draw[line width=0.5pt](0,1)-- (1,1);
			\draw[line width=0.5pt](1,2)-- (2,2);
			\draw[line width=0.5pt](2,3)-- (3,3);
			\draw[line width=0.5pt](3,4)-- (4,4);
			\draw[line width=0.5pt](4,5)-- (5,5);
			\draw[line width=0.5pt](5,6)-- (6,6);
			\draw[line width=0.5pt](6,7)-- (7,7);
			\draw[line width=0.5pt](7,8)-- (8,8);
			\draw[line width=0.5pt](8,9)-- (9,9);
			\draw[line width=0.5pt](9,10)-- (10,10);
			\foreach \x/\y in {8/-5, 8/-4, 9/-4, 10/-4, 11/-4, 11/-3, 11/-2, 12/-2, 12/-1, 13/-1, 14/-1}
			{
				\draw[fill=lightgray] (\x,-\y) -- (\x+1,-\y) -- (\x+1,-\y-1) -- (\x,-\y-1) -- cycle;
			}
			\foreach \x/\y in {8/-5, 8/-4, 9/-4, 10/-4, 11/-4, 11/-3, 11/-2, 12/-2, 12/-1, 13/-1, 14/-1}
			{
				\node at (\x+0.5,-\y-0.5) { \textcolor{white}{\small{1}}};
			}
		\end{tikzpicture}
	\end{equation}
	On the above diagrams we explicitly mark boxes of subset of unit width $\text{bs}^{(1)}$ with $1$ and boxes of subset $\text{bs}^{(2)}$ (width $2$) with $2$. We denote as $|\text{bs}^{(2)}|$ the number of boxes is subset of width 2. While $h(\text{bs}^{(1)})$ is a height of unit width subset. In our notation $i_{\Box_{\text{bs}}}, j_{\Box_{\text{bs}}}$ are coordinates of the head of the strip, i.e. the box for which $j_{\Box} - i_{\Box}$ has minimal value (for essentially double strip there are two such boxes, however there is no ambiguity because for both of them $\delta_{i_{\Box_{\text{bs}}},j_{\Box_{\text{bs}}}} = 1$). We provide a representative example of $\frac{\partial}{\partial \tau_2}$ action:
	\begin{equation}
		\frac{\partial}{\partial \tau_2}\text{SIB}_{\negthickspace \negthickspace \negthickspace \negthickspace \begin{tikzpicture}[scale=0.2]
				\foreach \i/\j in {0/0, 1/0, 2/0, 3/0, 4/0, 5/0, 1/1, 2/1, 3/1, 4/1, 5/1, 2/2, 3/2, 4/2, 3/3, 4/3}
				{
					\draw[thick] (\i,\j) -- (\i+1,\j) -- (\i+1,\j+1) -- (\i,\j+1) -- (\i,\j);
				}
		\end{tikzpicture}} = \text{SIB}_{ \negthickspace \begin{tikzpicture}[scale=0.2]
				\foreach \i/\j in {0/0, 1/0, 2/0, 3/0, 4/0, 5/0, 1/1, 2/1, 3/1, 4/1, 5/1}
				{
					\draw[thick] (\i,\j) -- (\i+1,\j) -- (\i+1,\j+1) -- (\i,\j+1) -- (\i,\j);
				}
		\end{tikzpicture}} + \text{SIB}_{ \negthickspace \negthickspace \negthickspace \begin{tikzpicture}[scale=0.2]
				\foreach \i/\j in {0/0, 1/0, 2/0, 3/0, 4/0, 5/0, 1/1, 2/1, 3/1, 2/2, 3/2}
				{
					\draw[thick] (\i,\j) -- (\i+1,\j) -- (\i+1,\j+1) -- (\i,\j+1) -- (\i,\j);
				}
		\end{tikzpicture}} - 2 \cdot \text{SIB}_{\negthickspace \negthickspace \negthickspace \negthickspace \begin{tikzpicture}[scale=0.2]
				\foreach \i/\j in {0/0, 1/0, 2/0, 3/0, 4/0, 1/1, 2/1, 3/1, 2/2, 3/2, 3/3}
				{
					\draw[thick] (\i,\j) -- (\i+1,\j) -- (\i+1,\j+1) -- (\i,\j+1) -- (\i,\j);
				}
		\end{tikzpicture}}
	\end{equation} 
	
	Virasoro operators $\hat{\mathcal{L}}_n$ ($n>0$) act by removing double strips as well as time-derivatives \eqref{Konts time derivative}, however in this case coefficients are not constants:
	\begin{tcolorbox}
		\begin{equation}
			\label{Konts Virasoro action}
			\hat{\mathcal{L}}_{n} \, \text{SIB}_{\lambda} = \frac{1}{4} \cdot \sum_{\substack{ \text{bs} \in \text{BS}^{*}(\lambda) \\ |\text{bs}| = 2k+1}} (-1)^{\frac{|\text{bs}^{(2)}|}{2} + h(\text{bs}^{(1)}) + 1} \cdot (2 - \delta_{i_{\Box_{\text{bs}}}, j_{\Box_{\text{bs}}}}) \cdot \left( k + j_{\Box_{\text{bs}}} - i_{\Box_{\text{bs}}} - \frac{|\text{bs}^{(2)}|}{2}\right) \cdot \text{SIB}_{\lambda - \text{bs}}
		\end{equation}  
	\end{tcolorbox}
	Note that the box removing rules for the above Virasoro operators and corresponding time-derivatives \eqref{Konts time derivative} are the same. While in case of GH model the box removing rules are different for time-derivatives \eqref{GH model time derivatives} and Virasoro operators \eqref{GH model Virasoro action} (in case of Virasoro operators there is an additional sum over two distinct border strips).

	\section{Conclusion}
	\label{sec::conclusion}
	In this paper we made a trial of systematic construction of superintegrable basis (SIB) in matrix models for particular examples of Gaussian Hermitian and cubic Kontsevich models. Our construction procedure is based on the main conjecture: \textit{superintegrable formulas are manifestations of underlying "hidden symmetry" algebras}. 
	
	Motivated by the example of Gaussian Hermitian model, which symmetry algebra $W_{1 + \infty}$ is well-known, we formulate a (conjectural) set of assumptions, that may allow one to reconstruct superintegrable polynomials in other models. These assumptions concern the shape of Young-like diagrams that enumerate SIB polynomials, Pierri rules for adding and removing boxes in these diagrams and existence of Hamiltonian operator that encodes SIB polynomials as its own eigenfunctions. 
	
	Our method represents first steps towards systematic construction of SIB polynomials. We applied the method for models where SIB polynomials were already known, while the true success of the method would come after a discovery of SIB polynomials in more complicated models (e.g. generalized Kontsevich models \cite{Kharchev:1991cu, Mironov:2021lbx, Morozov:2020ccp} or super-eigenvalue models \cite{Alvarez-Gaume:1991vno, Akemann:1997ca, McArthur:1993hw, Ciosmak:2016wpx, Wang:2020hko }). 
	
	We list here disadvantages of our method and possible modifications:
	\begin{itemize}
		\item In our method we used box adding/removing rules only for the simplest time-variables and higher grading operators only for single-row sector. However, as a result we observe that all time-derivatives \eqref{GH model time derivatives}, \eqref{Konts time derivative} and Virasoro operators \eqref{GH model Virasoro action}, \eqref{Konts Virasoro action} remove boxes via special border strip rules for all diagrams. 
		
		In case of more complicated models the above simple reasoning may not allow to completely fix the form of SIB polynomials. It may be useful to impose box removing rules for all time-derivatives and Virasoro operators at the very beginning.
		
		\item In our method we impose very special condition of finite spin for Hamiltonian operator \eqref{Hamiltonian HGM}, \eqref{Hamiltonian Konts}. It may be wrong for other cases, and the finite spin condition may be replaced with more universal one (e.g. self-adjoint property with respect to proper scalar product, "sum over box" property of eigenvalues, etc.).
	\end{itemize}

	\section*{Acknowledgments}
	
	We are indebted for illuminating discussions to D.Galakhov, A.Mironov, A.Morozov, A.Popolitov, A.Sleptsov. 
	
	The work was partially funded within the state assignment of the Institute for Information Transmission Problems of RAS (N.T.). Our work is partly supported by the grants of the Foundation for the Advancement of Theoretical Physics and Mathematics ``BASIS'' (N.T.).

	\bibliographystyle{utphys}
	\bibliography{biblio}
	
	\section*{Appendix A. Another approach.}
	
	In both Hermitian and Kontsevich cases the resulting Hamiltonians have a very special property -- the eigenvalue $\kappa_{\lambda}$ corresponding to the diagram $\lambda$ is given by sum over the all boxes:
	\begin{equation}
		\kappa_{\lambda} = \sum_{\Box \in \lambda} \Omega_{\Box} 
	\end{equation}
	where $\Omega_{\Box}$ depends on the coordinates of the box (see \eqref{Ham eigvals} and \eqref{eigval Kontsevich}). Following an algorithm from \cite{Galakhov:2025phf} we can construct infinite set of commuting operators, starting from our Hamiltonians \eqref{Schur Hamiltonian} and \eqref{Kontsevich Hamiltonian}. We briefly remind the algorithm here. On the first step we consider two operators:
	\begin{equation}
		\hat{E}_{0} \, \text{SIB}_{\lambda} = \sum_{\Box \in \text{Add}(\lambda)} C_{\lambda, \lambda + \Box} \, \text{SIB}_{\lambda + \Box}
	\end{equation}
	\begin{equation}
		\hat{F}_{0} \, \text{SIB}_{\lambda} = \sum_{\Box \in \text{Rem}(\lambda)} \tilde{C}_{\lambda, \lambda - \Box} \, \text{SIB}_{\lambda - \Box}
	\end{equation}
	In our cases the role of $\hat{E}_{0},\hat{F}_{0}$ operators play the smallest time variable and corresponding derivative $p_1, \frac{\partial}{\partial p_1}$ and $\tau_0, \frac{\partial}{\partial \tau_0}$. On the second step we consider a tower auxiliary operators:
	\begin{align}
		\begin{aligned}
			\hat{E}_{0} &\hspace{15mm} \hat{F}_{0} \\
			\hat{E}_{1} = \Big[ \hat{H},\hat{E}_{0} \Big] &\hspace{15mm} \hat{F}_{1} = \Big[ \hat{F}_{0}, \hat{H} \Big] \\
			\hat{E}_{2} = \Big[ \hat{H}, \Big[ \hat{H},\hat{E}_{0} \Big]\Big] &\hspace{15mm} \hat{F}_{2} = \Big[\Big[ \hat{F}_{0}, \hat{H} \Big], \hat{H} \Big] \\
			\ldots &\hspace{15mm} \ldots
		\end{aligned}
	\end{align}
	that have the following action on polynomials:
	\begin{equation}
		\hat{E}_{k} \, \text{SIB}_{\lambda} = \sum_{\Box \in \text{Add}(\lambda)} \left( \Omega_{\Box}  \right)^{k} \cdot C_{\lambda, \lambda + \Box} \, \text{SIB}_{\lambda + \Box}
	\end{equation}
	\begin{equation}
		\hat{F}_{k} \, \text{SIB}_{\lambda} = \sum_{\Box \in \text{Rem}(\lambda)} \left( \Omega_{\Box}  \right)^{k} \cdot \tilde{C}_{\lambda, \lambda - \Box} \, \text{SIB}_{\lambda - \Box}
	\end{equation}
	On the final step we consider the following operators:
	\begin{equation}
		\hat{\mathcal{H}}_{a+b} := \Big[ \hat{E}_{a}, \hat{F}_{b} \Big]
	\end{equation}
	These operators have remarkable properties -- they depend only on the sum of indices $a+b$, form commuting family and commute with starting Hamiltonian $\hat{H}$:
	\begin{equation}
		\Big[ \hat{\mathcal{H}}_{a}, \hat{\mathcal{H}}_{b} \Big] = 0 \hspace{15mm} \Big[ \hat{\mathcal{H}}_{a}, \hat{H} \Big] = 0
	\end{equation}
	Therefore, our families of polynomials $\text{SIB}_{\lambda}$ are their eigenfunctions:
	\begin{equation}
		\hat{\mathcal{H}}_{a} \, \text{SIB}_{\lambda} = \mathcal{E}_{\lambda,a} \cdot \text{SIB}_{\lambda}
	\end{equation}
	Basing on the above reasoning we propose alternative approach to obtain $\text{SIB}_{\lambda}$ polynomials in Hermitian and Kontsevich cases:
	\begin{itemize}
		\item We start from fixed-spin ansatz for Hamiltonian \eqref{Hamiltonian HGM} and \eqref{Hamiltonian Konts}, that depend on unknown coefficients $(A_{a,b}, B_{a,b})$ and $(A_{a,b,c}, B_{a,b,c}, C_{a,b,c,d},G_{a})$ respectively.
		\item Using ansatz for Hamiltonians we compute auxiliary operators $\hat{E}_{k}$, $\hat{F}_{k}$ and then $\hat{\mathcal{H}}_{a,b}$.
		\item We impose the following conditions:
		\begin{equation}
			\hat{\mathcal{H}}_{a,b} = \hat{\mathcal{H}}_{a+b}
		\end{equation}
		\begin{equation}
			\Big[ \hat{\mathcal{H}}_{a}, \hat{H} \Big] = 0
		\end{equation}
		\begin{equation}
			\Big[ \hat{\mathcal{H}}_{a}, \hat{\mathcal{H}}_{b} \Big] = 0 
		\end{equation}
		that determine special relations on coefficients $(A_{a,b}, B_{a,b})$ and $(A_{a,b,c}, B_{a,b,c}, C_{a,b,c,d},G_{a})$. 
		\item We observe that these nonlinear system of equations have unique solution up to rescaling of time variables. SIB polynomials are restored up to overall scaling as a common set of eigenfunctions.
	\end{itemize}
\end{document}